\documentclass[sn-mathphys,Numbered]{sn-jnl} % Math and Physical Sciences Reference Style
\usepackage{fix-cm} % Fix size substitutions with differences

\usepackage{multirow}%
\usepackage{amsmath,amssymb,amsfonts}%
\usepackage[title]{appendix}%
\usepackage{xcolor}%
\usepackage{manyfoot}%
\usepackage{booktabs}%
\usepackage{longtable}%
\usepackage{graphicx} % Required for inserting images
\usepackage[version=3]{mhchem}
\usepackage{csquotes}
\usepackage{amsmath}	% Advanced maths commands
\usepackage{siunitx}
\usepackage{hyperref}
\usepackage{array}
\usepackage{ragged2e} % Provides \RaggedRight
\hypersetup{
colorlinks=true,
allcolors=blue,
}
\usepackage[percent]{overpic}
\usepackage{pdflscape}
    
\begin{document}

\title{Exploring Synergies between Twinkle and Ariel: a Pilot Study}

\author*[  1,2]{\fnm{Andrea} \sur{Bocchieri}\footnotemark[2]}\email{andrea.bocchieri@uniroma1.it}%\thanks{These authors contributed equally to this work}

\author[  3]{\fnm{Luke} \sur{Booth}\footnote[2]{Equal contribution}}

\author[3,4,1,5]{\fnm{Lorenzo V.} \sur{Mugnai}}

\affil*[1]{\orgdiv{Dipartimento di Fisica}, \orgname{La Sapienza Università di Roma}, \orgaddress{\street{Piazzale Aldo Moro 5}, \city{Roma}, \postcode{00185}, \country{Italy}}}

\affil*[2]{\orgdiv{INAF}, \orgname{Osservatorio Astrofisico di Arcetri}, \orgaddress{\street{Largo Enrico Fermi 5}, \city{Firenze}, \postcode{50125}, \country{Italy}}}

\affil[3]{\orgdiv{School of Physics and Astronomy}, \orgname{Cardiff University}, \orgaddress{\street{Queens Buildings, The Parade}, \city{Cardiff}, \postcode{CF24 3AA}, \country{UK}}}

\affil[4]{\orgdiv{Department of Physics and Astronomy}, \orgname{University College London}, \orgaddress{\street{Gower Street}, \city{London}, \postcode{WC1E 6BT}, \country{UK}}}

\affil[5]{\orgdiv{INAF}, \orgname{Osservatorio Astronomico di Palermo}, \orgaddress{\street{Piazza del Parlamento 1}, \city{Palermo}, \postcode{I-90134}, \country{Italy}}}

\date{Received Month XX, YYYY; accepted Month XX, YYYY}

\abstract{Launching in 2027 and 2029, respectively, \textit{Twinkle} and \textit{Ariel} will conduct the first large-scale homogeneous spectroscopic surveys of the atmospheres of hundreds of diverse exoplanets around a range of host star types for statistical understanding. This will fundamentally transition the field to an era of population-level characterisation. 
In this pilot study, we aim to explore possible synergies between \textit{Twinkle} and \textit{Ariel} to determine for instance whether prior \textit{Twinkle} observations can substantially inform the target selection and observing strategy of \textit{Ariel}. 
This study primarily aims to encourage further investigation by both consortium communities by showing what a potential scientific synergy would look like on a promising scientific case that requires further exploration. 
For this aim, we select a small subset of `cool' planets that are also particularly well-suited to be observed by \textit{Twinkle} and therefore \textit{Ariel}. By using representative noise estimates for both missions, we compute the number of visits required for an observation. Then, we simulate and retrieve transmission spectra of each target, assuming gaseous, H$_2$/He-dominated atmospheres and various atmospheric models to test different scenarios. 
For all candidates, we find that atmospheric parameters are generally retrieved well within 1--$\sigma$ to input values, with \textit{Ariel} typically achieving tighter constraints.  
We also find that retrieved values may depend on the tier when \textit{Ariel} can achieve Tier~3 in a single visit, due to the information loss that may occur in binning.
We demonstrate that for a small subset of cool gaseous planets, exploitable synergies exist between \textit{Twinkle} and \textit{Ariel} observations and \textit{Twinkle} may very well provide a vantage point to plan \textit{Ariel} observations. The true extent of the potential synergies, far beyond our considered sample, will be determined by the final target lists. 
Once Twinkle is operational and its performance is known, it could reliably inform Ariel’s target prioritization and Ariel's capabilities which are already well-established can help define optimal targets and observational approaches for Twinkle. Therefore, further exploration of potential synergies is highly warranted especially after Twinkle is operational and Ariel's launch date approaches.}


\maketitle

\section{Introduction}
\label{sec:introduction}
First detected in the early and mid-1990s, exoplanets are now known to be commonplace and a ubiquitous outcome of star formation, with current estimates suggesting that on average a star hosts at least one planet~\citep{Cassan2012, Populations_Batalha_2014PNAS}. With over 5400 confirmed detections around host stars that vary from giants to dwarfs, binaries to stellar remnants, the field of exoplanet science is continually growing. Many thousands of candidate planets await validation via follow-up observations and statistical methods, whilst new ground- and space-based observatories such as PLATO~\citep{Rauer2024}, extremely large telescopes (E-ELT~\citep{E-ELT_Quanz_2015, E-ELT_Neichel_2018}, TMT~\citep{TMT_Skidmore_2015} and others), \textit{Ariel}~\citep{Tinetti2018}, \textit{Twinkle} \cite{Twinkle_Edwards_2019, Twinkle_SPIEPoster_Stotesbury_2022}, and concepts such as LIFE~\citep{LIFE1_Quanz_2022} show a continued investment in the field. \\

Though significant advances in understanding have been made over the last two decades, there are still many open questions on the formation and evolution of exoplanets and the nature of atmospheric processes. To help resolve these questions, thorough stellar characterization and precisely measured planetary radius and masses are necessary. By far the most prevalent detection method to date is the transit method, and as such, current data inventories are strongly biased towards planets with radius measurements. Whilst $\sim 25\%$ of known planets % 1606/5967=26.9% (07/08/2025)
have measured masses, only a small subset of transiting planets ($<20\%$) 
%mass error < 10% -- 398 / 4167
%mass error < 20% -- 906 / 4467 (07/08/2025)
%mass error < 25% -- 840 / 4167
have masses measured with a precision of $< 20\%$ of the reported value\footnote{based on calculations made using the NASA Exoplanet Archive composite parameters database (accessed 2023.11.23)}. The ability to combine both radius and mass values has been leveraged to obtain initial mean bulk density measurements, which can provide broad constraints on global structure and composition. These have been used to suggest physical interpretations for observed phenomena such as the Kepler radius valley (a.k.a. Fulton Gap)~\citep{RadiusValley_Fulton_2017} and the hot Neptune desert~\citep{NeptunianDesert_Szab_2011, NeptunianDesert_Mazeh_2016, NeptunianDesert_LTT9779b_Edwards2023}, although it is widely accepted that atmospheric data are required to begin breaking degeneracies~\citep{InteriorDegeneracies_Rogers&Seager_2010}. 

At present, the number of planets with atmospheric data from spectroscopy is low, accounting for $<$4\% of the known population. The $\sim$150 unique planets\footnote{\url{http://research.iac.es/proyecto/exoatmospheres/table.php}} with atmospheric spectra sparsely sample a wide parameter space in radius and temperature, with the bulk of these datasets coming from ground-based programmes, the \textit{Hubble Space Telescope} (HST), \textit{Spitzer} and, more recently, \textit{James Webb Space Telescope} (JWST) observations. JWST has enabled the study of individual planet atmospheres in great depth, both below the radius valley (e.g., 
55\,Cnc\,e \citep{Hu2024}, 
WASP-47\,e \citep{WASP-47e_JWSTprop_Zieba_2023}, % JWST Proposal ONLY ? No published paper
LHS\,3844\,b \citep{LHS3844b_JWSTprop_Zieba_2023}, % JWST Proposal ONLY -- phase-curve ? No published paper
LHS\,1140\,b \citep{Cadieux2024}, 
LP791-18\,d \citep{LP791-18d_JWSTprop_Benneke_2024}) %\citep{}) % BAAS - methane detection with NIRSpec/PRISM. No published paper ? yet ?
and above it (e.g., K2-18\,b \citep{K2-18b_CH4_DMS_2023}, TOI-270\,d \citep{TOI-270d_JWSTNIRSpec_Holmberg&Madhusudhan_2024A&A}, GJ\,3470\,b \citep{GJ3470b_JWSTNIRCam_Beatty2024arXiv}, WASP-107\,b \citep{WASP-107_JWSTNIRSpec_Sing_2024arXiv}, WASP-39\,b \citep{Rustamkulov2023}, WASP-43\,b \citep{Bell2023}). Its early achievements include definitive evidence of photochemistry on a hot Jupiter, and more recently the first detection of a secondary atmosphere on a rocky world. These results and more will continue to shape this rapidly evolving field. However, JWST is not dedicated to observing exoplanets and therefore it is not well-suited to large-scale systematic surveys. The lessons learned from JWST are invaluable for the upcoming dedicated missions that will perform the first spectroscopic surveys of exoplanets: \textit{Twinkle} and \textit{Ariel}. Key advantages of these missions is their wide, continuous spectral coverage in the visible to mid-infrared, obtained in a single shot. Thus, these missions will produce homogeneous data across their respective observing samples of 100s of exoplanetary targets, extending comparative planetology beyond the Solar System and reveal underlying trends in the observed population.
% WASP-47e, LHS\,3844\,b -- Sebastian Zieba
% LP719-18\,c -- Pierre-Alexis Roy -- https://craq-astro.ca/rencontre/2024/Roy_PierreAlexis.pdf
% TOI-178 -- Matthew Hooton

%Upcoming space missions such as \textit{Twinkle} and \textit{Ariel} will conduct the first systematic surveys of a large and diverse sample of exoplanets to extend comparative planetology to the Galaxy scale and reveal underlying trends. Each will provide simultaneous coverage in a single observation across their respective wavelength ranges... 
% 1) In order to obtain wide spectral coverage, multiple instrument filters are commonly used, ... SOMETHING ABOUT multiple visits to the target, vertical offsets and stitching the spectrum back together. 
% 2) brightest targets (can) saturate JWST modes.
To optimize the scientific output and prepare the data exploitation for these two missions, significant work has been devoted to atmospheric studies~\citep{Twinkle_Edwards_2019,Changeat2020,Mugnai2021a,Bocchieri2023} of simulated exoplanetary populations. Potential synergies are widely recognized and here we discuss a preliminary attempt at leveraging the capabilities of both missions to characterise the atmospheres of a small test sample. 

The paper is structured as follows; Section~\ref{sec:introduction} outlines the technical specifications of the \textit{Twinkle} and \textit{Ariel} missions and discusses the scientific motivation for this study. In Section~\ref{sec:methods}, target selection methods are outlined and a brief review of the literature is included for each selected planet. We also detail the forward model creation, the noise models employed, and the self-retrievals performed on each of the simulated spectra. We include our findings in Section~\ref{sec:results}, then discuss their implications and opportunities arising from them in Section~\ref{sec:discussion}. Finally, we conclude and summarise this pilot study in Section~\ref{sec:conclusions}.

\subsection{Ariel}

\textit{Ariel}~\citep{Tinetti2018, Tinetti2021} is a pioneering ESA space mission that will revolutionize our understanding of exoplanets by surveying their atmospheres and compositions with spectroscopy. As part of ESA’s Cosmic Vision programme, \textit{Ariel} will launch in 2029 and operate from the L2 point, where it will perform a chemical survey of about 1000 warm and hot exoplanets over a wide range of sizes, masses, temperatures, and host star properties. \textit{Ariel} is designed to measure atmospheric signals from planets with better than 20--100~ppm post-processing stability relative to the host star, depending on the target brightness, for a single transit observation. 
% An observation is considered to last 2.5 times T$_{14}$, the time between the first and the last contact between the planetary and stellar disks, to collect data both in and out of transit for the light curve fit and the transit depth estimation~\citep{Mugnai2020}.
\textit{Ariel} will exploit its unique capability to cover the entire 0.5 to 7.8-$\mu$m spectral range in one shot, capturing the peak emission of these planets and detecting many important molecular species~\citep{Encrenaz2015}. 

The \textit{Ariel} payload~\citep{Eccleston2024} consists of a Telescope Assembly incorporating an off-axis Cassegrain telescope with a 1-m class primary mirror~\citep{Pace2024} that feeds two instruments: the Fine Guidance System (FGS)~\citep{Rataj2019} and the \textit{Ariel} InfraRed Spectrometer (AIRS)~\citep{Martignac2022}. FGS provides three photometric channels (VIS-Phot, 0.5–0.6 $\mu$m; FGS1, 0.6–0.80 $\mu$m; FGS2, 0.80–1.1 $\mu$m) and a low-resolution Near-InfraRed Spectrometer (NIRSpec, 1.1–1.95 $\mu$m and $R \geq$ 15). AIRS is a broad-band, low-to-medium-resolution near-infrared spectrometer operating between 1.95 $\mu$m and 7.8 $\mu$m, with two independent channels (CH0, 1.95–3.9 $\mu$m and $R \geq$ 100; CH1, 3.9–7.8 $\mu$m and $R \geq$ 30). Both instruments use Teledyne's HxRG Mercury-Cadmium-Telluride (MCT) detectors that have high quantum efficiency and low noise. The AIRS detectors are the only items that require active cooling via an active Ne JT cooler. The payload design has been optimized using lessons learned from previous space missions to achieve high photometric stability and spectral accuracy while mitigating systematic errors or allowing their removal post-processing~\citep{Bocchieri:2025}. Therefore, the Ariel payload design reaches photon noise limited performances on all targets of the mission~\citep{Mugnai2020}.

After each observation, the resulting spectrum from each spectrometer is binned during data analysis to optimize the signal-to-noise ratio (S/N). Therefore, by implementing different binning options, the mission adopts a four-tier observation strategy to optimize the science return. We refer to~\citep[][their Section~3.3]{ArielMRS_Edwards2019} for the definition of each tier. We report these definitions in Table~\ref{tab:ariel-tiers} for convenience.

\begin{table}[]
\caption{Tiers definition in the Ariel observing strategy}
\label{tab:ariel-tiers}
\begin{tabular}{@{}lcccc@{}}
\toprule
\textbf{Spectrometer channel} & \textbf{Wavelength range} & \textbf{Tier 1} & \textbf{Tier 2} & \textbf{Tier 3} \\ \midrule
NIRSpec & 1.1 -- 1.95\,\textmu m & R $\sim$ 1 & R $\sim$ 10 & R $\sim$ 20 \\
AIRS CH0 & 1.95 -- 3.9\,\textmu m & R $\sim$ 3 & R $\sim$ 50 & R $\sim$ 100 \\
AIRS CH1 & 3.9 -- 7.8\,\textmu m & R $\sim$ 1 & R $\sim$ 10 & R $\sim$ 30 \\ \bottomrule
\end{tabular}
\end{table}

This optimized instrumental setup and observation strategy will enable \textit{Ariel} to address fundamental questions such as: ``What are exoplanets made of?'', ``How do they form and evolve?'' and ``What are the physical processes shaping planetary atmospheres?''~\citep{Changeat2020, Mugnai2021a, Turrini2018}. By studying a large and diverse sample of exoplanets, \textit{Ariel} will also reveal statistical trends and correlations among their properties, shedding light on their origin and evolution~\citep{ArielMRS_Edwards2019}. 
The data acquired by \textit{Ariel} will be processed by dedicated pipelines that will extract the spectra of the exoplanet atmospheres from the transit or eclipse signals~\citep{Pearson2022}. The data products will be made available to the scientific community through a public archive hosted by ESA~\citep{Puig2018}. \textit{Ariel}'s data policy is that Tier~1 data will be made available to the community immediately, after due quality controls; for tiers~2 and~3, after a short proprietary time period of 6 months.

\subsection{Twinkle}
% Set for first light in 2025, the 0.45m \textit{Twinkle} space telescope is expected to characterise the atmospheres of 10s to 100s of exoplanets located between $\pm 40^\circ$ of the ecliptic during its nominal four-year exoplanet survey. The two spectroscopic channels at 0.5-2.4$\mu$m and 2.4-4.5$\mu$m reach peak spectral resolutions of R=70 and R=50 respectively and can be operated simultaneously, providing continuous coverage between 0.5-4.5$\mu$m and expanding on the wavelength coverage of HST/WFC3 by a factor of X. The use of reliable heritage components --- FAST-TRACKED, LEO -> reduced costs --- means it will be operational approximately half a decade before Ariel, and as such with similar spectral resolution and overlapping wavelength coverage will be a useful precursor/barometer/test-case to inform final \textit{Ariel} target list. [REVISE WORDING HERE]. 

Expected to launch in late 2027 and operate from a Low-Earth Orbit (LEO) at an altitude of 1200~km with an orbital period of $\sim105$~minutes, the \textit{\textit{Twinkle} Space Telescope}~\citep{Twinkle_Edwards_2019} is a commercial venture by Blue Skies Space Ltd. (BSSL). Set to study both solar-system and extra-solar objects, the mission is expected to characterize the atmospheres of 10s to 100s of exoplanets during its nominal three-year exoplanet survey\footnote{\url{http://bssl.space/twinkle/}}, providing the first large-scale, homogeneous, exoplanet spectroscopic survey. 

The spacecraft consists of a 0.45-m diameter primary mirror with an actively cooled inner sanctum, in which a spectrometer with two channels is housed. CH0 0.5-2.4~$\mu$m and CH1 2.4-4.5~$\mu$m, that will reach peak spectral resolving powers of $R = 70$ and $R = 50$, respectively~\citep{Twinkle_SPIEPoster_Stotesbury_2022, Twinkle_Stotesbury_2024SPIE}. These channels will operate simultaneously, providing continuous coverage between 0.5--4.5~$\mu$m and expanding on the wavelength coverage of HST/WFC3 by just over a factor of 4. 

The mission's Sun-synchronous polar orbital configuration allows for the observation of targets between $\pm 40^\circ$ of the ecliptic, whilst the use of reliable, established heritage components enables reduced costs relative to traditional mission pathways. Consequently, \textit{Twinkle} represents an alternate roadmap for space-based research missions. Another main difference with Twinkle is that its mission philosophy is not requirement-driven but is heavily dependent on the as-built performance which will be predicted through modelling and assessed in-flight. So, while this requires Twinkle to be more flexible in its observation plan, which will be revised if necessary once its in-flight performance is checked, it also enables the mission to speed up schedule and lower costs. Twinkle's accelerated design and launch programme ($\lesssim$10~years from conception to launch) will allow it to be operational approximately 2.5\,years before Ariel. 
Twinkle’s data release policy is set to feature public release of any data included in publications released in the time period during the primary mission, with a full public release of all data obtained during the primary mission to occur within 6 months of this end date (approx Q1 2031).

Twinkle's similar spectral resolution and overlapping wavelength coverage to Ariel will make it a useful precursor to inform the final \textit{Ariel} target list as the data obtained during the primary mission will be published early-on during Ariel's in-flight mission life. It is noteworthy that Ariel's spectral range extends further in the infrared, however Twinkle's visible spectrometer provides higher resolution data than Ariel's three photometers between 0.5 and 1.1 micron. This wavelength range is predominantly useful for the characterization of the host star activity, scattering slopes and optical features of hot exoplanets, and phase curves of the same. 
A thorough study of how this higher-resolution data in the visible is of use for Twinkle and how that could provide complementary information for Ariel observations and their interpretation is currently outside the scope of the present work, which focuses on just six sub-1000\,K planets. \bigskip

% With the details of the mission still to be finalised, we assume a conservative 75\% observing efficiency due to Earth-occultation from low-Earth orbit.

Table~\ref{tab:mission-summary} provides a synoptic comparison of both missions, including additional details, for convenience. Note that some of Twinkle's performance specifications are currently N/A as the mission relies heavily on as-built performance.

\begin{table}[!h]
\centering
\caption{Comparison of Twinkle and Ariel Space Telescope Instruments as required specifications.}
\begin{minipage}{\textwidth}
\begin{tabular}{>{\RaggedRight}p{3.2cm}>{\RaggedRight}p{3.2cm}>{\RaggedRight}p{3.2cm}>{\RaggedRight}p{3cm}}
\toprule
\textbf{Parameter} & \textbf{Twinkle} & \textbf{Ariel} & \textbf{Notes} \\
\midrule
\multicolumn{4}{l}{\textbf{Mission Overview}} \\
\midrule
Primary Science Goal & Exoplanet \& Solar System object studies & Chemical survey of exoplanet atmospheres & Ariel focused on statistical sample. \\
Scheduled Launch Date & Q3 2027 & Q4 2029 & $\sim$2\,year gap between missions. \\
Mission Duration & 3 (+3.5) years & 4 (+2) years & Extended mission in () \\
Orbit & Sun-synchronous, LEO ($\sim$1200 km) & L2 & Different thermal environments. \\
Observing strategy & N/A & 4-Tiers & See Table 1 \\
\midrule
\multicolumn{4}{l}{\textbf{Instrument Specifications}} \\
\midrule
Telescope Collecting Area & 0.45\,m & 0.64\,m & Ariel has larger light collecting area. \\
Wavelength Coverage & 0.5-4.5\,\textmu m & 0.5-7.8\,\textmu m & Ariel extends further into mid-IR. \\
Spectral Resolution & 20$<$R$<$70 (0.5-2.4\,\textmu m) 20$<$R$<$50 (2.4-4.5\,\textmu m) & $R\geq15$ (1.1-1.95\,\textmu m) $R\geq100$ (1.95-3.9\,\textmu m) $R\geq30$ (3.9-7.8\,\textmu m) & Twinkle has higher resolution in the 0.5-1.1\,\textmu m range. \\
Straylight & N/A & $<$ 1\% & \\
\midrule
\multicolumn{4}{l}{\textbf{Detector Performance}\footnote{More information in the public domain at \url{https://www.teledyne-si.com/en-us/Products-and-Services\_/Documents/Infrared\%20and\%20Visible\%20FPAs/TSI-0855\%20H2RG\%20Brochure-25Feb2022.pdf}}
} \\
\midrule
Detector Type & Teledyne H2RG (MCT) & Teledyne HxRG MCT & \\
Pixel size & 18\,\textmu m & 18\,\textmu m & \\
Quantum Efficiency & $>$60\% & $>$60\% & Varies by channel.\\
Read Noise & 20 e$^{-}$ & 20 e$^{-}$ & [CDS/pixel]. \\
Dark Current & 5 e$^{-}$/s/pixel & 5 e$^{-}$/s/pixel & \\
Full Well Capacity & $\sim$ 80 ke$^{-}$ & 100 ke$^{-}$ (FGS), 85-50 ke$^{-}$ (AIRS) & \\
\midrule
\multicolumn{4}{l}{\textbf{Stability Requirements}} \\
\midrule
Detector Temperature & $<$~90~K & 65~K (FGS), 42~K (AIRS) & AIRS requires active Ne JT cooling. \\
% Temperature Stability & N/A & XX mK & \\
Gain Noise & N/A & 40 ppm / $\sqrt{h}$  &  Worst-case. \\
% Voltage Stability & N/A & XX $\mu$V & \\
Jitter noise & N/A & 20 ppm (const. $>1$ hr) & \\
Stability post-processing & N/A & 20-100\,ppm & Depends on the target's brightness.\\
\midrule
\multicolumn{4}{l}{\textbf{Observational Capabilities}} \\
\midrule
Field of Regard & Within $\pm40^{\circ}$ of ecliptic & Whole sky in 6 months & \\
SNR Goal & $>$5 & $>$7 in AIRS (each tier) & Ariel has 4 tiers. \\
Targets (Expected) & 10s to 100s exoplanets & $\sim$1000\,exoplanets & \\
Figure of Merit & N/A & 0.165 m$^2$ (NIRSpec), 0.132 m$^2$ (AIRS) & [A$_{tel}$ $\cdot$ QE $\cdot$ transm.]. \\
Nyquist sampled signals & N/A & $\checkmark$ & ($>$ 2 pix. for FWHM)\\
\midrule
\multicolumn{4}{l}{\textbf{Data \& Operations}} \\
\midrule
% Science Data Rate & N/A & $\sim$2.3 GB/day & \\
Observing Efficiency & $\sim$75\% (conservative estimate) & $\sim$85\% & Twinkle limited by Earth occultation. \\
Data Availability & Publicly available after conclusion of initial 3~year mission. & Tier~1: immediate; Tier~2 and 3: 6 months; Tier~4: 1 year& \\
\bottomrule
\end{tabular}
\end{minipage}
\label{tab:mission-summary}
\end{table}

\subsection{Science Case}

The investigation of cool and temperate gaseous exoplanets presents an intriguing scientific opportunity to advance planet formation, evolution and migration theories through improved constraints on atmospheric metallicity across distinct planetary populations, particularly in the context of utilizing both the \textit{Twinkle} and \textit{Ariel} missions. These planets are not proposed to be systematically investigated with \textit{Ariel}, while \textit{Twinkle} has a dedicated survey for cool gaseous planets (\textcolor{blue}{\textit{Twinkle} group paper (in prep)}; \citep{TwinkleCoolGaseousSurvey_Booth_2024}), making it even more compelling to explore their characterization using \textit{Twinkle} and how it can inform future \textit{Ariel} observations. 

One key advantage offered by \textit{Twinkle} and subsequently \textit{Ariel} is their broad, continuous spectral coverage, which is essential for resolving degeneracies in atmospheric composition and thermodynamics \citep[e.g.,][]{Yip2020a,Changeat2020b}. While instruments like HST/WFC3 are primarily sensitive to water (H$_2$O) and methane (CH$_4$) features in the near-infrared, the recent launch of JWST, with its 6.5-m primary mirror and spectral coverage from 0.6 to 28.5 $\mu$m, has opened uncharted territory in atmospheric characterization. This was evidenced by recent inferences of CO$_2$ and SO$_2$ in the atmosphere of the hot Jupiter WASP-39b~\citep{Rustamkulov2022, Feinstein2022, Ahrer2022b, Tsai2022}, and the 5-$\sigma$ detection of \ce{CH_4} in K2-18~b~\citep{K2-18b_CH4_DMS_2023}, the latter of which appears to solve the long-standing \enquote{missing methane problem}~\citep{Benneke2019,Tsiaras2019,Fortney2020}. However, JWST is a multi-purpose observatory and, although it allocates a considerable amount of observing time to exoplanet observations, conducting systematic and unbiased surveys of exoplanet atmospheres is not a principal focus. In contrast, although less sensitive, \textit{Twinkle} and \textit{Ariel} will enable a systematic characterization of exoplanet atmospheres methodically, with each survey capable of producing a large and homogeneous dataset across extended wavelengths, sufficient to detect many major expected molecular species such as H$_2$O, CO$_2$, CH$_4$, NH$_3$, HCN, H$_2$S, TiO, and VO~\citep[e.g.,][]{Tinetti2013, Encrenaz2015}. This comprehensive approach will 
%pave the way for addressing and potentially resolving these degeneracies \commentLB{degeneracies between what ?}, and 
help us gain further insight into the true nature of exoplanets. 

Understanding the properties of cool gaseous exoplanets (those with equilibrium temperatures below 1000~K) is crucial for refining our understanding of atmospheric metallicity and extending existing trends found in the literature. By identifying and exploiting potential synergies between \textit{Twinkle} and \textit{Ariel}, we can optimize the scientific return, with insights gained from \textit{Twinkle} observations able to provide valuable support for proposing targeted observations with \textit{Ariel}. While a few sub-700~K planets are part of a possible realization of the \textit{Ariel} Mission Reference Sample~\citep{ArielMRS_Edwards2019}, observing these colder objects requires a significant amount of mission time. Therefore, it is essential to construct a well-justified scientific case that integrates the observational capabilities of \textit{Ariel} with potential preliminary observations conducted by Twinkle. Therefore, we focus specifically on cool gaseous exoplanets~\citep[e.g.,][]{Encrenaz2022}, but stress that other planetary classes may also benefit from such a coordinated approach.
% This coordinated approach will maximize the scientific impact and contribute to a more comprehensive understanding of temperate exoplanets.

\section{Methods}
\label{sec:methods}
In this pilot study exploring a possible \textit{Twinkle}-\textit{Ariel} synergy, we simulate atmospheric transmission spectra for both missions for a small selection of targets with sub-$1000$~K equilibrium temperatures, as described in Section~\ref{subs:target-selection}. We employed different atmospheric models to simulate the high-resolution forward model spectra to investigate the representativity of results under a variety of chemical and thermodynamic conditions. This is described in Section~\ref{subs:forwardmodels}, where we also explain how we produced the transmission spectra `as observed' by \textit{Twinkle} and \textit{Ariel}. Finally, in Section~\ref{subs:retrievals}, we present the methodology utilized to perform spectral retrievals for each target and atmospheric model, as well as the metrics used to present the results.

\subsection{Target Selection}
\label{subs:target-selection}
Both \textit{Twinkle} and \textit{Ariel} will observe hundreds of exoplanetary spectra of suitable targets for transmission spectroscopy and their target lists are expected to partially overlap. 
Rather than extending our selection of targets to all planets that are in both missions' candidate target lists, in this pilot study, we investigate only six planets according to pre-defined criteria. We select planets with equilibrium temperatures of $<1000$~K from the proposed \textit{Twinkle} cool gaseous planet survey (\textcolor{blue}{\textit{Twinkle} group paper (in prep)}, \citep{TwinkleCoolGaseousSurvey_Booth_2024}) that can achieve an expected S/N on the spectrum sufficient for detailed characterization within a few visits. 
The planetary and stellar parameters used throughout this study are sourced from the 2019 realization of the Ariel Mission Reference Sample~\citep{ArielMRS_Edwards2019} and are reported in Table~\ref{tab:targets}.
A ``visit'' is defined here as the time spent by the telescope on a given target during the transit event (lasting the time between the first and last contact, T$_{14}$) and the baseline before and after the transit lasting 1.5 T$_{14}$ which is adequate for the light curve fit and the transit depth estimation. This gives a total visit duration of 2.5 T$_{14}$. 
From this, we define an ``observation'' as the combined set of visits required to achieve the appropriate S/N threshold. 

Specifically, we use the methodology outlined in~\cite{Mugnai2020} to calculate the S/N, thus assuming that the noise between visits in a given observation is uncorrelated. The noise estimates for an observation are obtained by rescaling the noise estimated for a single visit by the square root of the number of visits required to meet the specified S/N threshold. 
In the S/N calculation, we consider the median noise over each spectroscopic channel; the signal is computed assuming their observable atmospheres extend to $5$ scale heights for a H$_2$/He-dominated atmosphere. As in~\cite{Mugnai2020}, we require a S/N of at least $7$ in one of the spectroscopic channels when spectra are binned on the spectral grid of \textit{Twinkle} and \textit{Ariel} in Tier~3 (see Table~\ref{tab:ariel-tiers}). 
Clearly, for Twinkle, this condition is fulfilled first in CH0 for all planets in the sample, as can be readily seen from visual inspection of Figure~\ref{fig:noise-curves} (see later). 
For Ariel, we consider only the AIRS channels for this criterion. 
The S/N condition is met first in AIRS-CH0 for all planets bar GJ~3470~b and GJ~436~b. 
Note that meeting the requisite S/N in Tier~3 ensures that it is also met in Tier~2 of the \textit{Ariel} mission. In this work, we only consider Tier~2 and 3, as Tier~1 is not designed for the same quantitative atmospheric characterization.

We require the S/N threshold to be reached in less than 10 visits for both missions. Although this criteria is arbitrary and is mainly set to constrain the resulting sample size, it retains physical motivation, corresponding to a reasonable portion of a space-mission lifetime and whilst being below the largest number of co-added transits from a single instrument in presently published transmission spectra~\citep{HST_11transits_GJ9827d_Roy_2023, HST_15transits_GJ1214b_Kreidberg_2014}. Notably, for \textit{Twinkle} we impose a conservative estimate of 75\% efficiency on the observation (see later), to account for the loss of signal during Earth-occultation events caused by \textit{Twinkle}'s low-Earth-orbit. Due to the higher sensitivity of \textit{Ariel}, the resulting target selection is mostly \textit{Twinkle}-driven. 

In the following paragraphs, we briefly describe each selected target. Table~\ref{tab:targets} summarizes their properties in terms of stellar and fundamental planetary parameters. The table also contains the expected number of required visits for each planet for both missions and the median S/N achieved in the brightest spectroscopic channel. 

\paragraph*{\textbf{GJ~436~b}}
A well-studied, archetypical warm Neptune, GJ~436~b has a 2.64-day orbit around an M2.5-dwarf. Extensively studied from the ground with high-resolution cross-correlation spectroscopy, optical and NIR spectral datapoints from HST/STIS, HST/WFC3 and \textit{Spitzer}~\citep{GJ436b_LymanAlpha_2014,GJ436b_HSTspectrum_2018}, multiple atmospheric species including \ce{CO}, \ce{CO_2}, \ce{CH_4} and \ce{H_2O} have been detected in its atmosphere~\citep{GJ436b_DaysideEmission_2011}. This, combined with significant atmospheric escape driven by XUV irradiation inferred from the detection of hydrogen Lyman-alpha makes GJ 436 b an intriguing target for future atmospheric study.
% Hu 2014 abstract -- CH4-poor, CO-rich atmosphere. --> results of Stevenson et al 2013 paper, Spitzer dayside emission spectroscopy.
% HST WFC3 observation not spectral / used to obtain atmospheric transmission spectrum ?... Stevenson et al Nov2014.
% GJ 436b RV, RM signal etc -- (Polar orbit & migration) -- Bourrier+ 2023

\paragraph*{\textbf{GJ~3470~b}}
Discovered in 2012 orbiting an M1.5 dwarf~\citep{GJ3470b_Detection_2012}, GJ~3470~b is a short-period, warm Neptune. Recent claims by a citizen science group have claimed detection of multiple large planetary companions on exterior orbits, however these claims are currently debated~\citep{GJ3470b_NoExtraPlanets_2023}. Bulk-density calculations suggest a large \ce{H}/\ce{He} dominated atmosphere, the presence of which was confirmed, along with \ce{H_2O} using HST and Spitzer observations conducted using STIS, WFC3 and IRAC spanning 0.55 to 4.5~$\mu$m~\citep{GJ3470b_HSTTransmisionSpectra_2019}. More recently using Lyman-alpha and helium I triplet lines, atmospheric outflow and escape have been inferred~\citep{GJ3470b_HeliumTriplet_2020, GJ3470b_HeliumTriplet_2021}, shedding further light on atmospheric dynamics in planets at the edges of the hot-Neptune desert.

\paragraph*{\textbf{K2-141~c}}
A warm, short-period, sub-Jovian planet first detected in 2018 around a K4-dwarf, K2-141~c has a poorly constrained radius due to its grazing transit configuration~\citep{K2-141c_DISCOVERY_Malavolta2018}. Not-detected in the radial-velocity search conducted with HARPS-N, so far it has only been possible to place an upper limit on the mass of K2-141~c~\citep{K2-141c_DISCOVERY_Malavolta2018}. Consequently, there have been no spectroscopic observations to date, and the planet will remain a challenging target until an improved mass measurement can be made. It further remains a challenging target due to its grazing transit configuration, which increases the difficulty of obtaining precise and accurate transmission spectroscopy measurements. 
%making accurate and precise measurements of the transit depth, and hence wavelength-dependent transit depth in the form of a transmission spectrum.
We retain this planet in our sample, assuming a non-grazing configuration in our radiometric estimates, and therefore the results shown in this work for this planet are not necessarily representative.

\paragraph*{\textbf{WASP-69~b}} A Jovian-sized planet with a Saturn-like mass, WASP-69 b was discovered in 2014 transiting a K5-dwarf with a short, 3.868-day orbital period~\citep{WASP69b_Discovery_2014}. With a host of atmospheric species detected simultaneously from ground-based high-resolution spectroscopy~\citep{WASP69b_HighResSpectroscopy_2022} and a water detection from a 2016 HST/WFC3 transmission spectrum~\citep{WASP69b_HSTspectrum_2018}, this planet is readily amenable to future atmospheric characterisation.

\paragraph*{\textbf{WASP-80~b}}
A similar planet to WASP-69 b, with a Jovian-like radius ($0.952\ M_\text{J}$) and Saturn-like mass, WASP-80~b orbits a K-dwarf with an orbital period of 3.068 days~\citep{WASP80b_HSTspectrum_2022}. Combined HST STIS and WFC3 spectra, supplemented by Spitzer IRAC spectro-photometric datapoints at $3.6\ \mu$m and $4.5\ \mu$m, revealed a strong absorption feature at $1.4\ \mu$m and evidence for Rayleigh scattering in the optical. From this, the presence of \ce{H_2O} and haze particles in the atmosphere of WASP-80~b were inferred~\citep{WASP80b_HSTspectrum_2022}, with three further molecules, \ce{CH_4}, \ce{HCN} and \ce{NH_3} being detected at high significance using ground-based high-resolution cross-correlation spectroscopy~\citep{WASP80b_HighResSpectroscopy_2022}.
% HST spectra suggests enhanced metallicity.
% HRCCS suggests solar comp + diseq chem

\paragraph*{\textbf{WASP-107~b}}
Discovered in 2017, WASP-107~b is an inflated near-Jovian-radius planet with a Neptunian-like mass, orbiting a K-dwarf star on a 5.721-day orbit~\citep{WASP107b_LowDensity_2021}. The high significance detection of \ce{H_2O} from a 2017 HST/WFC3 spectrum and resultant modelling conducted indicates a solar to low-super-solar atmospheric metallicity composition, with potential hints of methane depletion from the subsolar C/O ratio \cite{WASP107b_H2Odetection_2018}.

\subsection{Noise estimates}
\label{subs:sensitivity-curves}
\begin{figure*}
    \centering
        \includegraphics[width=0.93\textwidth]{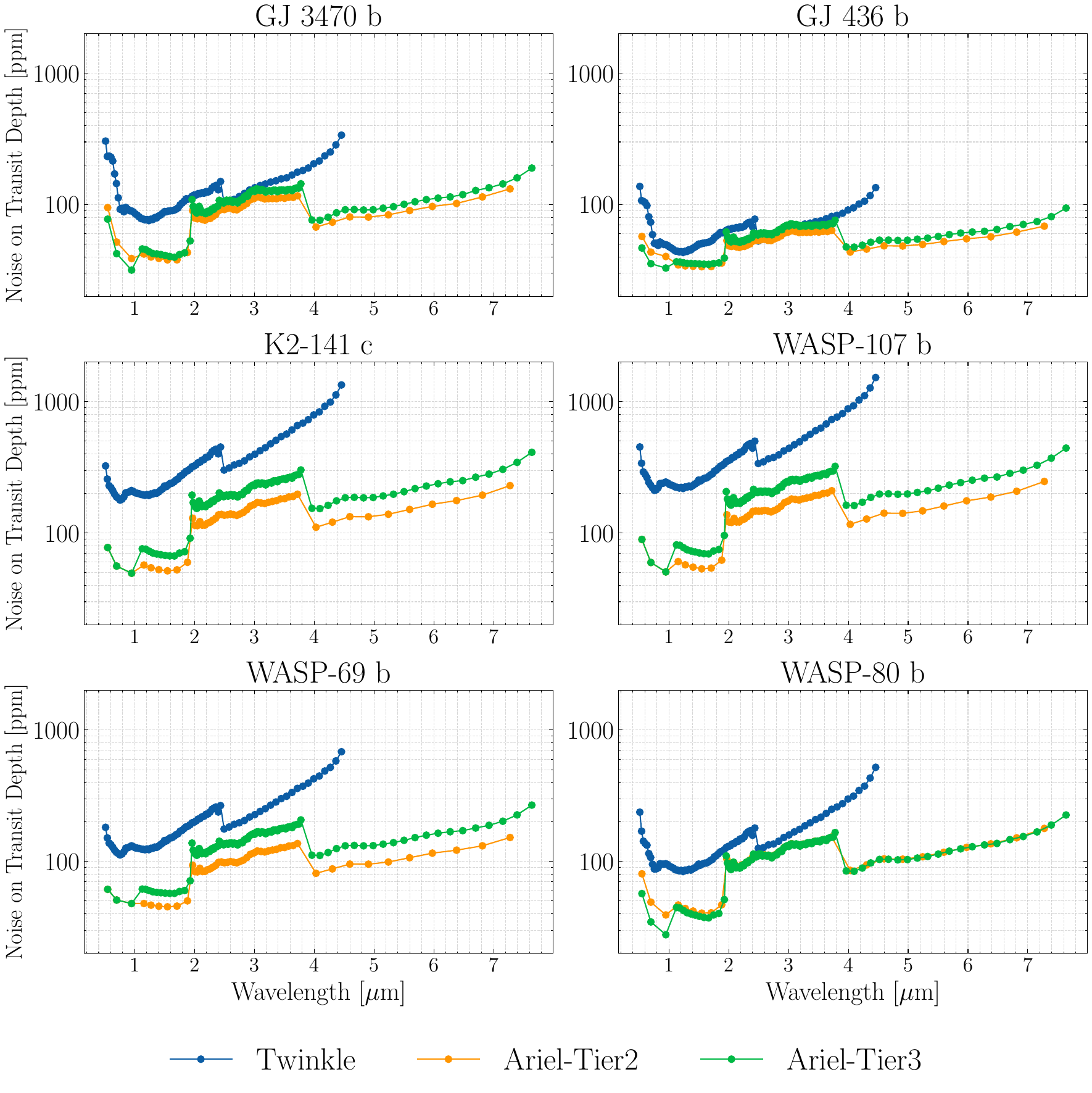}
    \caption{Expected noise estimates for a transit observation (an observation is the combined set of visits required to achieve the S/N threshold, see text) conducted with \textit{Twinkle} (blue) and \textit{Ariel} (orange: Tier 2; green: Tier 3). The data are binned to the corresponding spectral grids on the horizontal axis. %\commentLM{Here I would write again the the y-axis is the relative noise estimated of a transit observation (or for an hour time scale), and that the data are binned according to the spectral resolution on the x-axis. This is to help the reader }
    }
    \label{fig:noise-curves}
\end{figure*}

To simulate spectra `as observed' with \textit{Twinkle} and \textit{Ariel}, we use radiometric estimates of the total noise on an observation, obtained for \textit{Twinkle} from the radiometric tool, \textit{TwinkleRad}~\citep{Twinkle_SPIEPoster_Stotesbury_2022} [via B. Edwards, private communication] and for \textit{Ariel} using the online radiometric noise simulator\footnote{Found on \url{https://exodb.space/}, with access currently restricted to \textit{Ariel} Consortium members.}, ArielRad\footnote{The software versions used are ExoRad2 v2.1.111, ArielRad-payloads v0.0.17, and ArielRad v2.4.26.}~\citep{Mugnai2020}. Both simulators are adapted from the generic point source radiometric model ExoRad2~\citep{Mugnai2023} using the \textit{Ariel} and \textit{Twinkle} payload configurations, ensuring that our simulation framework is consistent. 
Alongside the payload configuration, ExoRad2 takes as input the description of the target point source and foregrounds (e.g., zodiacal emission) and estimates the total optical efficiency, by combining the optical elements and the foregrounds. Then, ExoRad2 computes the radiometric performance estimates for each photometric channel and spectral bin, including the measured signal from the target and corresponding photon noise, as well as additional noise sources (e.g., read noise, dark current noise). In the case of \textit{Ariel}, ArielRad includes margins for correlated and time-dependent noise sources (e.g., pointing jitter, obtained from external estimate with the time-domain simulator ExoSim2~\citep{Mugnai:2025} adapted to \textit{Ariel}). For reference, these are listed in Table~\ref{tab:mission-summary} under ``Stability Requirements''. In any case, \textit{Ariel}'s performance is photon noise-limited across all planets in our sample and the read noise is only relevant in the red-most part of AIRS-CH1.
With regards to Twinkle, the detailed noise budget is still considered proprietary and therefore we are only able to comment on the total noise budget summarized in Figure~\ref{fig:noise-curves}, see later for details. 

%  Given the descriptions of an observational target and
% the instrumentation, ExoRad 2.0 estimates several performance metrics for each photometric
% channel and spectral bin. These include the total optical efficiency, the measured signal from
% the target, the saturation times, the read noise, the photon noise, the dark current noise, the
% zodiacal emission, the instrument-self emission and the sky foreground emission.

To account for the loss of data caused by LEO Earth-occultation events during scheduled \textit{Twinkle} observations, we assume an observing efficiency of 75\% and rescale the noise estimates by the number of observations required to achieve the desired S/N at the full spectral resolution of the instrument. The observing efficiency chosen here represents a conservative estimate (priv. communication) of the increased noise from systematic sources that we are not modelling in this paper. Ongoing work by the Blue Skies Space team on combining the knowledge of the systematic and radiometric effects will allow this effective loss of observing efficiency to be quantified in a later publication. We note that whilst Twinkle will intersect the South Atlantic Anomaly (SAA) during these LEO Earth-occultation events, the spacecraft platform has shown past reliable performance and features multiple-redundancy systems, allowing data collected to be used. This will be verified during the commissioning phase.
% With \textit{Ariel} designed to achieve a base photometric precision of 20-100 ppm on the 1-hour timescale relevant for transits, achieved in-part due to its L2 orbit, such rescaling is unnecessary and consequently when combined with a 4x larger light collecting area than Twinkle, %\textit{Ariel} (0.64m$^2$) vs \textit{Twinkle} (0.16m$^2$) 
% \textit{Ariel} requires a fewer number of transits to achieve the desired S/N for the majority of targets.  
In contrast, \textit{Ariel}'s L2 orbit allows continuous observation of the target in a thermally and photometrically stable condition. Importantly, \textit{Ariel}'s four times larger light collecting area than \textit{Twinkle} means that it will require fewer visits than \textit{Twinkle} to achieve the desired S/N threshold of $\geq 7$ for most targets.

The numbers of transits required to achieve the S/N threshold are listed in Table~\ref{tab:targets} and ``sensitivity curves'' for every planet and configuration (\textit{Twinkle}, and \textit{Ariel} in Tier~2 and Tier~3) are presented in Figure~\ref{fig:noise-curves}. These curves represent the computed noise on the transit depth measurement vs wavelength obtained in one observation (defined as the combined set of ``visits''). Later, we attach these noise estimates to the computed binned forward model transmission spectra on the respective wavelength grids of both missions (see Figure~\ref{fig:fitted-spectrum-model1}). 

We note here that for \textit{Ariel}, there are three targets; K2-141~c, WASP-69~b, and WASP-107~b, for which a single observation is enough to achieve the desired S/N in Tier~3 and therefore in Tier~2. Having the same number of observations but different spectral grids provides an interesting opportunity to compare results across tiers.
When the S/N requirement is met for a higher tier with the same number of transits as a lower tier, binning to the lower tier might alter resulting inferences on planetary and atmospheric properties.
Consequently, this study may also shed light on the information loss occurring when using too-conservative binning.

\subsection{Forward Models}
\label{subs:forwardmodels}
To generate the synthetic, high-resolution atmospheric transmission spectra of the planets, we utilized the TauREx~3 atmospheric retrieval framework~\citep{AlRefaie2021}. We produced four sets of models that differ in atmospheric chemistry, temperature-pressure (T-P) profiles, or the presence of clouds. Using different models enables us to explore various scenarios for the atmospheres, the knowledge of which was largely limited at the time of submission of this publication. Recent observations of cool gaseous planets by JWST have begun to shed more light on this population, but are still limited to a few keystone planets (e.g., K2-18~b, \cite{K2-18b_CH4_DMS_2023}; TOI-270~d, \cite{TOI-270d_JWSTNIRSpec_Holmberg&Madhusudhan_2024A&A, TOI-270d_JWST-NIRISS/SOSS+NIRSpec/G395H_CH4CO2H2O_Bjorn_2024arXiv}; WASP-107~b, \cite{WASP-107_JWSTNIRSpec_Sing_2024arXiv,WASP-107b_JWSTNIRCam/MIRI_NoCH4_Dyrek_2024Natur,WASP-107b_JWSTNIRCam/MIRI_CH4_Welbanks_2024Natur}), with occasionally differing inferences.

The scientific rationale behind choosing four different models is to test whether \textit{Ariel} and Twinkle can give consistent results in a controlled experiment where the spectra exhibit different spectral shapes. 
% We are interested in analyzing how well we can extract the relevant information using \textit{Twinkle} or \textit{Ariel} observations under a variety of conditions and investigating to what extent their different designs may provide complementary clues to interpret the spectra. 
We could envisage an alternative approach involving a single, detailed forward atmospheric model retrieved with different models with \textit{Twinkle} and \textit{Ariel} to investigate, e.g., whether the differences in the retrievals between the two missions are consistent as a function of the model used. However, this falls beyond the scope of this pilot paper, and would also be misleading in light of the large model space that thoroughly testing this would require~\citep{Welbanks:2025} as well as our limited knowledge about any potential systematics on Twinkle's side. 

All our atmospheric models implement primordial (H$_{2}$/He) gaseous atmospheres with collision-induced absorption (CIA) and Rayleigh scattering. Models~1 and 2 add traces of molecules typically found at high abundances under equilibrium chemistry in $<$1000 K atmospheres: CH$_4$, H$_2$O, NH$_3$, CO$_2$, and CO. These were injected with isochemical abundances (constant with altitude) of 100 ppm. In comparison to model~1, which assumes an isothermal atmosphere, model~2 employs a Guillot TP profile~\citep{Guillot2010}. Models~3 and 4 revert to the assumption of an isothermal atmosphere, but differ from Model~1 by assuming chemical equilibrium (implemented via ACE chemistry)~\citep{Agundez2012, Agundez2020}, where molecular opacities and volume mixing ratios (VMRs) depend solely on pressure, temperature, C/O ratio, and metallicity. Model~3 implements a clear atmosphere, while model~4 includes grey opaque clouds with an arbitrarily chosen top pressure of $0.05$ bar. These clouds have two main effects: (i) they attenuate the spectral signature of the molecular absorbers, and (ii) they introduce a degeneracy with the planetary radius~\citep{Changeat2020b}, which is defined as the surface below which the atmosphere becomes opaque at all wavelengths. Table~\ref{tab:models} summarizes the atmospheric models, while Table~\ref{tab:opacities} provides a comprehensive list of the opacities and CIA used in this study.

\begin{landscape}
\begin{table*}
\caption{Summary of the selected targets' properties and predicted number of transit visits required for an observation. \label{tab:targets}}
\begin{tabular}{@{}c|ccccc|cccc|ccc@{}}
\toprule
Planet Name &
  $R_\text{p}$ &
  $M_\text{p}$ &
  $T_\text{p}$ &
  P &
  $T_\text{dur}$ &
  $R_\text{s}$ &
  Mag K &
  $T_\text{s}$ &
  $M_\text{s}$ &  & \# visits &  \\
 &
    {[}$R_\oplus${]} &
    {[}$M_\oplus${]} &
    {[}K{]} &
    {[}days{]} &
    {[}hrs{]} &
    {[}$R_\odot${]} &
     &
    {[}K{]} &
    {[}$M_\odot${]} & \textit{Twinkle} & Ariel/T2 & Ariel/T3 \\
\midrule
GJ 436 b   & 4.08  & 25.40  & 708 & 2.64 & 1.02 & 0.46 & 6.07 & 3,586 & 0.47  & 4&2&3\\
GJ~3470~b  & 4.48  & 12.58  & 703 & 3.34 & 1.66 & 0.55 & 7.99 & 3,600 & 0.54  & 6&2&3\\
K2-141~c   & 6.85  & 7.40   & 720 & 7.75 & 2.37 & 0.68 & 8.40 & 4,599 & 0.71  & 1&1&1\\
WASP-69 b  & 12.18 & 82.63  & 929 & 3.87 & 1.79 & 0.86 & 7.46 & 4,700 & 0.98  & 1&1&1\\
WASP-80~b  & 10.96 & 171.62 & 799 & 3.07 & 1.80 & 0.59 & 8.35 & 4,143 & 0.58  & 8&2&4\\
WASP-107~b & 10.31 & 30.51  & 720 & 5.72 & 2.41 & 0.67 & 8.64 & 4,425 & 0.68  & 1&1&1\\ 
\bottomrule
\end{tabular}
\end{table*}

\begin{table*}[htbp]
\centering
\caption{The atmospheric models assumed for each planet. We list molecules with cross-sections available in TauREx\,3 and included in the analysis.\label{tab:models}}
\begin{tabular}{@{}l|ccccc@{}}
    \toprule
                     & \textbf{T-P profile} & \textbf{\begin{tabular}[c]{@{}c@{}}Atmospheric \\ Layers\end{tabular}} & \textbf{Chemistry} & \textbf{Molecules}     & \textbf{Clouds}  \\ \midrule
    \textbf{Model 1} & Isothermal   & 100   & Constant    & \ce{H_2O}, \ce{CH_4}, \ce{NH_3}, \ce{CO_2}, \ce{CO}  & No \\
    \textbf{Model 2} & Guillot      & 100   & Constant    & Same molecules as above  & No \\
     & & & & \ce{OH}, \ce{H_2O}, \ce{H_2O_2}, \ce{O_2}, \ce{HNO_3},  \ce{HCN} & \\
    \textbf{Model 3} & Isothermal   & 100   & ACE                 & \ce{CN}, \ce{NH}, \ce{NH_3}, \ce{H_2CO}, \ce{CO}, \ce{CO_2},  & No \\
    & & & & \ce{CH}, \ce{CH_3}, \ce{CH_4}, \ce{C_2H_2}, \ce{C_2H_4},  \ce{NO}  &  \\
    \textbf{Model 4} & Isothermal   & 100   & ACE                & Same molecules as above & Yes \\
    & & & & & {[}5000~Pa{]}  \\ \bottomrule
    \end{tabular}
\end{table*}
\end{landscape}
%H2O, OH, H2O2, O2, HNO3, H2CO, HCN, CO, CO2 CH, CH3, CH4, CN, NO, C2H2, C2H4, NH, NH3

\begin{table}
\centering
\caption{List of opacity contributions used in this work and their references.\label{tab:opacities}}%
\begin{tabular}{@{}cc@{}}
    \toprule
    \textbf{Opacity} & \textbf{Reference(s)} \\
    \midrule
    H$_2$-H$_2$ & ~\citet{Abel2011,Fletcher2018} \\
    H$_2$-He & ~\citet{Abel2012} \\
    OH & ~\citet{Yousefi2018}\\
    H$_2$O & ~\citet{Polyansky2018} \\
    H$_2$O$_2$ & ~\citet{Al-Refaie2016}\\
    O$_2$ & ~\citet{Gordon2017}\\
    HNO$_3$ & ~\citet{Pavlyuchko2015}\\
    HCN & ~\citet{Barber2014}\\
    CN & ~\citet{Brooke2014}\\
    NH & ~\citet{Fernando2018}\\
    NH$_3$  & ~\citet{Coles2019} \\
    H$_2$CO & ~\citet{Al-Refaie2015}\\
    CO  & ~\citet{Li2015} \\
    CO$_2$  & ~\citet{Rothman2010} \\
    CH & ~\citet{Masseron2014}\\
    CH$_3$ & ~\citet{Adam2019}\\
    CH$_4$ & ~\citet{Yurchenko2017} \\
    C$_2$H$_2$ & ~\citet{Chubb2020}\\
    C$_2$H$_4$ & ~\citet{Mant2018}\\
    NO & ~\citet{Hargreaves2019}\\
    \bottomrule    
\end{tabular}
\end{table}

%\tags{Paragraph below moved to sensitivty curves sub-section}
%To simulate spectra `as observed' with Twinkle, we used the radiometric estimates of the total noise on an observation, obtained using the \textit{Twinkle} radiometric tool, \textit{TwinkleRad} [Billy Edwards via private communication]. We assumed an observing efficiency of 75\% to account for the loss of data during a scheduled observation caused by LEO Earth-occultation events and we rescaled the noise estimates by the number of observations required to achieve the desired S/N at the full spectral resolution of the instrument. 
Each high-resolution forward model spectrum obtained with TauREx~3 was binned to the spectral grid of \textit{Ariel} in both Tier~2 and Tier~3 modes and Twinkle's nominal spectral grid, with corresponding expected errorbars for each bin being attached. Here, errorbars were calculated as described in Section~\ref{subs:sensitivity-curves}. We chose not to scatter the data according to the noise, because we aim to compare retrievals between \textit{Ariel} and Twinkle, investigating possible biases and inherent correlations between parameters: scattering the data would defeat this purpose, introducing susceptibilities to the random noise realizations. Moreover, assuming that there exists sufficient stability and redundancy in the information content, we expect that the retrieved mean values would not exhibit significant differences compared to using scattered spectra~\citep{Feng2018, Changeat2019, Changeat2020}. 

%\tags{move section, as written, to Section \ref{subs:sensitivity-curves}}
%We repeated the same procedure for Ariel, using the radiometric noise simulator ArielRad~\citep{Mugnai2020} to produce up-to-date uncertainty estimates for the mission. \comment{We used the ArielRad online available for Consortium members.} 
%For Ariel, we produced spectra observed in both Tier~2 and~3 of the mission to compare the results. It should be noted that for K2-141~c, WASP-69~b, and WASP-107~b, a single observation is enough to achieve the desired S/N in Tier~3. Therefore, in these cases, the only difference between adopting one Tier definition or the other is the spectral grid. Comparing the results across Tiers in this condition may shed light on the information loss occurring in binning. 

% To ensure reproducibility, we list in Table~\ref{tab:versions} the software versions. 

% \begin{table}
%     \centering
%     \begin{tabular}{c@{ }c@{ }}
%     \hline
%     \bf{Code}     & \bf{Version} \\ 
%     \hline
%     TauREx~3          & 3.1.1-alpha      \\    
%     ArielRad          & 2.4.26           \\
%     ExoRad            & 2.1.111          \\
%     ArielRad-Payloads & 0.0.17           \\ 
%     \hline
%     \end{tabular}
%     \caption{Code versions used in this study.}
%     \label{tab:versions}
% \end{table}

\subsection{Retrievals}
\label{subs:retrievals}
To investigate the combined effects of binning and wavelength coverage and explore whether prior \textit{Twinkle} observations are likely to yield informative results for Ariel, we perform self-retrievals on the forward-modelled spectra described in Section~\ref{subs:forwardmodels}, that is, we assume the same atmospheric models in the fitting procedure for each corresponding model. Here we again utilise TauREx~3\footnote{TauRex version 3.1.4-alpha}. The free parameters of the retrievals depend on the model and are summarized in Table~\ref{tab:params-and-priors}, along with the priors assumed in the retrieval:

\begin{table}
\centering
\caption{Fit parameters and their priors for retrievals conducted in this work.\label{tab:params-and-priors}}%
\begin{tabular}{@{}ccccc@{}}
\toprule
\textbf{Parameters} & \textbf{Units} & \textbf{Priors} & \textbf{Scale} & \textbf{Model} \\
\midrule
$R_\text{P}$ & $R_\text{J}$ & $\pm$10\% & linear & 1, 2, 3, 4 \\
CH$_4$ & VMR & 10$^{-9}$; 10$^{-1}$ & log & 1, 2  \\
CO$_2$ & VMR & 10$^{-9}$; 10$^{-1}$ & log & 1, 2  \\
H$_2$O & VMR & 10$^{-9}$; 10$^{-1}$ & log & 1, 2  \\
NH$_3$ & VMR & 10$^{-9}$; 10$^{-1}$ & log  & 1, 2 \\
CO & VMR & 10$^{-9}$; 10$^{-1}$ & log & 1, 2  \\
T& K& 300; 2000 & linear & 1, 3, 4  \\
T$_\text{irr}$& K& $\pm$50\% & linear & 2  \\
Z$_\text{ace}$& scalar & default & default & 3, 4  \\
C/O$_\text{ace}$& scalar & default & default & 3, 4  \\
P$_\text{clouds}$& Pa & default & default & 4  \\
\bottomrule
\end{tabular}
% \footnotetext{.}
\end{table}

\begin{itemize}
    \item Model~1. We fit the planet's radius, the isothermal atmospheric temperature, and each of the molecular mixing ratios. We use broad logarithmic uniform priors ranging from 10$^{-9}$ to 10$^{-1}$ for the latter. 
    \item Model~2. We fit the same parameters as in model~1, replacing the isothermal temperature with the effective temperature characterizing the irradiation intensity~\citep{Guillot2010}.
    \item Model~3. We fit the planet's radius, the isothermal atmospheric temperature, the ACE metallicity ($Z_\text{ACE}$), and the ACE C/O ratio.
    \item Model~4. We fit the same parameters as in model~4, and in addition, we fit the grey cloud pressure level.
\end{itemize}

We set the evidence tolerance to $0.5$ and sample the parameter space through $1500$ live points using the \texttt{MultiNest} algorithm\footnote{v3.11, Release April 2018}~\citep{Feroz2009, Buchner2023}. 

\section{Results} %\commentLM{see also Conclusion comment}
\label{sec:results}
\subsection{Fixed Chemistry Models}
\label{subs: Fixed Constant Chem Results}
\begin{figure*}
    \centering
    % \begin{overpic}[width=0.92\textwidth]{figures/model1-20231103_1500_lp_chem_cst_5mols_bound-9/o-c_all.pdf}  % copied into main dir as model1-o-c_all.pdf
    \begin{overpic}[width=0.92\textwidth]{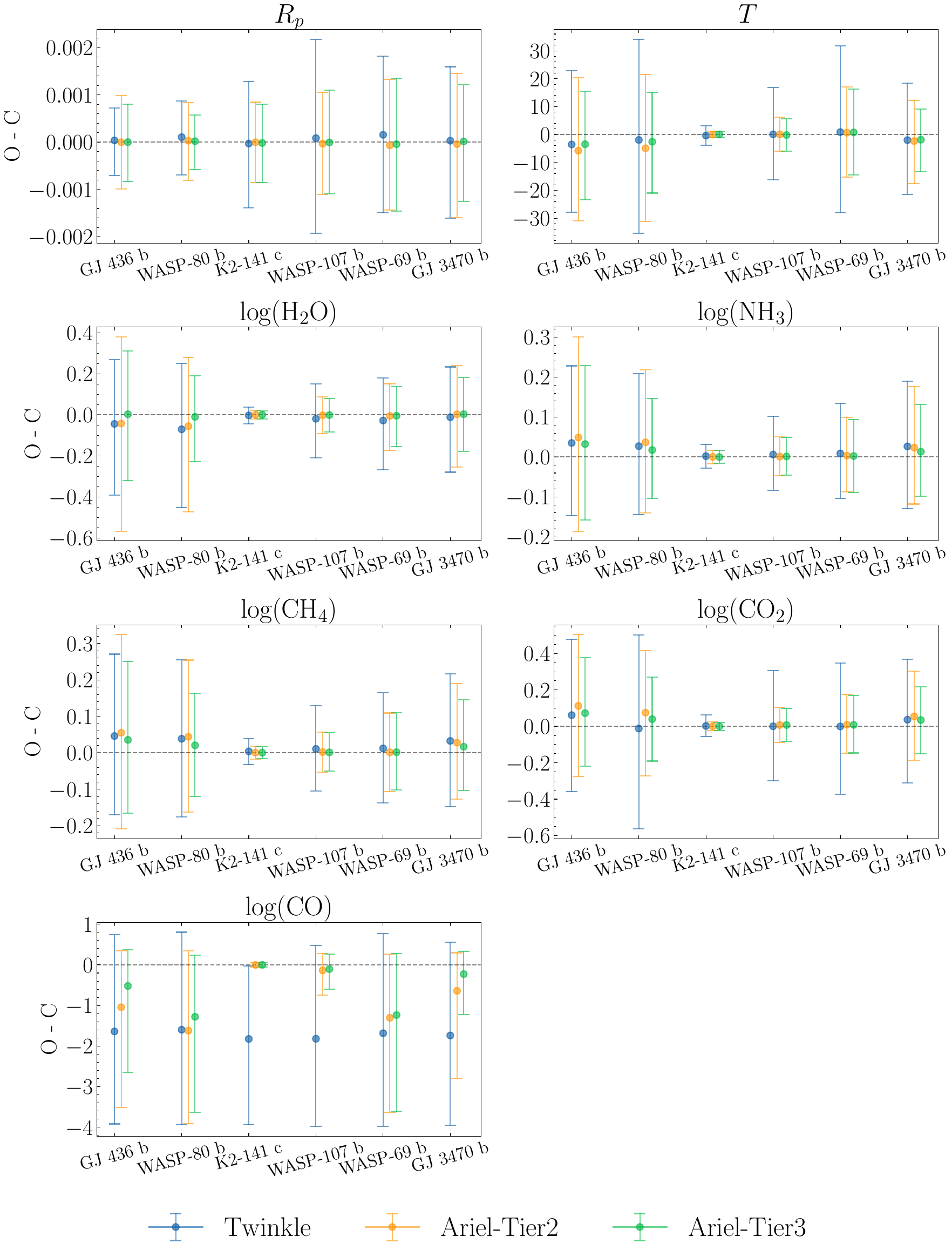}
    %\hspace{1.15cm}
    %\includegraphics[width=0.5\columnwidth]{figures/model1-20231103_1500_lp_chem_cst_5mols_bound-9/WASP-69 b_posteriors-COv1.png}
    %\includegraphics[width=0.5\columnwidth]{figures/model1-20231103_1500_lp_chem_cst_5mols_bound-9/WASP-107~b_posteriors-COv1.png}
    \put(41,6){
    \includegraphics[scale=0.20]{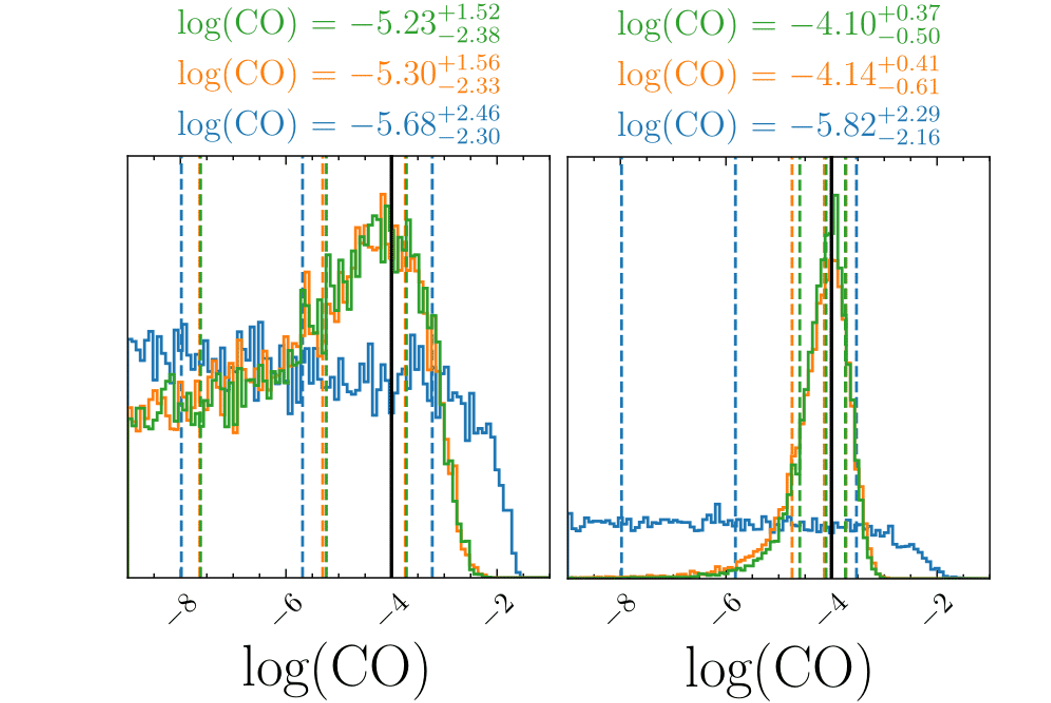}
    }
    \end{overpic}
    \caption{O-C results for the retrieved parameters w.r.t. their ground truths in model~1, which consists of a constant-with-altitude chemistry and isothermal T-P profile. 
    % Panels are labelled left-to-right, top-to-bottom (a-h). 
    % Panels (a-g) show retrieval results for each free parameter in model~1. 
    Values for each planet are displayed as the retrieved median value minus forward-model input value, with errorbars spanning 16th-84th quantiles shown. 
    The lower right panel shows CO posterior distributions for WASP-69~b (left) and WASP-107~b (right), singled out from the rest of the sample for visual purposes.   
    This showcases an example of different results between \textit{Ariel} and Twinkle: in Twinkle's case, only upper-limits can be obtained due to the unconstrained posterior distribution for both planets, while \textit{Ariel} is able to constrain the abundance of CO within 1 dex only for WASP-107~b.
    % only upper-limits can be obtained due to the non-Gaussian behavior of the posterior distribution, and where well-constrained values are obtained in the simulated \textit{Ariel} retrievals.
     }
    \label{fig:model1-results}
\end{figure*}

For retrievals conducted on both constant-with-altitude chemistry forward models (1 and 2), we find that planetary and atmospheric parameters are well-retrieved. In all cases, bar CO as discussed below, the truth value for each parameter is encompassed within the 1-$\sigma$  retrieval confidence intervals (defined as spanning the 16th to 84th quartiles). This is shown for model~1 in Figure~\ref{fig:model1-results}, with model 2 results showing similar behaviour. We make all summary results, along with individual corner plots and retrieved spectra available for the interested reader~\citep{Bocchieri2024}.

\begin{figure*}
    \centering
    \includegraphics[width=0.93\textwidth]{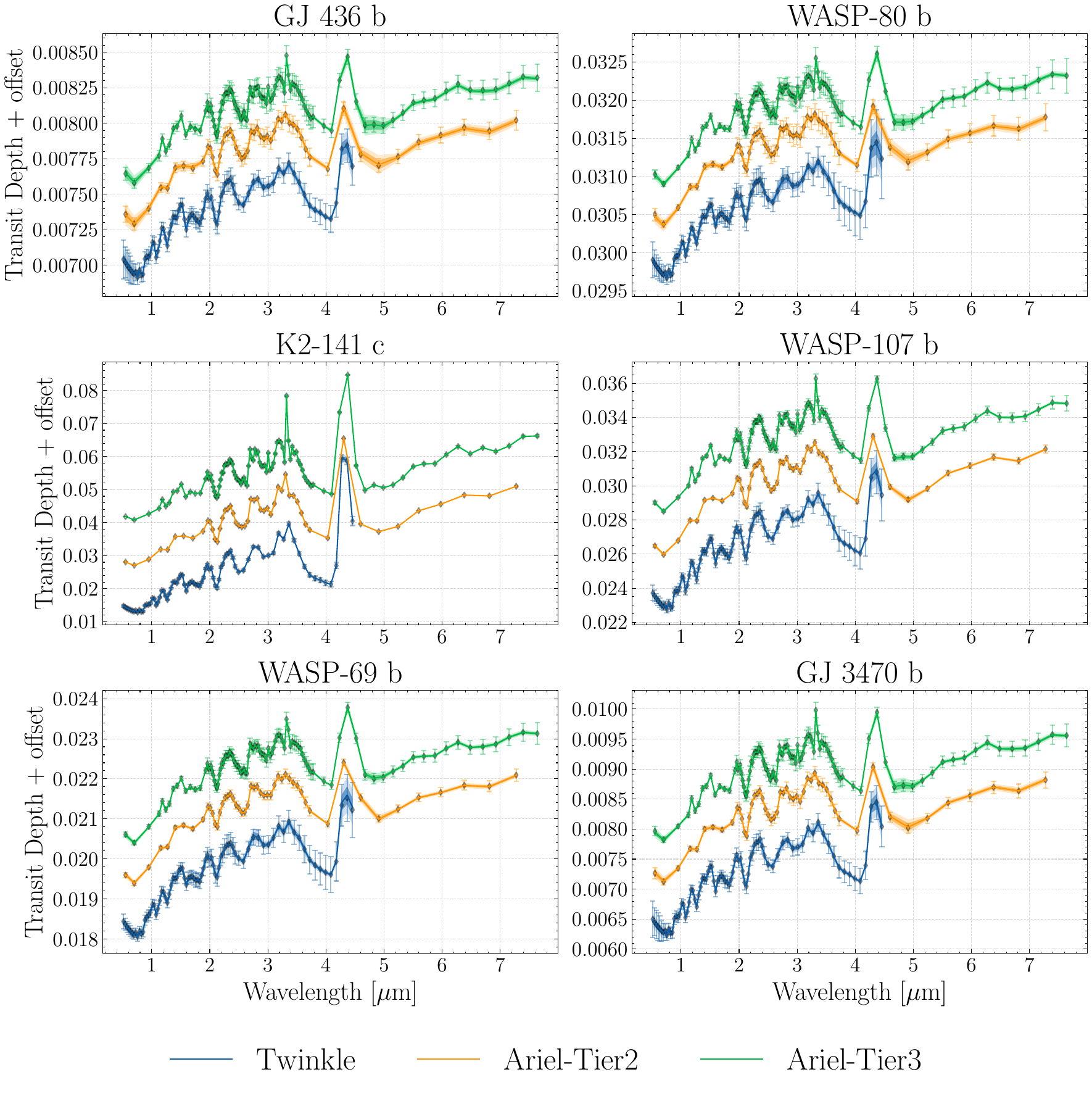}
    \caption{Retrieved spectra (solid lines with shaded 1- and 2-$\sigma$ confidence intervals) obtained from fitting the observed spectra (data points with errorbars) given each simulated instrument mode (blue: \textit{Twinkle}, orange: \textit{\textit{Ariel} Tier~2}, and green: \textit{\textit{Ariel} Tier~3}). The results for all planets in this study are shown, in model~1: cloud-free atmosphere with constant-with-altitude chemistry and isothermal T-P profile. Note: an arbitrary vertical offset between each spectrum is imposed for better visual representation.}
    \label{fig:fitted-spectrum-model1}
\end{figure*}

As can be seen in Figure~\ref{fig:model1-results}, \ce{CO} is generally retrieved only with an upper limit on its abundance, resulting in non-Gaussian marginalized posteriors and correspondingly large retrieval confidence intervals, as highlighted in the lower-right panel of the figure. We attribute the retrieval of an upper limit rather than a well-bounded VMR to a combination of: i) few or no spectral datapoints close to the strongest CO absorption feature at 4.7~$\mu$m (as can be seen in Figure~\ref{fig:fitted-spectrum-model1}) and ii) masking of short wavelength (1.6~$\mu$m and 2.34~$\mu$m) CO absorption features by molecules with larger cross-sections. This inability to constrain \ce{CO} well is more prominent for \textit{Twinkle} due to the wavelength cut-off at 4.5~$\mu$m and the larger errorbars on spectral datapoints in this region, which are evidenced in Figure~\ref{fig:noise-curves}. Additionally, we note that as \textit{Ariel} has the capability of optimizing the spectral grid during post-processing, constraints on \ce{CO} may be improved in specific cases.  %\commentLM{I did not understand this section. I had to read it twice and pass it through google. I suggest reshaping it in a less convoluted way.}

Whilst retrieved parameters are more well-constrained by simulated \textit{Ariel} observations than those by \textit{Twinkle}, the relative difference in errorbar magnitude is typically no larger than a factor of 2 for all retrieved parameters (bar \ce{CO}, as alluded to above). We note that this matches our prior expectations and attribute it to the larger mirror size of \textit{Ariel} relative to \textit{Twinkle}, as described in Section~\ref{subs:sensitivity-curves}. Though simulated \textit{Ariel} observations are able to provide tighter constraints on VMRs, for all included molecules bar CO retrieval confidence intervals are $<$ 1~dex. This combined with the good agreement found between retrieved parameter values obtained from simulated \textit{Twinkle} and \textit{Ariel} spectra, irrespective of observational tier, illustrates that the additional coverage provided by \textit{Ariel} at wavelengths longer than $4.5~\mu$m does not noticeably impact retrieved abundances in the studied cases and provides an excellent preliminary motivation to use the inferences from the analysis of early spectra taken by \textit{Twinkle} (scheduled to launch before \textit{Ariel}) to inform \textit{Ariel}'s observation plan.

% \begin{figure*}
%     \centering
% 	\includegraphics[width=\columnwidth]{figures/model1-20231103_1500_lp_chem_cst_5mols_bound-9/WASP-69 b_spectrum.pdf}
% 	\includegraphics[width=\columnwidth]{figures/model1-20231103_1500_lp_chem_cst_5mols_bound-9/WASP-80 b_spectrum.pdf}
% 	\includegraphics[width=\columnwidth]{figures/model1-20231103_1500_lp_chem_cst_5mols_bound-9/GJ 3470 b_spectrum.pdf}
% 	\includegraphics[width=\columnwidth]{figures/model1-20231103_1500_lp_chem_cst_5mols_bound-9/K2-141 c_spectrum.pdf}
% 	\includegraphics[width=\columnwidth]{figures/model1-20231103_1500_lp_chem_cst_5mols_bound-9/GJ 436 b_spectrum.pdf}
% 	\includegraphics[width=\columnwidth]{figures/model1-20231103_1500_lp_chem_cst_5mols_bound-9/WASP-107 b_spectrum.pdf}
%     \caption{Fitted spectra for each simulated instrument mode (blue: \textit{Twinkle}, orange: \textit{\textit{Ariel} Tier~2} and green: \textit{\textit{Ariel} Tier~3}) from model 1 retrievals: cloud-free atmosphere with constant-with-altitude chemistry and an isothermal T-P profile conducted on all planets in this study.}
%     \label{fig:fitted-spectrum-model1}
% \end{figure*}

%\newpage
\subsection{Equilibrium Chemistry Models}
\label{subs: Eq Chem Results}
\begin{figure*}
    \centering
    \includegraphics[width=0.93\textwidth]{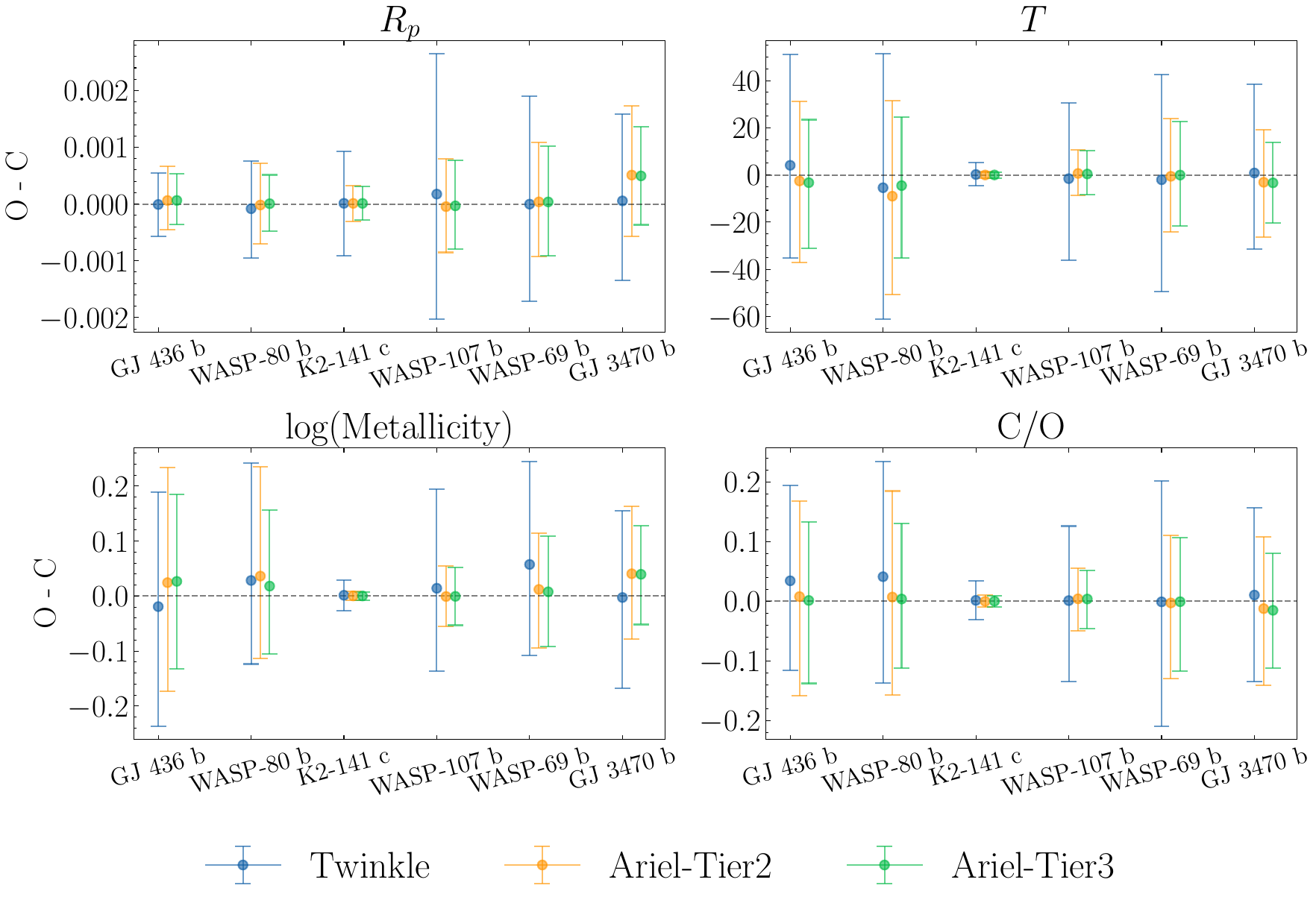}
    \caption{O-C results for the retrieved parameters w.r.t. their ground truths in model~3, a cloud-free atmosphere under equilibrium chemistry. Values are displayed for each planet as retrieved median value minus forward-model input value, with errorbars spanning 16th-84th quantiles ranges.}
    \label{fig:model3-results}
\end{figure*}

\begin{figure*}
    \centering
    \includegraphics[width=\textwidth]{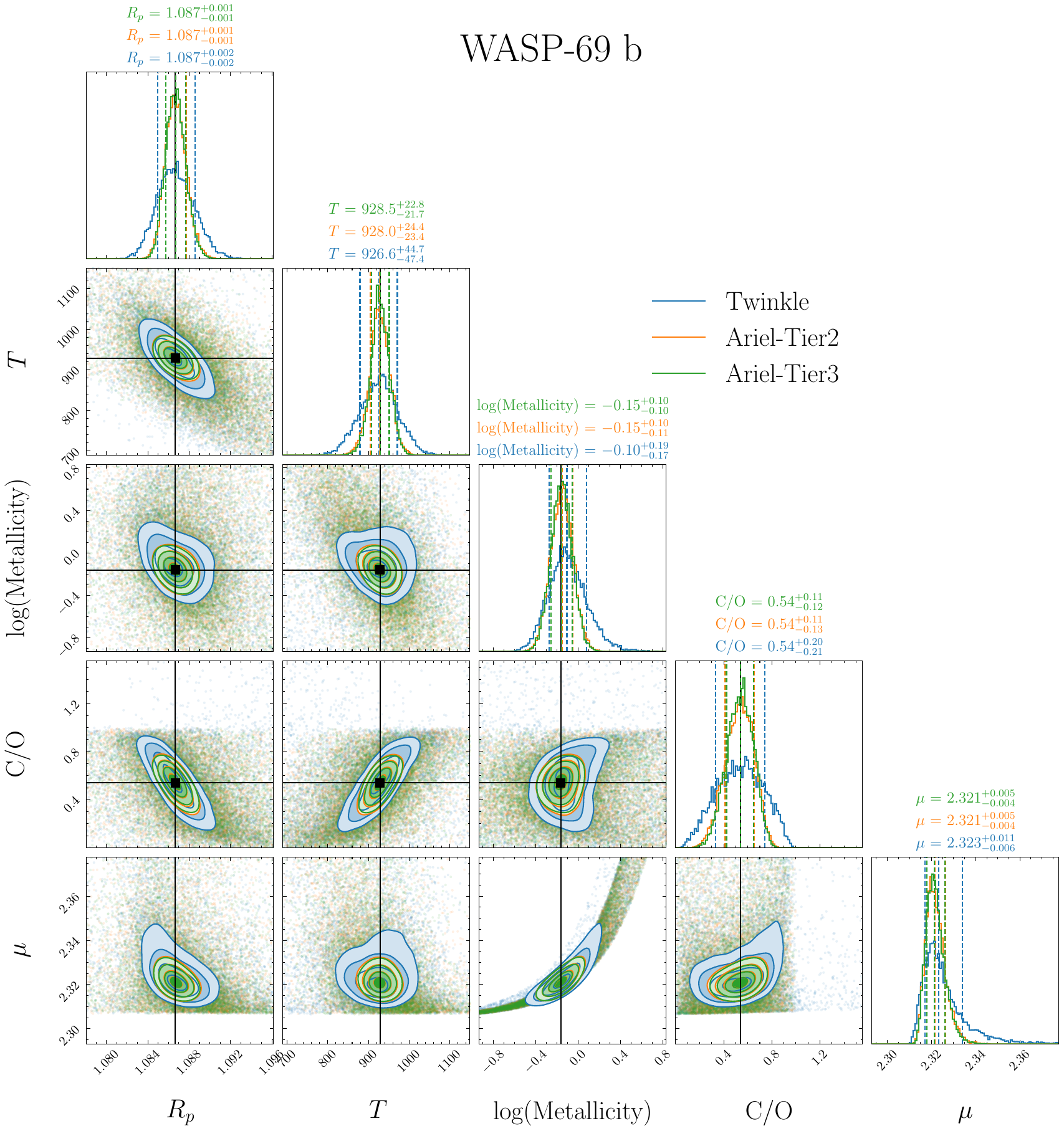}
    \caption{WASP-69~b corner plot for cloud-free, equilibrium chemistry atmospheric retrievals with an isothermal T-P profile (model 3). Contour plots and posterior distributions are shown for each simulated instrument mode (blue: \textit{Twinkle}, orange: \textit{Ariel} Tier~2 and green: \textit{Ariel} Tier~3), with black lines showing forward-model input truth values.}
    \label{fig:model3-corner-WASP-69b}
\end{figure*}

For retrievals conducted on simulated spectra produced by both equilibrium chemistry forward models (3 and 4) we find that retrieved atmospheric metallicity, (retrieved in log-space as log(\textit{Z})), and C/O ratios encompass truth values well within their 1-$\sigma$ confidence intervals, defined as above in Section~\ref{subs: Fixed Constant Chem Results}. This is shown in Figure~\ref{fig:model3-results} which summarises our retrieval results for model 3, whilst Figure~\ref{fig:model3-corner-WASP-69b} shows the nested sampling corner plot with posterior distributions for WASP-69~b in this model regime. We provide similar figures for model 4, and include retrieval spectra and corner plots for each planet in our sample in the supplementary material~\citep{Bocchieri2024}. 

As can be seen in Figure~\ref{fig:model3-corner-WASP-69b}, the 1-$\sigma$ confidence intervals on all parameters are well-constrained, with magnitudes $<$~0.25~dex for both log(\textit{Z}) and C/O.
In addition, as in models 1 and 2, we again find that there is good agreement between \textit{Ariel} and \textit{Twinkle}. As such, when combined with the precision of the retrievals, our results suggest that under the conditions imposed for the underlying forward models, searching for trends in metallicity (such as those of \cite{MZTrend_Wakeford2018, MZTrend_Welbanks2019} and \cite{NoMZTrend_Edwards2022_HST70}) against other planetary, stellar, or system parameters will be feasible with both \textit{Ariel} and \textit{Twinkle}. Furthermore, in instances where individual planets are not observed by both instruments, this ability to reliably and accurately retrieve atmospheric metallicity may lead to a larger observed population from which trends can be elucidated. \\
%\textcolor{green}{As an example, Figure~\ref{fig:model3-corner-WASP-69b} shows likelihood-dominated behaviour in the marginalized posterior histogram of each retrieval parameter\footnote{The interested reader may refer to plots provided in the supplimentary material\footnotemark[6]{} for all results.}.}
%\commentLM{well... if he has a free evening to enjoy all the plots while having a glass of red wine in the bathtub... see my comment in appendix}.}

Model 4 differs from model 3 by the inclusion of a grey cloud deck at 5000~Pa. This is well-retrieved alongside $R_\text{p}$, $T_\text{p}$, log(\textit{Z}) and the C/O ratio, with the truth values for each parameter falling within the 1-$\sigma$ confidence intervals. Although retrieved C/O and metallicity are minimally affected for the planets included in this study, where clouds are present, confidence intervals can be up to a factor of $\sim$2 larger for log(\textit{Z}). We caution that for higher-altitude cloud decks, spectral features are expected to be further truncated, which naturally impacts the ability to retrieve accurate molecular abundances and thereby atmospheric metallicity and C/O. Consequently, the minimal effect seen in our results on retrieved atmospheric metallicity and C/O ratio may not hold true. However, since factors that govern the formation mechanism, location and composition of clouds and hazes remain an active area of study, poorly constrained by current observations, we leave investigation of these effects to further work that is beyond the scope of this paper. 

% \begin{figure*}
% 	\includegraphics[width=\columnwidth]{figures/model3-20231103_1500_lp_chem_eq_tp_const/planet_radius.pdf}
% 	\includegraphics[width=\columnwidth]{figures/model3-20231103_1500_lp_chem_eq_tp_const/T.pdf}
% 	\includegraphics[width=\columnwidth]{figures/model3-20231103_1500_lp_chem_eq_tp_const/ace_co.pdf}
% 	\includegraphics[width=\columnwidth]{figures/model3-20231103_1500_lp_chem_eq_tp_const/log_ace_metallicity.pdf}
%     \caption{Panels are labelled left-to-right, top-to-bottom (a-d), showing retrieval results for each free parameter in model 3: cloud-free atmosphere under equilibrium chemistry. Values are displayed for each planet as forward-model input minus retrieved median, with errorbars spanning 16th-84th quartile ranges.}
%     \label{fig:model3-results}
% \end{figure*}

%\begin{figure*}
% 	\includegraphics[width=\columnwidth]{figures/model4-20231103_1500_lp_chem_eq_tp_const_with_clouds/planet_radius.pdf}
% 	\includegraphics[width=\columnwidth]{figures/model4-20231103_1500_lp_chem_eq_tp_const_with_clouds/T.pdf}
% 	\includegraphics[width=\columnwidth]{figures/model4-20231103_1500_lp_chem_eq_tp_const_with_clouds/ace_co.pdf}
% 	\includegraphics[width=\columnwidth]{figures/model4-20231103_1500_lp_chem_eq_tp_const_with_clouds/log_ace_metallicity.pdf}
% 	\includegraphics[width=\columnwidth]{figures/model4-20231103_1500_lp_chem_eq_tp_const_with_clouds/log_clouds_pressure.pdf}
%     \caption{Panels are labelled left-to-right, top-to-bottom (a-e), showing retrieval results for each free parameter in model 4: cloudy atmosphere with equilibrium chemistry and an isothermal T-P profile. Values are displayed for each planet as retrieved median minus forward-model input, with errorbars spanning 16th-84th quartile ranges.}
%     \label{fig:model4-results}
%\end{figure*}

\section{Discussion}
\label{sec:discussion}
Cool gaseous planets are an underexplored class of objects with great potential to expand our knowledge of atmospheric chemistry. Due to the inherently more challenging nature of observing cool planets, leveraging synergies between \textit{Ariel} and \textit{Twinkle} would be highly beneficial to uniformly characterize this class of objects and reveal possible trends. Whilst this pilot study does not factor in the presence of systematics, which is expected from \textit{Twinkle}'s low-Earth orbit, the use of conservative radiometric estimates (which include budgets for known instrumental effects such as line-of-sight jitter and the efficiency budget; see Section~\ref{subs:target-selection}) for \textit{Twinkle} may partially offset any negative impact. To the knowledge of the authors, any systematic trends are expected to be similar to those present in HST/WFC3 observations, barring the strong ``hooks'' and ``ramps'' from detector charge trapping / persistence, which will be mitigated. Assuming a 75\% loss across the Twinkle wavelength range might, however, not be representative of in-flight conditions given that systematics could be more prominent in certain wavelength ranges. Also, the assumption that post-processing noise between transits is uncorrelated can only be assessed during flight.

Throughout Section~\ref{sec:results} we demonstrate that for all planets in our \textit{Twinkle}-driven sample, median values of retrieval parameters are in excellent agreement across the instruments and observing modes simulated. We additionally show that input values used to generate the underlying model spectra are recovered within the 1-$\sigma$ confidence intervals. This is found to be true across the variety of atmospheric models that encompass two chemistry regimes: constant-with-altitude chemistry and equilibrium chemistry; two temperature-pressure profiles: isothermal and Guillot-like and two simple cloud scenarios: cloudless and deep grey clouds. Though this is a non-exhaustive list of atmospheric models, our results indicate, for our sample, observations of a given planet conducted by either \textit{Ariel} or \textit{Twinkle} will lead to consistent inferences on atmospheric properties. To robustly confirm this, this work could be extended to incorporate more complex T-P profiles, cloud physics and chemical networks. \\ % or be extended to planets in warmer temperature regimes. \\ % **LIMITS OF STUDY MENTIONED HERE**
% INCLUDE LINKING SENTENCE HERE -- TESS bright candidates possible with both Ariel and Twinkle...

Based on our findings, we posit that \textit{Twinkle} could be an effective precursor to \textit{Ariel} by potentially 1) informing the decision-making process on the \textit{Ariel} target list, and 2) expanding the total inventory of atmospheric spectroscopic data beyond what could be gained from independent operations.
%3) by leveraging the simultaneous operation of both telescopes. 
Both factors combined will be key to optimizing the total scientific output and legacy of both missions.
%%% \commentLM{First, I like this paragraph. Good job! I have a silly question: given that Twinkle seems to have better resolution in the visible (figure 3), can't it also help Ariel by characterising better the star and the haze area of the spectrum?} %%%
% **IMPLICATIONS OF STUDY MENTIONED ABOVE, WITH THINKING AND JUSTIFICATION INCLUDED BELOW**
This is particularly true given the differing but complementary sky coverages offered by both telescopes, with \textit{Twinkle} set to observe targets within $\pm 40^\circ$ of the ecliptic plane and \textit{Ariel} having continuous viewing zones at the ecliptic poles in addition to full sky coverage~\citep{ArielMRS_Edwards2022}. Consequently, where \textit{Twinkle} is able to observe \textit{Ariel} targets with only a marginal performance deficit (as shown in this work) %\textit{Ariel} performs marginally better (both in Tier~2 and 3), the results are overall consistent with Twinkle.
it opens up several possibilities to optimize scheduled observing time. For cool gaseous planets, this could include the observation of similar targets outside \textit{Twinkle}'s field of regard (FoR), or more challenging targets within. Here, some proportion of this population within \textit{Twinkle}'s FoR may be more feasible to observe with \textit{Ariel}, thanks to its larger mirror size. Combining both aforementioned factors would enable the total inventory of atmospheric spectroscopic data for $<$1000~K gaseous planets to be enlarged. 

This ideology is further punctuated by the inevitable discovery of additional new \textit{TESS} planets prior to the launch of each respective mission, many of which will orbit bright stars and are therefore highly amenable targets for transmission spectroscopy. New \textit{TESS} planets will include a substantial number in the cool gaseous planet regime, with at least 15 such planets confirmed by \textit{TESS} photometry and follow-up of \textit{TESS} \textit{mono-} and \textit{duo-}transit candidates \citep[e.g.,][]{TOI-1420b_TESS_LowDensity_2023, TOI-199b_TESS_CharacterisedwarmSaturn_2023, Mdwarf_TOI-3785b_Powers_2023,Mdwarf_TOI-904bc_Harris_2023, Mdwarf_TOI-5344b_Han_2023, HIP9618_TESS+CHEOPS_2023, TOI-5678b_TESS+CHEOPS+HARPS_2023, TOI-4600bc_TESS+ground-based_2023} since the creation of the target lists that underpin this work (the 2019 realization of the \textit{Ariel} Mission Reference Sample~\citep{ArielMRS_Edwards2019}\footnote{since updated to include new exoplanets discovered by \textit{TESS}~\citep{ArielMRS_Edwards2022} and continuously kept up to date with potential \textit{Ariel} targets at \url{https://github.com/arielmission-space/Mission\_Candidate\_Sample}.} and 2022 \textit{Twinkle} cool gaseous planets survey target list \citep{TwinkleCoolGaseousSurvey_Booth_2024}). With an increasing number of planets amenable to transmission spectroscopy continuing to populate this regime, the exploitation of synergies to maximise scientific output is imperative. % OR "would be a significant boon to the wider exoplanet community".

\section{Conclusions}
\label{sec:conclusions}
% the conclusion can summarize what we did, what we found and why it's important to further study this plus mention the impact for Ariel and Twinkle communities & the impact for exoplanet field as a whole
%\commentLM{We need a conclusion. I suggest reshaping a little the results and discussions. In my opinion, there is some confusion on what goes where (it always bugs me too). 1) Results should only list the direct results of your applied method. 2) Discussion is where you mention limits, implications, and justifications for your results (this is where you put your thinking). 3) Conclusions are where you summarise the importance of your study. Remember, the average reader reads the abstract, the conclusion, the figures, and maybe then everything else, so you need a good summarising conclusion section.}
This pilot study is a first attempt at exploring potential synergies between \textit{Twinkle} and \textit{Ariel} in the coming years. To investigate this, we selected a subsample of cool gaseous planets present in \textit{Twinkle}'s current proposed target list \citep{TwinkleCoolGaseousSurvey_Booth_2024} and a possible realization of the \textit{Ariel} Mission Reference Sample of~\citet{ArielMRS_Edwards2019}, which can be characterized within 10~visits. Our resulting selection is a biased, \textit{Twinkle}-driven sample of individual planetary targets that both missions are capable of observing at sufficient signal-to-noise. % **(1) MOTIVATION AND TARGET LIST
For each target within this work, we produce four synthetic spectra, spanning a combination of two chemistry regimes, two temperature-pressure profiles and two cloud scenarios. We bin these to spectra ``as observed'' by \textit{Ariel} and \textit{Twinkle} using representative radiometric noise estimates and spectral grids. % **(2) BRIEF, VAGUE MENTION OF MODELS
Our results showcase the possibility of confidently retrieving the relevant atmospheric parameters, without obvious biases. Within their 1-$\sigma$ confidence intervals, the retrieved median values readily encompass the input values used to generate the underlying spectra. Furthermore, we find excellent agreement for a given parameter and planet across the two telescopes simulated. Consequently, whilst \textit{Twinkle} requires additional transits (and therefore telescope time) to observe almost all targets, under the conditions imposed by our study, for our sample the performance deficit between \textit{Twinkle} and \textit{Ariel} is sufficiently small such that for all atmospheric properties, bar VMR(CO), inferred values are near-equivalent and only the associated uncertainties are improved by \textit{Ariel} observations.  % ** (3) KEY FINDING(s)
This motivates the suggestion that for planets present in the target lists of both missions, total scientific output could be optimized by exploiting their synergies. Conceivable methods include the use of \textit{Ariel} to observe planetary atmospheres that are unfeasible with \textit{Twinkle} and the use of \textit{Twinkle}'s prior observations to inform \textit{Ariel}'s observing schedule. Further optimization methods could also be conceived and implemented by both consortia. % ** (4) INTERPRETATION 
We strongly encourage future studies to further explore synergies that may exist between \textit{Ariel} and \textit{Twinkle}, as well as in conjunction with other instrumentation (e.g., JWST and ground-based telescopes). Future studies could also investigate known systematic effects, as well as the ability to combine data sets from the two space telescopes and what benefits could be gained from this. % ** (5) FUTURE WORK

\section*{Contributions}
Andrea Bocchieri and Luke Booth contributed with equal merits to the conception, scientific motivation, analysis, figure creation, and writing of the paper. All authors provided comments on the analysis. Andrea Bocchieri and Luke Booth edited the final manuscript. All authors read and approved the final manuscript. 

\section*{Acknowledgements}
The work of Andrea Bocchieri and Lorenzo V.~Mugnai was supported by the Italian Space Agency (ASI) with \textit{Ariel} grant n. 2021.5.HH.0. The work of Luke Booth was supported by a STFC doctoral training grant. The authors further wish to thank Ahmed Al-Refaie for his continued support during the retrieval analysis, and thank Matt Griffin and Subhajit Sarkar for their review of the paper prior to submission. Finally, we thank the BSSL team for their inputs and clarifications regarding Twinkle.

\section*{Data Availability}
The data used in this work will be shared on reasonable request made to the author(s).

\section*{Funding}
Open access funding provided by Università degli Studi di Roma La Sapienza within the CRUI-CARE Compact Agreement.

%%%%%%%%%%%%%%%%%%%% REFERENCES %%%%%%%%%%%%%%%%%%

% The best way to enter references is to use BibTeX:
\bibliography{aa}

%% BioMed_Central_Bib_Style_v1.01

\begin{thebibliography}{120}
% BibTex style file: bmc-mathphys.bst (version 2.1), 2014-07-24
\ifx \bisbn   \undefined \def \bisbn  #1{ISBN #1}\fi
\ifx \binits  \undefined \def \binits#1{#1}\fi
\ifx \bauthor  \undefined \def \bauthor#1{#1}\fi
\ifx \batitle  \undefined \def \batitle#1{#1}\fi
\ifx \bjtitle  \undefined \def \bjtitle#1{#1}\fi
\ifx \bvolume  \undefined \def \bvolume#1{\textbf{#1}}\fi
\ifx \byear  \undefined \def \byear#1{#1}\fi
\ifx \bissue  \undefined \def \bissue#1{#1}\fi
\ifx \bfpage  \undefined \def \bfpage#1{#1}\fi
\ifx \blpage  \undefined \def \blpage #1{#1}\fi
\ifx \burl  \undefined \def \burl#1{\textsf{#1}}\fi
\ifx \doiurl  \undefined \def \doiurl#1{\url{https://doi.org/#1}}\fi
\ifx \betal  \undefined \def \betal{\textit{et al.}}\fi
\ifx \binstitute  \undefined \def \binstitute#1{#1}\fi
\ifx \binstitutionaled  \undefined \def \binstitutionaled#1{#1}\fi
\ifx \bctitle  \undefined \def \bctitle#1{#1}\fi
\ifx \beditor  \undefined \def \beditor#1{#1}\fi
\ifx \bpublisher  \undefined \def \bpublisher#1{#1}\fi
\ifx \bbtitle  \undefined \def \bbtitle#1{#1}\fi
\ifx \bedition  \undefined \def \bedition#1{#1}\fi
\ifx \bseriesno  \undefined \def \bseriesno#1{#1}\fi
\ifx \blocation  \undefined \def \blocation#1{#1}\fi
\ifx \bsertitle  \undefined \def \bsertitle#1{#1}\fi
\ifx \bsnm \undefined \def \bsnm#1{#1}\fi
\ifx \bsuffix \undefined \def \bsuffix#1{#1}\fi
\ifx \bparticle \undefined \def \bparticle#1{#1}\fi
\ifx \barticle \undefined \def \barticle#1{#1}\fi
\bibcommenthead
\ifx \bconfdate \undefined \def \bconfdate #1{#1}\fi
\ifx \botherref \undefined \def \botherref #1{#1}\fi
\ifx \url \undefined \def \url#1{\textsf{#1}}\fi
\ifx \bchapter \undefined \def \bchapter#1{#1}\fi
\ifx \bbook \undefined \def \bbook#1{#1}\fi
\ifx \bcomment \undefined \def \bcomment#1{#1}\fi
\ifx \oauthor \undefined \def \oauthor#1{#1}\fi
\ifx \citeauthoryear \undefined \def \citeauthoryear#1{#1}\fi
\ifx \endbibitem  \undefined \def \endbibitem {}\fi
\ifx \bconflocation  \undefined \def \bconflocation#1{#1}\fi
\ifx \arxivurl  \undefined \def \arxivurl#1{\textsf{#1}}\fi
\csname PreBibitemsHook\endcsname

%%% 1
\bibitem[\protect\citeauthoryear{{Cassan} et~al.}{2012}]{Cassan2012}
\begin{barticle}
\bauthor{\bsnm{{Cassan}}, \binits{A.}},
\bauthor{\bsnm{{Kubas}}, \binits{D.}},
\bauthor{\bsnm{{Beaulieu}}, \binits{J.-P.}},
\bauthor{\bsnm{al.}}:
\batitle{{One or more bound planets per Milky Way star from microlensing observations}}.
\bjtitle{\nat}
\bvolume{481}(\bissue{7380}),
\bfpage{167}--\blpage{169}
(\byear{2012})
\doiurl{10.1038/nature10684}
{\href{https://arxiv.org/abs/1202.0903}{{arXiv:1202.0903}}}
{[astro-ph.EP]}
\end{barticle}
\endbibitem

%%% 2
\bibitem[\protect\citeauthoryear{{Batalha}}{2014}]{Populations_Batalha_2014PNAS}
\begin{barticle}
\bauthor{\bsnm{{Batalha}}, \binits{N.M.}}:
\batitle{{Exploring exoplanet populations with NASA's Kepler Mission}}.
\bjtitle{Proceedings of the National Academy of Science}
\bvolume{111}(\bissue{35}),
\bfpage{12647}--\blpage{12654}
(\byear{2014})
\doiurl{10.1073/pnas.1304196111}
{\href{https://arxiv.org/abs/1409.1904}{{arXiv:1409.1904}}}
{[astro-ph.EP]}
\end{barticle}
\endbibitem

%%% 3
\bibitem[\protect\citeauthoryear{{Rauer} et~al.}{2024}]{Rauer2024}
\begin{botherref}
\oauthor{\bsnm{{Rauer}}, \binits{H.}},
\oauthor{\bsnm{{Aerts}}, \binits{C.}},
\oauthor{\bsnm{{Cabrera}}, \binits{J.}},
\oauthor{\bsnm{al.}}:
{The PLATO Mission}.
arXiv e-prints,
2406--05447
(2024)
\doiurl{10.48550/arXiv.2406.05447}
{\href{https://arxiv.org/abs/2406.05447}{{arXiv:2406.05447}}}
{[astro-ph.IM]}
\end{botherref}
\endbibitem

%%% 4
\bibitem[\protect\citeauthoryear{{Quanz} et~al.}{2015}]{E-ELT_Quanz_2015}
\begin{barticle}
\bauthor{\bsnm{{Quanz}}, \binits{S.P.}},
\bauthor{\bsnm{{Crossfield}}, \binits{I.}},
\bauthor{\bsnm{{Meyer}}, \binits{M.R.}},
\bauthor{\bsnm{al.}}:
\batitle{{Direct detection of exoplanets in the 3-10 {\ensuremath{\mu}}m range with E-ELT/METIS}}.
\bjtitle{International Journal of Astrobiology}
\bvolume{14}(\bissue{2}),
\bfpage{279}--\blpage{289}
(\byear{2015})
\doiurl{10.1017/S1473550414000135}
{\href{https://arxiv.org/abs/1404.0831}{{arXiv:1404.0831}}}
{[astro-ph.IM]}
\end{barticle}
\endbibitem

%%% 5
\bibitem[\protect\citeauthoryear{{Neichel} et~al.}{2018}]{E-ELT_Neichel_2018}
\begin{bchapter}
\bauthor{\bsnm{{Neichel}}, \binits{B.}},
\bauthor{\bsnm{{Mouillet}}, \binits{D.}},
\bauthor{\bsnm{{Gendron}}, \binits{E.}},
\bauthor{\bsnm{al.}}:
\bctitle{{Overview of the European Extremely Large Telescope and its instrument suite}}.
In: \beditor{\bsnm{{Di Matteo}}, \binits{P.}},
\beditor{\bsnm{{Billebaud}}, \binits{F.}},
\beditor{\bsnm{{Herpin}}, \binits{F.}},
\beditor{\bsnm{al.}} (eds.)
\bbtitle{SF2A-2018: Proceedings of the Annual Meeting of the French Society of Astronomy and Astrophysics},
p.
(\byear{2018}).
\doiurl{10.48550/arXiv.1812.06639}
\end{bchapter}
\endbibitem

%%% 6
\bibitem[\protect\citeauthoryear{{Skidmore} et~al.}{2015}]{TMT_Skidmore_2015}
\begin{barticle}
\bauthor{\bsnm{{Skidmore}}, \binits{W.}},
\bauthor{\bsnm{{TMT International Science Development Teams}}},
\bauthor{\bsnm{{Science Advisory Committee}}, \binits{T.}}:
\batitle{{Thirty Meter Telescope Detailed Science Case: 2015}}.
\bjtitle{Research in Astronomy and Astrophysics}
\bvolume{15}(\bissue{12}),
\bfpage{1945}
(\byear{2015})
\doiurl{10.1088/1674-4527/15/12/001}
{\href{https://arxiv.org/abs/1505.01195}{{arXiv:1505.01195}}}
{[astro-ph.IM]}
\end{barticle}
\endbibitem

%%% 7
\bibitem[\protect\citeauthoryear{{Tinetti} et~al.}{2018}]{Tinetti2018}
\begin{barticle}
\bauthor{\bsnm{{Tinetti}}, \binits{G.}},
\bauthor{\bsnm{{Drossart}}, \binits{P.}},
\bauthor{\bsnm{{Eccleston}}, \binits{P.}},
\bauthor{\bsnm{al.}}:
\batitle{{A chemical survey of exoplanets with ARIEL}}.
\bjtitle{Experimental Astronomy}
\bvolume{46}(\bissue{1}),
\bfpage{135}--\blpage{209}
(\byear{2018})
\doiurl{10.1007/s10686-018-9598-x}
\end{barticle}
\endbibitem

%%% 8
\bibitem[\protect\citeauthoryear{{Edwards} et~al.}{2019}]{Twinkle_Edwards_2019}
\begin{barticle}
\bauthor{\bsnm{{Edwards}}, \binits{B.}},
\bauthor{\bsnm{{Rice}}, \binits{M.}},
\bauthor{\bsnm{{Zingales}}, \binits{T.}},
\bauthor{\bsnm{al.}}:
\batitle{{Exoplanet spectroscopy and photometry with the Twinkle space telescope}}.
\bjtitle{Experimental Astronomy}
\bvolume{47}(\bissue{1-2}),
\bfpage{29}--\blpage{63}
(\byear{2019})
\doiurl{10.1007/s10686-018-9611-4}
{\href{https://arxiv.org/abs/1811.08348}{{arXiv:1811.08348}}}
{[astro-ph.EP]}
\end{barticle}
\endbibitem

%%% 9
\bibitem[\protect\citeauthoryear{{Stotesbury} et~al.}{2022}]{Twinkle_SPIEPoster_Stotesbury_2022}
\begin{bchapter}
\bauthor{\bsnm{{Stotesbury}}, \binits{I.}},
\bauthor{\bsnm{{Edwards}}, \binits{B.}},
\bauthor{\bsnm{{Lavigne}}, \binits{J.-F.}},
\bauthor{\bsnm{al.}}:
\bctitle{{Twinkle: a small satellite spectroscopy mission for the next phase of exoplanet science}}.
In: \beditor{\bsnm{{Coyle}}, \binits{L.E.}},
\beditor{\bsnm{{Matsuura}}, \binits{S.}},
\beditor{\bsnm{{Perrin}}, \binits{M.D.}} (eds.)
\bbtitle{Space Telescopes and Instrumentation 2022: Optical, Infrared, and Millimeter Wave}.
\bsertitle{Society of Photo-Optical Instrumentation Engineers (SPIE) Conference Series},
vol. \bseriesno{12180},
p. \bfpage{1218033}
(\byear{2022}).
\doiurl{10.1117/12.2641373}
\end{bchapter}
\endbibitem

%%% 10
\bibitem[\protect\citeauthoryear{{Quanz} et~al.}{2022}]{LIFE1_Quanz_2022}
\begin{barticle}
\bauthor{\bsnm{{Quanz}}, \binits{S.P.}},
\bauthor{\bsnm{{Ottiger}}, \binits{M.}},
\bauthor{\bsnm{{Fontanet}}, \binits{E.}},
\bauthor{\bsnm{al.}}:
\batitle{{Large Interferometer For Exoplanets (LIFE). I. Improved exoplanet detection yield estimates for a large mid-infrared space-interferometer mission}}.
\bjtitle{\aap}
\bvolume{664},
\bfpage{21}
(\byear{2022})
\doiurl{10.1051/0004-6361/202140366}
{\href{https://arxiv.org/abs/2101.07500}{{arXiv:2101.07500}}}
{[astro-ph.EP]}
\end{barticle}
\endbibitem

%%% 11
\bibitem[\protect\citeauthoryear{{Fulton} et~al.}{2017}]{RadiusValley_Fulton_2017}
\begin{barticle}
\bauthor{\bsnm{{Fulton}}, \binits{B.J.}},
\bauthor{\bsnm{{Petigura}}, \binits{E.A.}},
\bauthor{\bsnm{{Howard}}, \binits{A.W.}},
\bauthor{\bsnm{al.}}:
\batitle{{The California-Kepler Survey. III. A Gap in the Radius Distribution of Small Planets}}.
\bjtitle{\aj}
\bvolume{154}(\bissue{3}),
\bfpage{109}
(\byear{2017})
\doiurl{10.3847/1538-3881/aa80eb}
{\href{https://arxiv.org/abs/1703.10375}{{arXiv:1703.10375}}}
{[astro-ph.EP]}
\end{barticle}
\endbibitem

%%% 12
\bibitem[\protect\citeauthoryear{{Szab{\'o}} and {Kiss}}{2011}]{NeptunianDesert_Szab_2011}
\begin{barticle}
\bauthor{\bsnm{{Szab{\'o}}}, \binits{G.M.}},
\bauthor{\bsnm{{Kiss}}, \binits{L.L.}}:
\batitle{{A Short-period Censor of Sub-Jupiter Mass Exoplanets with Low Density}}.
\bjtitle{\apjl}
\bvolume{727}(\bissue{2}),
\bfpage{44}
(\byear{2011})
\doiurl{10.1088/2041-8205/727/2/L44}
{\href{https://arxiv.org/abs/1012.4791}{{arXiv:1012.4791}}}
{[astro-ph.EP]}
\end{barticle}
\endbibitem

%%% 13
\bibitem[\protect\citeauthoryear{{Mazeh} et~al.}{2016}]{NeptunianDesert_Mazeh_2016}
\begin{barticle}
\bauthor{\bsnm{{Mazeh}}, \binits{T.}},
\bauthor{\bsnm{{Holczer}}, \binits{T.}},
\bauthor{\bsnm{{Faigler}}, \binits{S.}}:
\batitle{{Dearth of short-period Neptunian exoplanets: A desert in period-mass and period-radius planes}}.
\bjtitle{\aap}
\bvolume{589},
\bfpage{75}
(\byear{2016})
\doiurl{10.1051/0004-6361/201528065}
{\href{https://arxiv.org/abs/1602.07843}{{arXiv:1602.07843}}}
{[astro-ph.EP]}
\end{barticle}
\endbibitem

%%% 14
\bibitem[\protect\citeauthoryear{{Edwards} et~al.}{2023}]{NeptunianDesert_LTT9779b_Edwards2023}
\begin{barticle}
\bauthor{\bsnm{{Edwards}}, \binits{B.}},
\bauthor{\bsnm{{Changeat}}, \binits{Q.}},
\bauthor{\bsnm{{Tsiaras}}, \binits{A.}},
\bauthor{\bsnm{al.}}:
\batitle{{Characterizing a World Within the Hot-Neptune Desert: Transit Observations of LTT 9779 b with the Hubble Space Telescope/WFC3}}.
\bjtitle{\aj}
\bvolume{166}(\bissue{4}),
\bfpage{158}
(\byear{2023})
\doiurl{10.3847/1538-3881/acea77}
{\href{https://arxiv.org/abs/2306.13645}{{arXiv:2306.13645}}}
{[astro-ph.EP]}
\end{barticle}
\endbibitem

%%% 15
\bibitem[\protect\citeauthoryear{{Rogers} and {Seager}}{2010}]{InteriorDegeneracies_Rogers&Seager_2010}
\begin{barticle}
\bauthor{\bsnm{{Rogers}}, \binits{L.A.}},
\bauthor{\bsnm{{Seager}}, \binits{S.}}:
\batitle{{A Framework for Quantifying the Degeneracies of Exoplanet Interior Compositions}}.
\bjtitle{\apj}
\bvolume{712}(\bissue{2}),
\bfpage{974}--\blpage{991}
(\byear{2010})
\doiurl{10.1088/0004-637X/712/2/974}
{\href{https://arxiv.org/abs/0912.3288}{{arXiv:0912.3288}}}
{[astro-ph.EP]}
\end{barticle}
\endbibitem

%%% 16
\bibitem[\protect\citeauthoryear{{Hu} et~al.}{2024}]{Hu2024}
\begin{barticle}
\bauthor{\bsnm{{Hu}}, \binits{R.}},
\bauthor{\bsnm{{Bello-Arufe}}, \binits{A.}},
\bauthor{\bsnm{{Zhang}}, \binits{M.}},
\bauthor{\bsnm{al.}}:
\batitle{{A secondary atmosphere on the rocky exoplanet 55 Cancri e}}.
\bjtitle{\nat}
\bvolume{630}(\bissue{8017}),
\bfpage{609}--\blpage{612}
(\byear{2024})
\doiurl{10.1038/s41586-024-07432-x}
{\href{https://arxiv.org/abs/2405.04744}{{arXiv:2405.04744}}}
{[astro-ph.EP]}
\end{barticle}
\endbibitem

%%% 17
\bibitem[\protect\citeauthoryear{{Zieba} et~al.}{2023a}]{WASP-47e_JWSTprop_Zieba_2023}
\begin{botherref}
\oauthor{\bsnm{{Zieba}}, \binits{S.}},
\oauthor{\bsnm{{Kreidberg}}, \binits{L.}},
\oauthor{\bsnm{{Miguel}}, \binits{Y.}},
\oauthor{\bsnm{{Vanderburg}}, \binits{A.}},
\oauthor{\bsnm{{Zilinskas}}, \binits{M.}},
\oauthor{\bsnm{{van Buchem}}, \binits{C.}}:
{Exploring the boundary between rocky and gaseous planets with WASP-47 e}.
JWST Proposal. Cycle 2, ID. \#3615
(2023)
\end{botherref}
\endbibitem

%%% 18
\bibitem[\protect\citeauthoryear{{Zieba} et~al.}{2023b}]{LHS3844b_JWSTprop_Zieba_2023}
\begin{botherref}
\oauthor{\bsnm{{Zieba}}, \binits{S.}},
\oauthor{\bsnm{{Hu}}, \binits{R.}},
\oauthor{\bsnm{{Kreidberg}}, \binits{L.}},
\oauthor{\bsnm{{Morley}}, \binits{C.}},
\oauthor{\bsnm{{Tenthoff}}, \binits{M.}},
\oauthor{\bsnm{{Wohler}}, \binits{C.}},
\oauthor{\bsnm{{Wohlfarth}}, \binits{K.}}:
{The search for regolith on the airless exoplanet LHS 3844 b}.
JWST Proposal. Cycle 2, ID. \#4008
(2023)
\end{botherref}
\endbibitem

%%% 19
\bibitem[\protect\citeauthoryear{{Cadieux} et~al.}{2024}]{Cadieux2024}
\begin{botherref}
\oauthor{\bsnm{{Cadieux}}, \binits{C.}},
\oauthor{\bsnm{{Doyon}}, \binits{R.}},
\oauthor{\bsnm{{MacDonald}}, \binits{R.J.}},
\oauthor{\bsnm{al.}}:
{Transmission Spectroscopy of the Habitable Zone Exoplanet LHS 1140 b with JWST/NIRISS}.
arXiv e-prints,
2406--15136
(2024)
\doiurl{10.48550/arXiv.2406.15136}
{\href{https://arxiv.org/abs/2406.15136}{{arXiv:2406.15136}}}
{[astro-ph.EP]}
\end{botherref}
\endbibitem

%%% 20
\bibitem[\protect\citeauthoryear{{Benneke} et~al.}{2024}]{LP791-18d_JWSTprop_Benneke_2024}
\begin{botherref}
\oauthor{\bsnm{{Benneke}}, \binits{B.}},
\oauthor{\bsnm{{Piaulet Ghorayeb}}, \binits{C.}},
\oauthor{\bsnm{{Roy}}, \binits{P.-A.}}:
{Thermal emission of a cool, potentially volcanically active exo-Earth}.
JWST Proposal. Cycle 3, ID. \#6457
(2024)
\end{botherref}
\endbibitem

%%% 21
\bibitem[\protect\citeauthoryear{{Madhusudhan} et~al.}{2023}]{K2-18b_CH4_DMS_2023}
\begin{barticle}
\bauthor{\bsnm{{Madhusudhan}}, \binits{N.}},
\bauthor{\bsnm{{Sarkar}}, \binits{S.}},
\bauthor{\bsnm{{Constantinou}}, \binits{S.}},
\bauthor{\bsnm{al.}}:
\batitle{{Carbon-bearing Molecules in a Possible Hycean Atmosphere}}.
\bjtitle{\apjl}
\bvolume{956}(\bissue{1}),
\bfpage{13}
(\byear{2023})
\doiurl{10.3847/2041-8213/acf577}
{\href{https://arxiv.org/abs/2309.05566}{{arXiv:2309.05566}}}
{[astro-ph.EP]}
\end{barticle}
\endbibitem

%%% 22
\bibitem[\protect\citeauthoryear{{Holmberg} and {Madhusudhan}}{2024}]{TOI-270d_JWSTNIRSpec_Holmberg&Madhusudhan_2024A&A}
\begin{barticle}
\bauthor{\bsnm{{Holmberg}}, \binits{M.}},
\bauthor{\bsnm{{Madhusudhan}}, \binits{N.}}:
\batitle{{Possible Hycean conditions in the sub-Neptune TOI-270 d}}.
\bjtitle{\aap}
\bvolume{683},
\bfpage{2}
(\byear{2024})
\doiurl{10.1051/0004-6361/202348238}
{\href{https://arxiv.org/abs/2403.03244}{{arXiv:2403.03244}}}
{[astro-ph.EP]}
\end{barticle}
\endbibitem

%%% 23
\bibitem[\protect\citeauthoryear{{Beatty} et~al.}{2024}]{GJ3470b_JWSTNIRCam_Beatty2024arXiv}
\begin{botherref}
\oauthor{\bsnm{{Beatty}}, \binits{T.G.}},
\oauthor{\bsnm{{Welbanks}}, \binits{L.}},
\oauthor{\bsnm{{Schlawin}}, \binits{E.}},
\oauthor{\bsnm{al.}}:
{Sulfur Dioxide and Other Molecular Species in the Atmosphere of the Sub-Neptune GJ 3470 b}.
arXiv e-prints,
2406--04450
(2024)
\doiurl{10.48550/arXiv.2406.04450}
{\href{https://arxiv.org/abs/2406.04450}{{arXiv:2406.04450}}}
{[astro-ph.EP]}
\end{botherref}
\endbibitem

%%% 24
\bibitem[\protect\citeauthoryear{{Sing} et~al.}{2024}]{WASP-107_JWSTNIRSpec_Sing_2024arXiv}
\begin{barticle}
\bauthor{\bsnm{{Sing}}, \binits{D.K.}},
\bauthor{\bsnm{{Rustamkulov}}, \binits{Z.}},
\bauthor{\bsnm{{Thorngren}}, \binits{D.P.}},
\bauthor{\bsnm{al.}}:
\batitle{{A warm Neptune's methane reveals core mass and vigorous atmospheric mixing}}.
\bjtitle{\nat}
\bvolume{630}(\bissue{8018}),
\bfpage{831}--\blpage{835}
(\byear{2024})
\doiurl{10.1038/s41586-024-07395-z}
{\href{https://arxiv.org/abs/2405.11027}{{arXiv:2405.11027}}}
{[astro-ph.EP]}
\end{barticle}
\endbibitem

%%% 25
\bibitem[\protect\citeauthoryear{{Rustamkulov} et~al.}{2023}]{Rustamkulov2023}
\begin{barticle}
\bauthor{\bsnm{{Rustamkulov}}, \binits{Z.}},
\bauthor{\bsnm{{Sing}}, \binits{D.K.}},
\bauthor{\bsnm{{Mukherjee}}, \binits{S.}},
\bauthor{\bsnm{al.}}:
\batitle{{Early Release Science of the exoplanet WASP-39b with JWST NIRSpec PRISM}}.
\bjtitle{\nat}
\bvolume{614}(\bissue{7949}),
\bfpage{659}--\blpage{663}
(\byear{2023})
\doiurl{10.1038/s41586-022-05677-y}
{\href{https://arxiv.org/abs/2211.10487}{{arXiv:2211.10487}}}
{[astro-ph.EP]}
\end{barticle}
\endbibitem

%%% 26
\bibitem[\protect\citeauthoryear{{Bell} et~al.}{2023}]{Bell2023}
\begin{botherref}
\oauthor{\bsnm{{Bell}}, \binits{T.J.}},
\oauthor{\bsnm{{Kreidberg}}, \binits{L.}},
\oauthor{\bsnm{{Kendrew}}, \binits{S.}},
\oauthor{\bsnm{al.}}:
{A First Look at the JWST MIRI/LRS Phase Curve of WASP-43b}.
arXiv e-prints,
2301--06350
(2023)
\doiurl{10.48550/arXiv.2301.06350}
{\href{https://arxiv.org/abs/2301.06350}{{arXiv:2301.06350}}}
{[astro-ph.IM]}
\end{botherref}
\endbibitem

%%% 27
\bibitem[\protect\citeauthoryear{{Changeat} et~al.}{2020}]{Changeat2020}
\begin{barticle}
\bauthor{\bsnm{{Changeat}}, \binits{Q.}},
\bauthor{\bsnm{{Al-Refaie}}, \binits{A.}},
\bauthor{\bsnm{{Mugnai}}, \binits{L.V.}},
\bauthor{\bsnm{al.}}:
\batitle{{Alfnoor: A Retrieval Simulation of the Ariel Target List}}.
\bjtitle{\aj}
\bvolume{160}(\bissue{2}),
\bfpage{80}
(\byear{2020})
\doiurl{10.3847/1538-3881/ab9a53}
{\href{https://arxiv.org/abs/2003.01839}{{arXiv:2003.01839}}}
{[astro-ph.EP]}
\end{barticle}
\endbibitem

%%% 28
\bibitem[\protect\citeauthoryear{{Mugnai} et~al.}{2021}]{Mugnai2021a}
\begin{barticle}
\bauthor{\bsnm{{Mugnai}}, \binits{L.V.}},
\bauthor{\bsnm{{Al-Refaie}}, \binits{A.}},
\bauthor{\bsnm{{Bocchieri}}, \binits{A.}},
\bauthor{\bsnm{al.}}:
\batitle{{Alfnoor: Assessing the Information Content of Ariel's Low-resolution Spectra with Planetary Population Studies}}.
\bjtitle{\aj}
\bvolume{162}(\bissue{6}),
\bfpage{288}
(\byear{2021})
\doiurl{10.3847/1538-3881/ac2e92}
{\href{https://arxiv.org/abs/2110.00503}{{arXiv:2110.00503}}}
{[astro-ph.EP]}
\end{barticle}
\endbibitem

%%% 29
\bibitem[\protect\citeauthoryear{{Bocchieri} et~al.}{2023}]{Bocchieri2023}
\begin{barticle}
\bauthor{\bsnm{{Bocchieri}}, \binits{A.}},
\bauthor{\bsnm{{Mugnai}}, \binits{L.V.}},
\bauthor{\bsnm{{Pascale}}, \binits{E.}},
\bauthor{\bsnm{al.}}:
\batitle{{Detecting molecules in Ariel low resolution transmission spectra}}.
\bjtitle{Experimental Astronomy}
\bvolume{56}(\bissue{2-3}),
\bfpage{605}--\blpage{644}
(\byear{2023})
\doiurl{10.1007/s10686-023-09911-x}
{\href{https://arxiv.org/abs/2309.06817}{{arXiv:2309.06817}}}
{[astro-ph.EP]}
\end{barticle}
\endbibitem

%%% 30
\bibitem[\protect\citeauthoryear{{Tinetti} et~al.}{2021}]{Tinetti2021}
\begin{botherref}
\oauthor{\bsnm{{Tinetti}}, \binits{G.}},
\oauthor{\bsnm{{Eccleston}}, \binits{P.}},
\oauthor{\bsnm{{Haswell}}, \binits{C.}},
\oauthor{\bsnm{al.}}:
{Ariel: Enabling planetary science across light-years}.
arXiv e-prints,
2104--04824
(2021)
\doiurl{10.48550/arXiv.2104.04824}
{\href{https://arxiv.org/abs/2104.04824}{{arXiv:2104.04824}}}
{[astro-ph.IM]}
\end{botherref}
\endbibitem

%%% 31
\bibitem[\protect\citeauthoryear{{Encrenaz} et~al.}{2015}]{Encrenaz2015}
\begin{barticle}
\bauthor{\bsnm{{Encrenaz}}, \binits{T.}},
\bauthor{\bsnm{{Tinetti}}, \binits{G.}},
\bauthor{\bsnm{{Tessenyi}}, \binits{M.}},
\bauthor{\bsnm{al.}}:
\batitle{{Transit spectroscopy of exoplanets from space: how to optimize the wavelength coverage and spectral resolving power}}.
\bjtitle{Experimental Astronomy}
\bvolume{40}(\bissue{2-3}),
\bfpage{523}--\blpage{543}
(\byear{2015})
\doiurl{10.1007/s10686-014-9415-0}
\end{barticle}
\endbibitem

%%% 32
\bibitem[\protect\citeauthoryear{{Eccleston} et~al.}{2024}]{Eccleston2024}
\begin{bchapter}
\bauthor{\bsnm{{Eccleston}}, \binits{P.}},
\bauthor{\bsnm{{Caldwell}}, \binits{A.}},
\bauthor{\bsnm{{Bishop}}, \binits{G.}},
\bauthor{\bsnm{al.}}:
\bctitle{{The Ariel payload design post-PDR}}.
In: \beditor{\bsnm{{Coyle}}, \binits{L.E.}},
\beditor{\bsnm{{Matsuura}}, \binits{S.}},
\beditor{\bsnm{{Perrin}}, \binits{M.D.}} (eds.)
\bbtitle{Space Telescopes and Instrumentation 2024: Optical, Infrared, and Millimeter Wave}.
\bsertitle{Society of Photo-Optical Instrumentation Engineers (SPIE) Conference Series},
vol. \bseriesno{13092},
p. \bfpage{130921}
(\byear{2024}).
\doiurl{10.1117/12.3019713}
\end{bchapter}
\endbibitem

%%% 33
\bibitem[\protect\citeauthoryear{{Pace} et~al.}{2024}]{Pace2024}
\begin{bchapter}
\bauthor{\bsnm{{Pace}}, \binits{E.}},
\bauthor{\bsnm{{Adler Abreu}}, \binits{M.}},
\bauthor{\bsnm{{Alonso}}, \binits{G.}},
\bauthor{\bsnm{al.}}:
\bctitle{{The telescope assembly of the Ariel space mission: an updated overview}}.
In: \beditor{\bsnm{{Coyle}}, \binits{L.E.}},
\beditor{\bsnm{{Matsuura}}, \binits{S.}},
\beditor{\bsnm{{Perrin}}, \binits{M.D.}} (eds.)
\bbtitle{Space Telescopes and Instrumentation 2024: Optical, Infrared, and Millimeter Wave}.
\bsertitle{Society of Photo-Optical Instrumentation Engineers (SPIE) Conference Series},
vol. \bseriesno{13092},
p. \bfpage{130921}
(\byear{2024}).
\doiurl{10.1117/12.3018735}
\end{bchapter}
\endbibitem

%%% 34
\bibitem[\protect\citeauthoryear{{Rataj} et~al.}{2019}]{Rataj2019}
\begin{bchapter}
\bauthor{\bsnm{{Rataj}}, \binits{M.}},
\bauthor{\bsnm{{Wawer}}, \binits{P.}},
\bauthor{\bsnm{{Skup}}, \binits{K.}},
\bauthor{\bsnm{al.}}:
\bctitle{{Design of fine guidance system (FGS) for ARIEL mission}}.
In: \beditor{\bsnm{{Romaniuk}}, \binits{R.S.}},
\beditor{\bsnm{{Linczuk}}, \binits{M.}} (eds.)
\bbtitle{Photonics Applications in Astronomy, Communications, Industry, and High-Energy Physics Experiments 2019}.
\bsertitle{Society of Photo-Optical Instrumentation Engineers (SPIE) Conference Series},
vol. \bseriesno{11176},
p. \bfpage{111763}
(\byear{2019}).
\doiurl{10.1117/12.2536800}
\end{bchapter}
\endbibitem

%%% 35
\bibitem[\protect\citeauthoryear{{Martignac} et~al.}{2022}]{Martignac2022}
\begin{bchapter}
\bauthor{\bsnm{{Martignac}}, \binits{J.}},
\bauthor{\bsnm{{Amiaux}}, \binits{J.}},
\bauthor{\bsnm{{Berth{\'e}}}, \binits{M.}},
\bauthor{\bsnm{al.}}:
\bctitle{{AIRS: ARIEL IR spectrometer status}}.
In: \beditor{\bsnm{{Coyle}}, \binits{L.E.}},
\beditor{\bsnm{{Matsuura}}, \binits{S.}},
\beditor{\bsnm{{Perrin}}, \binits{M.D.}} (eds.)
\bbtitle{Space Telescopes and Instrumentation 2022: Optical, Infrared, and Millimeter Wave}.
\bsertitle{Society of Photo-Optical Instrumentation Engineers (SPIE) Conference Series},
vol. \bseriesno{12180},
p. \bfpage{1218012}
(\byear{2022}).
\doiurl{10.1117/12.2628920}
\end{bchapter}
\endbibitem

%%% 36
\bibitem[\protect\citeauthoryear{{Bocchieri} et~al.}{2025}]{Bocchieri:2025}
\begin{barticle}
\bauthor{\bsnm{{Bocchieri}}, \binits{A.}},
\bauthor{\bsnm{{Mugnai}}, \binits{L.V.}},
\bauthor{\bsnm{{Pascale}}, \binits{E.}},
\bauthor{\bsnm{al.}}:
\batitle{{De-jittering Ariel: An optimized algorithm}}.
\bjtitle{Experimental Astronomy}
\bvolume{59}(\bissue{3}),
\bfpage{31}
(\byear{2025})
\doiurl{10.1007/s10686-025-09999-3}
{\href{https://arxiv.org/abs/2504.12907}{{arXiv:2504.12907}}}
{[astro-ph.EP]}
\end{barticle}
\endbibitem

%%% 37
\bibitem[\protect\citeauthoryear{{Mugnai} et~al.}{2020}]{Mugnai2020}
\begin{barticle}
\bauthor{\bsnm{{Mugnai}}, \binits{L.V.}},
\bauthor{\bsnm{{Pascale}}, \binits{E.}},
\bauthor{\bsnm{{Edwards}}, \binits{B.}},
\bauthor{\bsnm{al.}}:
\batitle{{ArielRad: the Ariel radiometric model}}.
\bjtitle{Experimental Astronomy}
\bvolume{50}(\bissue{2-3}),
\bfpage{303}--\blpage{328}
(\byear{2020})
\doiurl{10.1007/s10686-020-09676-7}
{\href{https://arxiv.org/abs/2009.07824}{{arXiv:2009.07824}}}
{[astro-ph.IM]}
\end{barticle}
\endbibitem

%%% 38
\bibitem[\protect\citeauthoryear{{Edwards} et~al.}{2019}]{ArielMRS_Edwards2019}
\begin{barticle}
\bauthor{\bsnm{{Edwards}}, \binits{B.}},
\bauthor{\bsnm{{Mugnai}}, \binits{L.}},
\bauthor{\bsnm{{Tinetti}}, \binits{G.}},
\bauthor{\bsnm{{Pascale}}, \binits{E.}},
\bauthor{\bsnm{{Sarkar}}, \binits{S.}}:
\batitle{{An Updated Study of Potential Targets for Ariel}}.
\bjtitle{\aj}
\bvolume{157}(\bissue{6}),
\bfpage{242}
(\byear{2019})
\doiurl{10.3847/1538-3881/ab1cb9}
{\href{https://arxiv.org/abs/1905.04959}{{arXiv:1905.04959}}}
{[astro-ph.EP]}
\end{barticle}
\endbibitem

%%% 39
\bibitem[\protect\citeauthoryear{{Turrini} et~al.}{2018}]{Turrini2018}
\begin{barticle}
\bauthor{\bsnm{{Turrini}}, \binits{D.}},
\bauthor{\bsnm{{Miguel}}, \binits{Y.}},
\bauthor{\bsnm{{Zingales}}, \binits{T.}},
\bauthor{\bsnm{al.}}:
\batitle{{The contribution of the ARIEL space mission to the study of planetary formation}}.
\bjtitle{Experimental Astronomy}
\bvolume{46}(\bissue{1}),
\bfpage{45}--\blpage{65}
(\byear{2018})
\doiurl{10.1007/s10686-017-9570-1}
{\href{https://arxiv.org/abs/1804.06179}{{arXiv:1804.06179}}}
{[astro-ph.EP]}
\end{barticle}
\endbibitem

%%% 40
\bibitem[\protect\citeauthoryear{{Pearson} et~al.}{2022}]{Pearson2022}
\begin{barticle}
\bauthor{\bsnm{{Pearson}}, \binits{C.}},
\bauthor{\bsnm{{Malaguti}}, \binits{G.}},
\bauthor{\bsnm{{Sarkar}}, \binits{S.}},
\bauthor{\bsnm{al.}}:
\batitle{{The Ariel ground segment and instrument operations science data centre}}.
\bjtitle{Experimental Astronomy}
\bvolume{53}(\bissue{2}),
\bfpage{773}--\blpage{806}
(\byear{2022})
\doiurl{10.1007/s10686-020-09691-8}
\end{barticle}
\endbibitem

%%% 41
\bibitem[\protect\citeauthoryear{{Puig} et~al.}{2018}]{Puig2018}
\begin{barticle}
\bauthor{\bsnm{{Puig}}, \binits{L.}},
\bauthor{\bsnm{{Pilbratt}}, \binits{G.}},
\bauthor{\bsnm{{Heske}}, \binits{A.}},
\bauthor{\bsnm{al.}}:
\batitle{{The Phase A study of the ESA M4 mission candidate ARIEL}}.
\bjtitle{Experimental Astronomy}
\bvolume{46}(\bissue{1}),
\bfpage{211}--\blpage{239}
(\byear{2018})
\doiurl{10.1007/s10686-018-9604-3}
\end{barticle}
\endbibitem

%%% 42
\bibitem[\protect\citeauthoryear{{Stotesbury} et~al.}{2024}]{Twinkle_Stotesbury_2024SPIE}
\begin{bchapter}
\bauthor{\bsnm{{Stotesbury}}, \binits{I.}},
\bauthor{\bsnm{{Bradley}}, \binits{L.}},
\bauthor{\bsnm{{Wilcock}}, \binits{B.J.}},
\bauthor{\bsnm{{Tinetti}}, \binits{G.}},
\bauthor{\bsnm{{Tessenyi}}, \binits{M.}},
\bauthor{\bsnm{{Savini}}, \binits{G.}},
\bauthor{\bsnm{{Windred}}, \binits{P.}},
\bauthor{\bsnm{{Tennyson}}, \binits{J.}}:
\bctitle{{Twinkle: programme updates}}.
In: \beditor{\bsnm{{Coyle}}, \binits{L.E.}},
\beditor{\bsnm{{Matsuura}}, \binits{S.}},
\beditor{\bsnm{{Perrin}}, \binits{M.D.}} (eds.)
\bbtitle{Space Telescopes and Instrumentation 2024: Optical, Infrared, and Millimeter Wave}.
\bsertitle{Society of Photo-Optical Instrumentation Engineers (SPIE) Conference Series},
vol. \bseriesno{13092},
p. \bfpage{1309213}
(\byear{2024}).
\doiurl{10.1117/12.3018791}
\end{bchapter}
\endbibitem

%%% 43
\bibitem[\protect\citeauthoryear{{Booth} et~al.}{2024}]{TwinkleCoolGaseousSurvey_Booth_2024}
\begin{barticle}
\bauthor{\bsnm{{Booth}}, \binits{L.}},
\bauthor{\bsnm{{Sarkar}}, \binits{S.}},
\bauthor{\bsnm{{Griffin}}, \binits{M.}},
\bauthor{\bsnm{{Edwards}}, \binits{B.}}:
\batitle{{Cool gaseous exoplanets: surveying the new frontier with twinkle}}.
\bjtitle{\mnras}
(\byear{2024})
\doiurl{10.1093/mnras/stae461}
\end{barticle}
\endbibitem

%%% 44
\bibitem[\protect\citeauthoryear{{Yip} et~al.}{2020}]{Yip2020a}
\begin{barticle}
\bauthor{\bsnm{{Yip}}, \binits{K.H.}},
\bauthor{\bsnm{{Tsiaras}}, \binits{A.}},
\bauthor{\bsnm{{Waldmann}}, \binits{I.P.}},
\bauthor{\bsnm{al.}}:
\batitle{{Integrating Light Curve and Atmospheric Modeling of Transiting Exoplanets}}.
\bjtitle{\aj}
\bvolume{160}(\bissue{4}),
\bfpage{171}
(\byear{2020})
\doiurl{10.3847/1538-3881/abaabc}
{\href{https://arxiv.org/abs/1811.04686}{{arXiv:1811.04686}}}
{[astro-ph.EP]}
\end{barticle}
\endbibitem

%%% 45
\bibitem[\protect\citeauthoryear{{Changeat} et~al.}{2020}]{Changeat2020b}
\begin{barticle}
\bauthor{\bsnm{{Changeat}}, \binits{Q.}},
\bauthor{\bsnm{{Keyte}}, \binits{L.}},
\bauthor{\bsnm{{Waldmann}}, \binits{I.P.}},
\bauthor{\bsnm{al.}}:
\batitle{{Impact of Planetary Mass Uncertainties on Exoplanet Atmospheric Retrievals}}.
\bjtitle{\apj}
\bvolume{896}(\bissue{2}),
\bfpage{107}
(\byear{2020})
\doiurl{10.3847/1538-4357/ab8f8b}
{\href{https://arxiv.org/abs/1908.06305}{{arXiv:1908.06305}}}
{[astro-ph.EP]}
\end{barticle}
\endbibitem

%%% 46
\bibitem[\protect\citeauthoryear{{Rustamkulov} et~al.}{2023}]{Rustamkulov2022}
\begin{barticle}
\bauthor{\bsnm{{Rustamkulov}}, \binits{Z.}},
\bauthor{\bsnm{{Sing}}, \binits{D.K.}},
\bauthor{\bsnm{{Mukherjee}}, \binits{S.}},
\bauthor{\bsnm{al.}}:
\batitle{{Early Release Science of the exoplanet WASP-39b with JWST NIRSpec PRISM}}.
\bjtitle{\nat}
\bvolume{614}(\bissue{7949}),
\bfpage{659}--\blpage{663}
(\byear{2023})
\doiurl{10.1038/s41586-022-05677-y}
{\href{https://arxiv.org/abs/2211.10487}{{arXiv:2211.10487}}}
{[astro-ph.EP]}
\end{barticle}
\endbibitem

%%% 47
\bibitem[\protect\citeauthoryear{{Feinstein} et~al.}{2023}]{Feinstein2022}
\begin{barticle}
\bauthor{\bsnm{{Feinstein}}, \binits{A.D.}},
\bauthor{\bsnm{{Radica}}, \binits{M.}},
\bauthor{\bsnm{{Welbanks}}, \binits{L.}},
\bauthor{\bsnm{al.}}:
\batitle{{Early Release Science of the exoplanet WASP-39b with JWST NIRISS}}.
\bjtitle{\nat}
\bvolume{614}(\bissue{7949}),
\bfpage{670}--\blpage{675}
(\byear{2023})
\doiurl{10.1038/s41586-022-05674-1}
{\href{https://arxiv.org/abs/2211.10493}{{arXiv:2211.10493}}}
{[astro-ph.EP]}
\end{barticle}
\endbibitem

%%% 48
\bibitem[\protect\citeauthoryear{{Ahrer} et~al.}{2023}]{Ahrer2022b}
\begin{barticle}
\bauthor{\bsnm{{Ahrer}}, \binits{E.-M.}},
\bauthor{\bsnm{{Stevenson}}, \binits{K.B.}},
\bauthor{\bsnm{{Mansfield}}, \binits{M.}},
\bauthor{\bsnm{al.}}:
\batitle{{Early Release Science of the exoplanet WASP-39b with JWST NIRCam}}.
\bjtitle{\nat}
\bvolume{614}(\bissue{7949}),
\bfpage{653}--\blpage{658}
(\byear{2023})
\doiurl{10.1038/s41586-022-05590-4}
{\href{https://arxiv.org/abs/2211.10489}{{arXiv:2211.10489}}}
{[astro-ph.EP]}
\end{barticle}
\endbibitem

%%% 49
\bibitem[\protect\citeauthoryear{{Tsai} et~al.}{2023}]{Tsai2022}
\begin{barticle}
\bauthor{\bsnm{{Tsai}}, \binits{S.-M.}},
\bauthor{\bsnm{{Lee}}, \binits{E.K.H.}},
\bauthor{\bsnm{{Powell}}, \binits{D.}},
\bauthor{\bsnm{al.}}:
\batitle{{Photochemically produced SO$_{2}$ in the atmosphere of WASP-39b}}.
\bjtitle{\nat}
\bvolume{617}(\bissue{7961}),
\bfpage{483}--\blpage{487}
(\byear{2023})
\doiurl{10.1038/s41586-023-05902-2}
{\href{https://arxiv.org/abs/2211.10490}{{arXiv:2211.10490}}}
{[astro-ph.EP]}
\end{barticle}
\endbibitem

%%% 50
\bibitem[\protect\citeauthoryear{{Benneke} et~al.}{2019}]{Benneke2019}
\begin{barticle}
\bauthor{\bsnm{{Benneke}}, \binits{B.}},
\bauthor{\bsnm{{Knutson}}, \binits{H.A.}},
\bauthor{\bsnm{{Lothringer}}, \binits{J.}},
\bauthor{\bsnm{al.}}:
\batitle{{A sub-Neptune exoplanet with a low-metallicity methane-depleted atmosphere and Mie-scattering clouds}}.
\bjtitle{Nature Astronomy}
\bvolume{3},
\bfpage{813}--\blpage{821}
(\byear{2019})
\doiurl{10.1038/s41550-019-0800-5}
{\href{https://arxiv.org/abs/1907.00449}{{arXiv:1907.00449}}}
{[astro-ph.EP]}
\end{barticle}
\endbibitem

%%% 51
\bibitem[\protect\citeauthoryear{{Tsiaras} et~al.}{2019}]{Tsiaras2019}
\begin{barticle}
\bauthor{\bsnm{{Tsiaras}}, \binits{A.}},
\bauthor{\bsnm{{Waldmann}}, \binits{I.P.}},
\bauthor{\bsnm{{Tinetti}}, \binits{G.}},
\bauthor{\bsnm{al.}}:
\batitle{{Water vapour in the atmosphere of the habitable-zone eight-Earth-mass planet K2-18 b}}.
\bjtitle{Nature Astronomy}
\bvolume{3},
\bfpage{1086}--\blpage{1091}
(\byear{2019})
\doiurl{10.1038/s41550-019-0878-9}
{\href{https://arxiv.org/abs/1909.05218}{{arXiv:1909.05218}}}
{[astro-ph.EP]}
\end{barticle}
\endbibitem

%%% 52
\bibitem[\protect\citeauthoryear{{Fortney} et~al.}{2020}]{Fortney2020}
\begin{barticle}
\bauthor{\bsnm{{Fortney}}, \binits{J.J.}},
\bauthor{\bsnm{{Visscher}}, \binits{C.}},
\bauthor{\bsnm{{Marley}}, \binits{M.S.}},
\bauthor{\bsnm{al.}}:
\batitle{{Beyond Equilibrium Temperature: How the Atmosphere/Interior Connection Affects the Onset of Methane, Ammonia, and Clouds in Warm Transiting Giant Planets}}.
\bjtitle{\aj}
\bvolume{160}(\bissue{6}),
\bfpage{288}
(\byear{2020})
\doiurl{10.3847/1538-3881/abc5bd}
{\href{https://arxiv.org/abs/2010.00146}{{arXiv:2010.00146}}}
{[astro-ph.EP]}
\end{barticle}
\endbibitem

%%% 53
\bibitem[\protect\citeauthoryear{{Tinetti} et~al.}{2013}]{Tinetti2013}
\begin{barticle}
\bauthor{\bsnm{{Tinetti}}, \binits{G.}},
\bauthor{\bsnm{{Encrenaz}}, \binits{T.}},
\bauthor{\bsnm{{Coustenis}}, \binits{A.}}:
\batitle{{Spectroscopy of planetary atmospheres in our Galaxy}}.
\bjtitle{\aapr}
\bvolume{21},
\bfpage{63}
(\byear{2013})
\doiurl{10.1007/s00159-013-0063-6}
\end{barticle}
\endbibitem

%%% 54
\bibitem[\protect\citeauthoryear{{Encrenaz} et~al.}{2022}]{Encrenaz2022}
\begin{barticle}
\bauthor{\bsnm{{Encrenaz}}, \binits{T.}},
\bauthor{\bsnm{{Coustenis}}, \binits{A.}},
\bauthor{\bsnm{{Gilli}}, \binits{G.}},
\bauthor{\bsnm{al.}}:
\batitle{{Observability of temperate exoplanets with Ariel}}.
\bjtitle{Experimental Astronomy}
\bvolume{53}(\bissue{2}),
\bfpage{375}--\blpage{390}
(\byear{2022})
\doiurl{10.1007/s10686-021-09793-x}
\end{barticle}
\endbibitem

%%% 55
\bibitem[\protect\citeauthoryear{{Roy} et~al.}{2023}]{HST_11transits_GJ9827d_Roy_2023}
\begin{barticle}
\bauthor{\bsnm{{Roy}}, \binits{P.-A.}},
\bauthor{\bsnm{{Benneke}}, \binits{B.}},
\bauthor{\bsnm{{Piaulet}}, \binits{C.}},
\bauthor{\bsnm{al.}}:
\batitle{{Water Absorption in the Transmission Spectrum of the Water World Candidate GJ 9827 d}}.
\bjtitle{\apjl}
\bvolume{954}(\bissue{2}),
\bfpage{52}
(\byear{2023})
\doiurl{10.3847/2041-8213/acebf0}
{\href{https://arxiv.org/abs/2309.10845}{{arXiv:2309.10845}}}
{[astro-ph.EP]}
\end{barticle}
\endbibitem

%%% 56
\bibitem[\protect\citeauthoryear{{Kreidberg} et~al.}{2014}]{HST_15transits_GJ1214b_Kreidberg_2014}
\begin{barticle}
\bauthor{\bsnm{{Kreidberg}}, \binits{L.}},
\bauthor{\bsnm{{Bean}}, \binits{J.L.}},
\bauthor{\bsnm{{D{\'e}sert}}, \binits{J.-M.}},
\bauthor{\bsnm{al.}}:
\batitle{{Clouds in the atmosphere of the super-Earth exoplanet GJ1214b}}.
\bjtitle{\nat}
\bvolume{505}(\bissue{7481}),
\bfpage{69}--\blpage{72}
(\byear{2014})
\doiurl{10.1038/nature12888}
{\href{https://arxiv.org/abs/1401.0022}{{arXiv:1401.0022}}}
{[astro-ph.EP]}
\end{barticle}
\endbibitem

%%% 57
\bibitem[\protect\citeauthoryear{{Kulow} et~al.}{2014}]{GJ436b_LymanAlpha_2014}
\begin{barticle}
\bauthor{\bsnm{{Kulow}}, \binits{J.R.}},
\bauthor{\bsnm{{France}}, \binits{K.}},
\bauthor{\bsnm{{Linsky}}, \binits{J.}},
\bauthor{\bsnm{{Loyd}}, \binits{R.O.P.}}:
\batitle{{Ly{\ensuremath{\alpha}} Transit Spectroscopy and the Neutral Hydrogen Tail of the Hot Neptune GJ 436b}}.
\bjtitle{\apj}
\bvolume{786}(\bissue{2}),
\bfpage{132}
(\byear{2014})
\doiurl{10.1088/0004-637X/786/2/132}
{\href{https://arxiv.org/abs/1403.6834}{{arXiv:1403.6834}}}
{[astro-ph.EP]}
\end{barticle}
\endbibitem

%%% 58
\bibitem[\protect\citeauthoryear{{Lothringer} et~al.}{2018}]{GJ436b_HSTspectrum_2018}
\begin{barticle}
\bauthor{\bsnm{{Lothringer}}, \binits{J.D.}},
\bauthor{\bsnm{{Benneke}}, \binits{B.}},
\bauthor{\bsnm{{Crossfield}}, \binits{I.J.M.}},
\bauthor{\bsnm{al.}}:
\batitle{{An HST/STIS Optical Transmission Spectrum of Warm Neptune GJ 436b}}.
\bjtitle{\aj}
\bvolume{155}(\bissue{2}),
\bfpage{66}
(\byear{2018})
\doiurl{10.3847/1538-3881/aaa008}
{\href{https://arxiv.org/abs/1801.00412}{{arXiv:1801.00412}}}
{[astro-ph.EP]}
\end{barticle}
\endbibitem

%%% 59
\bibitem[\protect\citeauthoryear{{Madhusudhan} and {Seager}}{2011}]{GJ436b_DaysideEmission_2011}
\begin{barticle}
\bauthor{\bsnm{{Madhusudhan}}, \binits{N.}},
\bauthor{\bsnm{{Seager}}, \binits{S.}}:
\batitle{{High Metallicity and Non-equilibrium Chemistry in the Dayside Atmosphere of hot-Neptune GJ 436b}}.
\bjtitle{\apj}
\bvolume{729}(\bissue{1}),
\bfpage{41}
(\byear{2011})
\doiurl{10.1088/0004-637X/729/1/41}
{\href{https://arxiv.org/abs/1004.5121}{{arXiv:1004.5121}}}
{[astro-ph.SR]}
\end{barticle}
\endbibitem

%%% 60
\bibitem[\protect\citeauthoryear{{Bonfils} et~al.}{2012}]{GJ3470b_Detection_2012}
\begin{barticle}
\bauthor{\bsnm{{Bonfils}}, \binits{X.}},
\bauthor{\bsnm{{Gillon}}, \binits{M.}},
\bauthor{\bsnm{{Udry}}, \binits{S.}},
\bauthor{\bsnm{al.}}:
\batitle{{A hot Uranus transiting the nearby M dwarf GJ 3470. Detected with HARPS velocimetry. Captured in transit with TRAPPIST photometry}}.
\bjtitle{\aap}
\bvolume{546},
\bfpage{27}
(\byear{2012})
\doiurl{10.1051/0004-6361/201219623}
{\href{https://arxiv.org/abs/1206.5307}{{arXiv:1206.5307}}}
{[astro-ph.EP]}
\end{barticle}
\endbibitem

%%% 61
\bibitem[\protect\citeauthoryear{{Tarrants} and {Li}}{2023}]{GJ3470b_NoExtraPlanets_2023}
\begin{botherref}
\oauthor{\bsnm{{Tarrants}}, \binits{T.}},
\oauthor{\bsnm{{Li}}, \binits{A.}}:
{No Evidence for Additional Planets at GJ 3470 from TESS and Archival Radial Velocities}.
arXiv e-prints,
2305--02551
(2023)
\doiurl{10.48550/arXiv.2305.02551}
{\href{https://arxiv.org/abs/2305.02551}{{arXiv:2305.02551}}}
{[astro-ph.EP]}
\end{botherref}
\endbibitem

%%% 62
\bibitem[\protect\citeauthoryear{{Benneke} et~al.}{2019}]{GJ3470b_HSTTransmisionSpectra_2019}
\begin{barticle}
\bauthor{\bsnm{{Benneke}}, \binits{B.}},
\bauthor{\bsnm{{Knutson}}, \binits{H.A.}},
\bauthor{\bsnm{{Lothringer}}, \binits{J.}},
\bauthor{\bsnm{al.}}:
\batitle{{A sub-Neptune exoplanet with a low-metallicity methane-depleted atmosphere and Mie-scattering clouds}}.
\bjtitle{Nature Astronomy}
\bvolume{3},
\bfpage{813}--\blpage{821}
(\byear{2019})
\doiurl{10.1038/s41550-019-0800-5}
{\href{https://arxiv.org/abs/1907.00449}{{arXiv:1907.00449}}}
{[astro-ph.EP]}
\end{barticle}
\endbibitem

%%% 63
\bibitem[\protect\citeauthoryear{{Ninan} et~al.}{2020}]{GJ3470b_HeliumTriplet_2020}
\begin{barticle}
\bauthor{\bsnm{{Ninan}}, \binits{J.P.}},
\bauthor{\bsnm{{Stefansson}}, \binits{G.}},
\bauthor{\bsnm{{Mahadevan}}, \binits{S.}},
\bauthor{\bsnm{al.}}:
\batitle{{Evidence for He I 10830 {\r{A}} Absorption during the Transit of a Warm Neptune around the M-dwarf GJ 3470 with the Habitable-zone Planet Finder}}.
\bjtitle{\apj}
\bvolume{894}(\bissue{2}),
\bfpage{97}
(\byear{2020})
\doiurl{10.3847/1538-4357/ab8559}
{\href{https://arxiv.org/abs/1910.02070}{{arXiv:1910.02070}}}
{[astro-ph.EP]}
\end{barticle}
\endbibitem

%%% 64
\bibitem[\protect\citeauthoryear{{Lamp{\'o}n} et~al.}{2021}]{GJ3470b_HeliumTriplet_2021}
\begin{barticle}
\bauthor{\bsnm{{Lamp{\'o}n}}, \binits{M.}},
\bauthor{\bsnm{{L{\'o}pez-Puertas}}, \binits{M.}},
\bauthor{\bsnm{{Sanz-Forcada}}, \binits{J.}},
\bauthor{\bsnm{al.}}:
\batitle{{Modelling the He I triplet absorption at 10 830 {\r{A}} in the atmospheres of HD 189733 b and GJ 3470 b}}.
\bjtitle{\aap}
\bvolume{647},
\bfpage{129}
(\byear{2021})
\doiurl{10.1051/0004-6361/202039417}
{\href{https://arxiv.org/abs/2101.09393}{{arXiv:2101.09393}}}
{[astro-ph.EP]}
\end{barticle}
\endbibitem

%%% 65
\bibitem[\protect\citeauthoryear{{Malavolta} et~al.}{2018}]{K2-141c_DISCOVERY_Malavolta2018}
\begin{barticle}
\bauthor{\bsnm{{Malavolta}}, \binits{L.}},
\bauthor{\bsnm{{Mayo}}, \binits{A.W.}},
\bauthor{\bsnm{{Louden}}, \binits{T.}},
\bauthor{\bsnm{al.}}:
\batitle{{An Ultra-short Period Rocky Super-Earth with a Secondary Eclipse and a Neptune-like Companion around K2-141}}.
\bjtitle{\aj}
\bvolume{155}(\bissue{3}),
\bfpage{107}
(\byear{2018})
\doiurl{10.3847/1538-3881/aaa5b5}
{\href{https://arxiv.org/abs/1801.03502}{{arXiv:1801.03502}}}
{[astro-ph.EP]}
\end{barticle}
\endbibitem

%%% 66
\bibitem[\protect\citeauthoryear{{Anderson} et~al.}{2014}]{WASP69b_Discovery_2014}
\begin{barticle}
\bauthor{\bsnm{{Anderson}}, \binits{D.R.}},
\bauthor{\bsnm{{Collier Cameron}}, \binits{A.}},
\bauthor{\bsnm{{Delrez}}, \binits{L.}},
\bauthor{\bsnm{al.}}:
\batitle{{Three newly discovered sub-Jupiter-mass planets: WASP-69b and WASP-84b transit active K dwarfs and WASP-70Ab transits the evolved primary of a G4+K3 binary}}.
\bjtitle{\mnras}
\bvolume{445}(\bissue{2}),
\bfpage{1114}--\blpage{1129}
(\byear{2014})
\doiurl{10.1093/mnras/stu1737}
{\href{https://arxiv.org/abs/1310.5654}{{arXiv:1310.5654}}}
{[astro-ph.EP]}
\end{barticle}
\endbibitem

%%% 67
\bibitem[\protect\citeauthoryear{{Guilluy} et~al.}{2022}]{WASP69b_HighResSpectroscopy_2022}
\begin{barticle}
\bauthor{\bsnm{{Guilluy}}, \binits{G.}},
\bauthor{\bsnm{{Giacobbe}}, \binits{P.}},
\bauthor{\bsnm{{Carleo}}, \binits{I.}},
\bauthor{\bsnm{al.}}:
\batitle{{The GAPS Programme at TNG. XXXVIII. Five molecules in the atmosphere of the warm giant planet WASP-69b detected at high spectral resolution}}.
\bjtitle{\aap}
\bvolume{665},
\bfpage{104}
(\byear{2022})
\doiurl{10.1051/0004-6361/202243854}
{\href{https://arxiv.org/abs/2207.09760}{{arXiv:2207.09760}}}
{[astro-ph.EP]}
\end{barticle}
\endbibitem

%%% 68
\bibitem[\protect\citeauthoryear{{Tsiaras} et~al.}{2018}]{WASP69b_HSTspectrum_2018}
\begin{barticle}
\bauthor{\bsnm{{Tsiaras}}, \binits{A.}},
\bauthor{\bsnm{{Waldmann}}, \binits{I.P.}},
\bauthor{\bsnm{{Zingales}}, \binits{T.}},
\bauthor{\bsnm{al.}}:
\batitle{{A Population Study of Gaseous Exoplanets}}.
\bjtitle{\aj}
\bvolume{155}(\bissue{4}),
\bfpage{156}
(\byear{2018})
\doiurl{10.3847/1538-3881/aaaf75}
{\href{https://arxiv.org/abs/1704.05413}{{arXiv:1704.05413}}}
{[astro-ph.EP]}
\end{barticle}
\endbibitem

%%% 69
\bibitem[\protect\citeauthoryear{{Wong} et~al.}{2022}]{WASP80b_HSTspectrum_2022}
\begin{barticle}
\bauthor{\bsnm{{Wong}}, \binits{I.}},
\bauthor{\bsnm{{Chachan}}, \binits{Y.}},
\bauthor{\bsnm{{Knutson}}, \binits{H.A.}},
\bauthor{\bsnm{al.}}:
\batitle{{The Hubble PanCET Program: A Featureless Transmission Spectrum for WASP-29b and Evidence of Enhanced Atmospheric Metallicity on WASP-80b}}.
\bjtitle{\aj}
\bvolume{164}(\bissue{1}),
\bfpage{30}
(\byear{2022})
\doiurl{10.3847/1538-3881/ac7234}
{\href{https://arxiv.org/abs/2205.10765}{{arXiv:2205.10765}}}
{[astro-ph.EP]}
\end{barticle}
\endbibitem

%%% 70
\bibitem[\protect\citeauthoryear{{Carleo} et~al.}{2022}]{WASP80b_HighResSpectroscopy_2022}
\begin{barticle}
\bauthor{\bsnm{{Carleo}}, \binits{I.}},
\bauthor{\bsnm{{Giacobbe}}, \binits{P.}},
\bauthor{\bsnm{{Guilluy}}, \binits{G.}},
\bauthor{\bsnm{al.}}:
\batitle{{The GAPS Programme at TNG XXXIX. Multiple Molecular Species in the Atmosphere of the Warm Giant Planet WASP-80 b Unveiled at High Resolution with GIANO-B}}.
\bjtitle{\aj}
\bvolume{164}(\bissue{3}),
\bfpage{101}
(\byear{2022})
\doiurl{10.3847/1538-3881/ac80bf}
{\href{https://arxiv.org/abs/2207.09761}{{arXiv:2207.09761}}}
{[astro-ph.EP]}
\end{barticle}
\endbibitem

%%% 71
\bibitem[\protect\citeauthoryear{{Piaulet} et~al.}{2021}]{WASP107b_LowDensity_2021}
\begin{barticle}
\bauthor{\bsnm{{Piaulet}}, \binits{C.}},
\bauthor{\bsnm{{Benneke}}, \binits{B.}},
\bauthor{\bsnm{{Rubenzahl}}, \binits{R.A.}},
\bauthor{\bsnm{al.}}:
\batitle{{WASP-107b's Density Is Even Lower: A Case Study for the Physics of Planetary Gas Envelope Accretion and Orbital Migration}}.
\bjtitle{\aj}
\bvolume{161}(\bissue{2}),
\bfpage{70}
(\byear{2021})
\doiurl{10.3847/1538-3881/abcd3c}
{\href{https://arxiv.org/abs/2011.13444}{{arXiv:2011.13444}}}
{[astro-ph.EP]}
\end{barticle}
\endbibitem

%%% 72
\bibitem[\protect\citeauthoryear{{Kreidberg} et~al.}{2018}]{WASP107b_H2Odetection_2018}
\begin{barticle}
\bauthor{\bsnm{{Kreidberg}}, \binits{L.}},
\bauthor{\bsnm{{Line}}, \binits{M.R.}},
\bauthor{\bsnm{{Thorngren}}, \binits{D.}},
\bauthor{\bsnm{al.}}:
\batitle{{Water, High-altitude Condensates, and Possible Methane Depletion in the Atmosphere of the Warm Super-Neptune WASP-107b}}.
\bjtitle{\apjl}
\bvolume{858}(\bissue{1}),
\bfpage{6}
(\byear{2018})
\doiurl{10.3847/2041-8213/aabfce}
{\href{https://arxiv.org/abs/1709.08635}{{arXiv:1709.08635}}}
{[astro-ph.EP]}
\end{barticle}
\endbibitem

%%% 73
\bibitem[\protect\citeauthoryear{Mugnai et~al.}{2023}]{Mugnai2023}
\begin{barticle}
\bauthor{\bsnm{Mugnai}, \binits{L.V.}},
\bauthor{\bsnm{Bocchieri}, \binits{A.}},
\bauthor{\bsnm{Pascale}, \binits{E.}}:
\batitle{Exorad 2.0: The generic point source radiometric model}.
\bjtitle{Journal of Open Source Software}
\bvolume{8}(\bissue{89}),
\bfpage{5348}
(\byear{2023})
\doiurl{10.21105/joss.05348}
\end{barticle}
\endbibitem

%%% 74
\bibitem[\protect\citeauthoryear{{Mugnai} et~al.}{2025}]{Mugnai:2025}
\begin{barticle}
\bauthor{\bsnm{{Mugnai}}, \binits{L.V.}},
\bauthor{\bsnm{{Bocchieri}}, \binits{A.}},
\bauthor{\bsnm{{Pascale}}, \binits{E.}},
\bauthor{\bsnm{{Lorenzani}}, \binits{A.}},
\bauthor{\bsnm{{Papageorgiou}}, \binits{A.}}:
\batitle{{ExoSim 2: the new exoplanet observation simulator applied to the Ariel space mission}}.
\bjtitle{Experimental Astronomy}
\bvolume{59}(\bissue{1}),
\bfpage{9}
(\byear{2025})
\doiurl{10.1007/s10686-024-09976-2}
{\href{https://arxiv.org/abs/2501.12809}{{arXiv:2501.12809}}}
{[astro-ph.IM]}
\end{barticle}
\endbibitem

%%% 75
\bibitem[\protect\citeauthoryear{{Al-Refaie} et~al.}{2021}]{AlRefaie2021}
\begin{barticle}
\bauthor{\bsnm{{Al-Refaie}}, \binits{A.F.}},
\bauthor{\bsnm{{Changeat}}, \binits{Q.}},
\bauthor{\bsnm{{Waldmann}}, \binits{I.P.}},
\bauthor{\bsnm{{Tinetti}}, \binits{G.}}:
\batitle{{TauREx 3: A Fast, Dynamic, and Extendable Framework for Retrievals}}.
\bjtitle{\apj}
\bvolume{917}(\bissue{1}),
\bfpage{37}
(\byear{2021})
\doiurl{10.3847/1538-4357/ac0252}
{\href{https://arxiv.org/abs/1912.07759}{{arXiv:1912.07759}}}
{[astro-ph.IM]}
\end{barticle}
\endbibitem

%%% 76
\bibitem[\protect\citeauthoryear{{Benneke} et~al.}{2024}]{TOI-270d_JWST-NIRISS/SOSS+NIRSpec/G395H_CH4CO2H2O_Bjorn_2024arXiv}
\begin{botherref}
\oauthor{\bsnm{{Benneke}}, \binits{B.}},
\oauthor{\bsnm{{Roy}}, \binits{P.-A.}},
\oauthor{\bsnm{{Coulombe}}, \binits{L.-P.}},
\oauthor{\bsnm{{Radica}}, \binits{M.}},
\oauthor{\bsnm{{Piaulet}}, \binits{C.}},
\oauthor{\bsnm{{Ahrer}}, \binits{E.-M.}},
\oauthor{\bsnm{{Pierrehumbert}}, \binits{R.}},
\oauthor{\bsnm{{Krissansen-Totton}}, \binits{J.}},
\oauthor{\bsnm{{Schlichting}}, \binits{H.E.}},
\oauthor{\bsnm{{Hu}}, \binits{R.}},
\oauthor{\bsnm{{Yang}}, \binits{J.}},
\oauthor{\bsnm{{Christie}}, \binits{D.}},
\oauthor{\bsnm{{Thorngren}}, \binits{D.}},
\oauthor{\bsnm{{Young}}, \binits{E.D.}},
\oauthor{\bsnm{{Pelletier}}, \binits{S.}},
\oauthor{\bsnm{{Knutson}}, \binits{H.A.}},
\oauthor{\bsnm{{Miguel}}, \binits{Y.}},
\oauthor{\bsnm{{Evans-Soma}}, \binits{T.M.}},
\oauthor{\bsnm{{Dorn}}, \binits{C.}},
\oauthor{\bsnm{{Gagnebin}}, \binits{A.}},
\oauthor{\bsnm{{Fortney}}, \binits{J.J.}},
\oauthor{\bsnm{{Komacek}}, \binits{T.}},
\oauthor{\bsnm{{MacDonald}}, \binits{R.}},
\oauthor{\bsnm{{Raul}}, \binits{E.}},
\oauthor{\bsnm{{Cloutier}}, \binits{R.}},
\oauthor{\bsnm{{Acuna}}, \binits{L.}},
\oauthor{\bsnm{{Lafreni{\`e}re}}, \binits{D.}},
\oauthor{\bsnm{{Cadieux}}, \binits{C.}},
\oauthor{\bsnm{{Doyon}}, \binits{R.}},
\oauthor{\bsnm{{Welbanks}}, \binits{L.}},
\oauthor{\bsnm{{Allart}}, \binits{R.}}:
{JWST Reveals CH$_4$, CO$_2$, and H$_2$O in a Metal-rich Miscible Atmosphere on a Two-Earth-Radius Exoplanet}.
arXiv e-prints,
2403--03325
(2024)
\doiurl{10.48550/arXiv.2403.03325}
{\href{https://arxiv.org/abs/2403.03325}{{arXiv:2403.03325}}}
{[astro-ph.EP]}
\end{botherref}
\endbibitem

%%% 77
\bibitem[\protect\citeauthoryear{{Dyrek} et~al.}{2024}]{WASP-107b_JWSTNIRCam/MIRI_NoCH4_Dyrek_2024Natur}
\begin{barticle}
\bauthor{\bsnm{{Dyrek}}, \binits{A.}},
\bauthor{\bsnm{{Min}}, \binits{M.}},
\bauthor{\bsnm{{Decin}}, \binits{L.}},
\bauthor{\bsnm{{Bouwman}}, \binits{J.}},
\bauthor{\bsnm{{Crouzet}}, \binits{N.}},
\bauthor{\bsnm{{Molli{\`e}re}}, \binits{P.}},
\bauthor{\bsnm{{Lagage}}, \binits{P.-O.}},
\bauthor{\bsnm{{Konings}}, \binits{T.}},
\bauthor{\bsnm{{Tremblin}}, \binits{P.}},
\bauthor{\bsnm{{G{\"u}del}}, \binits{M.}},
\bauthor{\bsnm{{Pye}}, \binits{J.}},
\bauthor{\bsnm{{Waters}}, \binits{R.}},
\bauthor{\bsnm{{Henning}}, \binits{T.}},
\bauthor{\bsnm{{Vandenbussche}}, \binits{B.}},
\bauthor{\bsnm{{Ardevol Martinez}}, \binits{F.}},
\bauthor{\bsnm{{Argyriou}}, \binits{I.}},
\bauthor{\bsnm{{Ducrot}}, \binits{E.}},
\bauthor{\bsnm{{Heinke}}, \binits{L.}},
\bauthor{\bsnm{{van Looveren}}, \binits{G.}},
\bauthor{\bsnm{{Absil}}, \binits{O.}},
\bauthor{\bsnm{{Barrado}}, \binits{D.}},
\bauthor{\bsnm{{Baudoz}}, \binits{P.}},
\bauthor{\bsnm{{Boccaletti}}, \binits{A.}},
\bauthor{\bsnm{{Cossou}}, \binits{C.}},
\bauthor{\bsnm{{Coulais}}, \binits{A.}},
\bauthor{\bsnm{{Edwards}}, \binits{B.}},
\bauthor{\bsnm{{Gastaud}}, \binits{R.}},
\bauthor{\bsnm{{Glasse}}, \binits{A.}},
\bauthor{\bsnm{{Glauser}}, \binits{A.}},
\bauthor{\bsnm{{Greene}}, \binits{T.P.}},
\bauthor{\bsnm{{Kendrew}}, \binits{S.}},
\bauthor{\bsnm{{Krause}}, \binits{O.}},
\bauthor{\bsnm{{Lahuis}}, \binits{F.}},
\bauthor{\bsnm{{Mueller}}, \binits{M.}},
\bauthor{\bsnm{{Olofsson}}, \binits{G.}},
\bauthor{\bsnm{{Patapis}}, \binits{P.}},
\bauthor{\bsnm{{Rouan}}, \binits{D.}},
\bauthor{\bsnm{{Royer}}, \binits{P.}},
\bauthor{\bsnm{{Scheithauer}}, \binits{S.}},
\bauthor{\bsnm{{Waldmann}}, \binits{I.}},
\bauthor{\bsnm{{Whiteford}}, \binits{N.}},
\bauthor{\bsnm{{Colina}}, \binits{L.}},
\bauthor{\bsnm{{van Dishoeck}}, \binits{E.F.}},
\bauthor{\bsnm{{{\"O}stlin}}, \binits{G.}},
\bauthor{\bsnm{{Ray}}, \binits{T.P.}},
\bauthor{\bsnm{{Wright}}, \binits{G.}}:
\batitle{{SO$_{2}$, silicate clouds, but no CH$_{4}$ detected in a warm Neptune}}.
\bjtitle{\nat}
\bvolume{625}(\bissue{7993}),
\bfpage{51}--\blpage{54}
(\byear{2024})
\doiurl{10.1038/s41586-023-06849-0}
{\href{https://arxiv.org/abs/2311.12515}{{arXiv:2311.12515}}}
{[astro-ph.EP]}
\end{barticle}
\endbibitem

%%% 78
\bibitem[\protect\citeauthoryear{{Welbanks} et~al.}{2024}]{WASP-107b_JWSTNIRCam/MIRI_CH4_Welbanks_2024Natur}
\begin{barticle}
\bauthor{\bsnm{{Welbanks}}, \binits{L.}},
\bauthor{\bsnm{{Bell}}, \binits{T.J.}},
\bauthor{\bsnm{{Beatty}}, \binits{T.G.}},
\bauthor{\bsnm{{Line}}, \binits{M.R.}},
\bauthor{\bsnm{{Ohno}}, \binits{K.}},
\bauthor{\bsnm{{Fortney}}, \binits{J.J.}},
\bauthor{\bsnm{{Schlawin}}, \binits{E.}},
\bauthor{\bsnm{{Greene}}, \binits{T.P.}},
\bauthor{\bsnm{{Rauscher}}, \binits{E.}},
\bauthor{\bsnm{{McGill}}, \binits{P.}},
\bauthor{\bsnm{{Murphy}}, \binits{M.}},
\bauthor{\bsnm{{Parmentier}}, \binits{V.}},
\bauthor{\bsnm{{Tang}}, \binits{Y.}},
\bauthor{\bsnm{{Edelman}}, \binits{I.}},
\bauthor{\bsnm{{Mukherjee}}, \binits{S.}},
\bauthor{\bsnm{{Wiser}}, \binits{L.S.}},
\bauthor{\bsnm{{Lagage}}, \binits{P.-O.}},
\bauthor{\bsnm{{Dyrek}}, \binits{A.}},
\bauthor{\bsnm{{Arnold}}, \binits{K.E.}}:
\batitle{{A high internal heat flux and large core in a warm Neptune exoplanet}}.
\bjtitle{\nat}
\bvolume{630}(\bissue{8018}),
\bfpage{836}--\blpage{840}
(\byear{2024})
\doiurl{10.1038/s41586-024-07514-w}
{\href{https://arxiv.org/abs/2405.11018}{{arXiv:2405.11018}}}
{[astro-ph.EP]}
\end{barticle}
\endbibitem

%%% 79
\bibitem[\protect\citeauthoryear{{Welbanks} et~al.}{2025}]{Welbanks:2025}
\begin{botherref}
\oauthor{\bsnm{{Welbanks}}, \binits{L.}},
\oauthor{\bsnm{{Nixon}}, \binits{M.C.}},
\oauthor{\bsnm{{McGill}}, \binits{P.}},
\oauthor{\bsnm{al.}}:
{The Challenges of Detecting Gases in Exoplanet Atmospheres}.
arXiv e-prints,
2504--21788
(2025)
\doiurl{10.48550/arXiv.2504.21788}
{\href{https://arxiv.org/abs/2504.21788}{{arXiv:2504.21788}}}
{[astro-ph.EP]}
\end{botherref}
\endbibitem

%%% 80
\bibitem[\protect\citeauthoryear{{Guillot}}{2010}]{Guillot2010}
\begin{barticle}
\bauthor{\bsnm{{Guillot}}, \binits{T.}}:
\batitle{{On the radiative equilibrium of irradiated planetary atmospheres}}.
\bjtitle{\aap}
\bvolume{520},
\bfpage{27}
(\byear{2010})
\doiurl{10.1051/0004-6361/200913396}
{\href{https://arxiv.org/abs/1006.4702}{{arXiv:1006.4702}}}
{[astro-ph.EP]}
\end{barticle}
\endbibitem

%%% 81
\bibitem[\protect\citeauthoryear{{Ag{\'u}ndez} et~al.}{2012}]{Agundez2012}
\begin{barticle}
\bauthor{\bsnm{{Ag{\'u}ndez}}, \binits{M.}},
\bauthor{\bsnm{{Venot}}, \binits{O.}},
\bauthor{\bsnm{{Iro}}, \binits{N.}},
\bauthor{\bsnm{al.}}:
\batitle{{The impact of atmospheric circulation on the chemistry of the hot Jupiter HD 209458b}}.
\bjtitle{\aap}
\bvolume{548},
\bfpage{73}
(\byear{2012})
\doiurl{10.1051/0004-6361/201220365}
{\href{https://arxiv.org/abs/1210.6627}{{arXiv:1210.6627}}}
{[astro-ph.EP]}
\end{barticle}
\endbibitem

%%% 82
\bibitem[\protect\citeauthoryear{{Ag{\'u}ndez} et~al.}{2020}]{Agundez2020}
\begin{barticle}
\bauthor{\bsnm{{Ag{\'u}ndez}}, \binits{M.}},
\bauthor{\bsnm{{Mart{\'\i}nez}}, \binits{J.I.}},
\bauthor{\bsnm{{de Andres}}, \binits{P.L.}},
\bauthor{\bsnm{{Cernicharo}}, \binits{J.}},
\bauthor{\bsnm{{Mart{\'\i}n-Gago}}, \binits{J.A.}}:
\batitle{{Chemical equilibrium in AGB atmospheres: successes, failures, and prospects for small molecules, clusters, and condensates}}.
\bjtitle{\aap}
\bvolume{637},
\bfpage{59}
(\byear{2020})
\doiurl{10.1051/0004-6361/202037496}
{\href{https://arxiv.org/abs/2004.00519}{{arXiv:2004.00519}}}
{[astro-ph.SR]}
\end{barticle}
\endbibitem

%%% 83
\bibitem[\protect\citeauthoryear{Abel et~al.}{2011}]{Abel2011}
\begin{barticle}
\bauthor{\bsnm{Abel}, \binits{M.}},
\bauthor{\bsnm{Frommhold}, \binits{L.}},
\bauthor{\bsnm{Li}, \binits{X.}},
\bauthor{\bsnm{Hunt}, \binits{K.L.C.}}:
\batitle{Collision-induced absorption by h2 pairs: From hundreds to thousands of kelvin}.
\bjtitle{The Journal of Physical Chemistry A}
\bvolume{115}(\bissue{25}),
\bfpage{6805}--\blpage{6812}
(\byear{2011})
\doiurl{10.1021/jp109441f}
{\href{https://arxiv.org/abs/https://doi.org/10.1021/jp109441f}{{https://doi.org/10.1021/jp109441f}}}.
\bcomment{PMID: 21207941}
\end{barticle}
\endbibitem

%%% 84
\bibitem[\protect\citeauthoryear{{Fletcher} et~al.}{2018}]{Fletcher2018}
\begin{barticle}
\bauthor{\bsnm{{Fletcher}}, \binits{L.N.}},
\bauthor{\bsnm{{Gustafsson}}, \binits{M.}},
\bauthor{\bsnm{{Orton}}, \binits{G.S.}}:
\batitle{{Hydrogen Dimers in Giant-planet Infrared Spectra}}.
\bjtitle{\apjs}
\bvolume{235}(\bissue{1}),
\bfpage{24}
(\byear{2018})
\doiurl{10.3847/1538-4365/aaa07a}
{\href{https://arxiv.org/abs/1712.02813}{{arXiv:1712.02813}}}
{[astro-ph.EP]}
\end{barticle}
\endbibitem

%%% 85
\bibitem[\protect\citeauthoryear{Abel et~al.}{2012}]{Abel2012}
\begin{barticle}
\bauthor{\bsnm{Abel}, \binits{M.}},
\bauthor{\bsnm{Frommhold}, \binits{L.}},
\bauthor{\bsnm{Li}, \binits{X.}},
\bauthor{\bsnm{Hunt}, \binits{K.L.C.}}:
\batitle{Infrared absorption by collisional h2-he complexes at temperatures up to 9000 k and frequencies from 0 to 20,000 cm(-1)}.
\bjtitle{The Journal of Chemical Physics}
\bvolume{136}(\bissue{4}),
\bfpage{044319}
(\byear{2012})
\doiurl{10.1063/1.3676405}
{\href{https://arxiv.org/abs/https://doi.org/10.1063/1.3676405}{{https://doi.org/10.1063/1.3676405}}}
\end{barticle}
\endbibitem

%%% 86
\bibitem[\protect\citeauthoryear{{Yousefi} et~al.}{2018}]{Yousefi2018}
\begin{barticle}
\bauthor{\bsnm{{Yousefi}}, \binits{M.}},
\bauthor{\bsnm{{Bernath}}, \binits{P.F.}},
\bauthor{\bsnm{{Hodges}}, \binits{J.}},
\bauthor{\bsnm{al.}}:
\batitle{{A new line list for the A$^{2}${\ensuremath{\Sigma}}$^{+}$ -X$^{2}$ {\ensuremath{\Pi}} electronic transition of OH}}.
\bjtitle{\jqsrt}
\bvolume{217},
\bfpage{416}--\blpage{424}
(\byear{2018})
\doiurl{10.1016/j.jqsrt.2018.06.016}
\end{barticle}
\endbibitem

%%% 87
\bibitem[\protect\citeauthoryear{{Polyansky} et~al.}{2018}]{Polyansky2018}
\begin{barticle}
\bauthor{\bsnm{{Polyansky}}, \binits{O.L.}},
\bauthor{\bsnm{{Kyuberis}}, \binits{A.A.}},
\bauthor{\bsnm{{Zobov}}, \binits{N.F.}},
\bauthor{\bsnm{al.}}:
\batitle{{ExoMol molecular line lists XXX: a complete high-accuracy line list for water}}.
\bjtitle{\mnras}
\bvolume{480}(\bissue{2}),
\bfpage{2597}--\blpage{2608}
(\byear{2018})
\doiurl{10.1093/mnras/sty1877}
{\href{https://arxiv.org/abs/1807.04529}{{arXiv:1807.04529}}}
{[astro-ph.EP]}
\end{barticle}
\endbibitem

%%% 88
\bibitem[\protect\citeauthoryear{{Al-Refaie} et~al.}{2016}]{Al-Refaie2016}
\begin{barticle}
\bauthor{\bsnm{{Al-Refaie}}, \binits{A.F.}},
\bauthor{\bsnm{{Polyansky}}, \binits{O.L.}},
\bauthor{\bsnm{{Ovsyannikov}}, \binits{R.I.}},
\bauthor{\bsnm{al.}}:
\batitle{{ExoMol line lists - XV. A new hot line list for hydrogen peroxide}}.
\bjtitle{\mnras}
\bvolume{461}(\bissue{1}),
\bfpage{1012}--\blpage{1022}
(\byear{2016})
\doiurl{10.1093/mnras/stw1295}
{\href{https://arxiv.org/abs/1607.00498}{{arXiv:1607.00498}}}
{[astro-ph.EP]}
\end{barticle}
\endbibitem

%%% 89
\bibitem[\protect\citeauthoryear{{Gordon} et~al.}{2017}]{Gordon2017}
\begin{barticle}
\bauthor{\bsnm{{Gordon}}, \binits{I.E.}},
\bauthor{\bsnm{{Rothman}}, \binits{L.S.}},
\bauthor{\bsnm{{Hill}}, \binits{C.}},
\bauthor{\bsnm{al.}}:
\batitle{{The HITRAN2016 molecular spectroscopic database}}.
\bjtitle{\jqsrt}
\bvolume{203},
\bfpage{3}--\blpage{69}
(\byear{2017})
\doiurl{10.1016/j.jqsrt.2017.06.038}
\end{barticle}
\endbibitem

%%% 90
\bibitem[\protect\citeauthoryear{{Pavlyuchko} et~al.}{2015}]{Pavlyuchko2015}
\begin{barticle}
\bauthor{\bsnm{{Pavlyuchko}}, \binits{A.I.}},
\bauthor{\bsnm{{Yurchenko}}, \binits{S.N.}},
\bauthor{\bsnm{{Tennyson}}, \binits{J.}}:
\batitle{{A hybrid variational-perturbation calculation of the ro-vibrational spectrum of nitric acid}}.
\bjtitle{\jcp}
\bvolume{142}(\bissue{9}),
\bfpage{094309}
(\byear{2015})
\doiurl{10.1063/1.4913741}
{\href{https://arxiv.org/abs/1412.6744}{{arXiv:1412.6744}}}
{[physics.chem-ph]}
\end{barticle}
\endbibitem

%%% 91
\bibitem[\protect\citeauthoryear{{Barber} et~al.}{2014}]{Barber2014}
\begin{barticle}
\bauthor{\bsnm{{Barber}}, \binits{R.J.}},
\bauthor{\bsnm{{Strange}}, \binits{J.K.}},
\bauthor{\bsnm{{Hill}}, \binits{C.}},
\bauthor{\bsnm{al.}}:
\batitle{{ExoMol line lists - III. An improved hot rotation-vibration line list for HCN and HNC}}.
\bjtitle{\mnras}
\bvolume{437}(\bissue{2}),
\bfpage{1828}--\blpage{1835}
(\byear{2014})
\doiurl{10.1093/mnras/stt2011}
{\href{https://arxiv.org/abs/1311.1328}{{arXiv:1311.1328}}}
{[astro-ph.SR]}
\end{barticle}
\endbibitem

%%% 92
\bibitem[\protect\citeauthoryear{{Brooke} et~al.}{2014}]{Brooke2014}
\begin{barticle}
\bauthor{\bsnm{{Brooke}}, \binits{J.S.A.}},
\bauthor{\bsnm{{Ram}}, \binits{R.S.}},
\bauthor{\bsnm{{Western}}, \binits{C.M.}},
\bauthor{\bsnm{al.}}:
\batitle{{Einstein A Coefficients and Oscillator Strengths for the A $^{2}${\ensuremath{\Pi}}-X $^{2}${\ensuremath{\Sigma}}$^{+}$ (Red) and B$^{2}${\ensuremath{\Sigma}}$^{+}$-X$^{2}${\ensuremath{\Sigma}}$^{+}$ (Violet) Systems and Rovibrational Transitions in the X $^{2}${\ensuremath{\Sigma}}$^{+}$ State of CN}}.
\bjtitle{\apjs}
\bvolume{210}(\bissue{2}),
\bfpage{23}
(\byear{2014})
\doiurl{10.1088/0067-0049/210/2/23}
\end{barticle}
\endbibitem

%%% 93
\bibitem[\protect\citeauthoryear{{Fernando} et~al.}{2018}]{Fernando2018}
\begin{barticle}
\bauthor{\bsnm{{Fernando}}, \binits{A.M.}},
\bauthor{\bsnm{{Bernath}}, \binits{P.F.}},
\bauthor{\bsnm{{Hodges}}, \binits{J.N.}},
\bauthor{\bsnm{al.}}:
\batitle{{A new linelist for the A$^{3}${\ensuremath{\Pi}}-X$^{3}${\ensuremath{\Sigma}}$^{-}$ transition of the NH free radical}}.
\bjtitle{\jqsrt}
\bvolume{217},
\bfpage{29}--\blpage{34}
(\byear{2018})
\doiurl{10.1016/j.jqsrt.2018.05.021}
\end{barticle}
\endbibitem

%%% 94
\bibitem[\protect\citeauthoryear{{Coles} et~al.}{2019}]{Coles2019}
\begin{barticle}
\bauthor{\bsnm{{Coles}}, \binits{P.A.}},
\bauthor{\bsnm{{Yurchenko}}, \binits{S.N.}},
\bauthor{\bsnm{{Tennyson}}, \binits{J.}}:
\batitle{{ExoMol molecular line lists - XXXV. A rotation-vibration line list for hot ammonia}}.
\bjtitle{\mnras}
\bvolume{490}(\bissue{4}),
\bfpage{4638}--\blpage{4647}
(\byear{2019})
\doiurl{10.1093/mnras/stz2778}
{\href{https://arxiv.org/abs/1911.10369}{{arXiv:1911.10369}}}
{[astro-ph.SR]}
\end{barticle}
\endbibitem

%%% 95
\bibitem[\protect\citeauthoryear{{Al-Refaie} et~al.}{2015}]{Al-Refaie2015}
\begin{barticle}
\bauthor{\bsnm{{Al-Refaie}}, \binits{A.F.}},
\bauthor{\bsnm{{Yachmenev}}, \binits{A.}},
\bauthor{\bsnm{{Tennyson}}, \binits{J.}},
\bauthor{\bsnm{al.}}:
\batitle{{ExoMol line lists - VIII. A variationally computed line list for hot formaldehyde}}.
\bjtitle{\mnras}
\bvolume{448}(\bissue{2}),
\bfpage{1704}--\blpage{1714}
(\byear{2015})
\doiurl{10.1093/mnras/stv091}
{\href{https://arxiv.org/abs/1506.00172}{{arXiv:1506.00172}}}
{[astro-ph.GA]}
\end{barticle}
\endbibitem

%%% 96
\bibitem[\protect\citeauthoryear{{Li} et~al.}{2015}]{Li2015}
\begin{barticle}
\bauthor{\bsnm{{Li}}, \binits{G.}},
\bauthor{\bsnm{{Gordon}}, \binits{I.E.}},
\bauthor{\bsnm{{Rothman}}, \binits{L.S.}},
\bauthor{\bsnm{al.}}:
\batitle{{Rovibrational Line Lists for Nine Isotopologues of the CO Molecule in the X $^{1}${\ensuremath{\Sigma}}$^{+}$ Ground Electronic State}}.
\bjtitle{\apjs}
\bvolume{216}(\bissue{1}),
\bfpage{15}
(\byear{2015})
\doiurl{10.1088/0067-0049/216/1/15}
\end{barticle}
\endbibitem

%%% 97
\bibitem[\protect\citeauthoryear{{Rothman} et~al.}{2010}]{Rothman2010}
\begin{barticle}
\bauthor{\bsnm{{Rothman}}, \binits{L.S.}},
\bauthor{\bsnm{{Gordon}}, \binits{I.E.}},
\bauthor{\bsnm{{Barber}}, \binits{R.J.}},
\bauthor{\bsnm{al.}}:
\batitle{{HITEMP, the high-temperature molecular spectroscopic database}}.
\bjtitle{\jqsrt}
\bvolume{111},
\bfpage{2139}--\blpage{2150}
(\byear{2010})
\doiurl{10.1016/j.jqsrt.2010.05.001}
\end{barticle}
\endbibitem

%%% 98
\bibitem[\protect\citeauthoryear{{Masseron} et~al.}{2014}]{Masseron2014}
\begin{barticle}
\bauthor{\bsnm{{Masseron}}, \binits{T.}},
\bauthor{\bsnm{{Plez}}, \binits{B.}},
\bauthor{\bsnm{{Van Eck}}, \binits{S.}},
\bauthor{\bsnm{al.}}:
\batitle{{CH in stellar atmospheres: an extensive linelist}}.
\bjtitle{\aap}
\bvolume{571},
\bfpage{47}
(\byear{2014})
\doiurl{10.1051/0004-6361/201423956}
{\href{https://arxiv.org/abs/1410.4005}{{arXiv:1410.4005}}}
{[astro-ph.SR]}
\end{barticle}
\endbibitem

%%% 99
\bibitem[\protect\citeauthoryear{{Adam} et~al.}{2019}]{Adam2019}
\begin{barticle}
\bauthor{\bsnm{{Adam}}, \binits{A.Y.}},
\bauthor{\bsnm{{Yachmenev}}, \binits{A.}},
\bauthor{\bsnm{{Yurchenko}}, \binits{S.N.}},
\bauthor{\bsnm{al.}}:
\batitle{{Variationally Computed IR Line List for the Methyl Radical CH3}}.
\bjtitle{Journal of Physical Chemistry A}
\bvolume{123}(\bissue{22}),
\bfpage{4755}--\blpage{4763}
(\byear{2019})
\doiurl{10.1021/acs.jpca.9b02919}
{\href{https://arxiv.org/abs/1905.05504}{{arXiv:1905.05504}}}
{[physics.chem-ph]}
\end{barticle}
\endbibitem

%%% 100
\bibitem[\protect\citeauthoryear{{Yurchenko} et~al.}{2017}]{Yurchenko2017}
\begin{barticle}
\bauthor{\bsnm{{Yurchenko}}, \binits{S.N.}},
\bauthor{\bsnm{{Amundsen}}, \binits{D.S.}},
\bauthor{\bsnm{{Tennyson}}, \binits{J.}},
\bauthor{\bsnm{al.}}:
\batitle{{A hybrid line list for CH$_{4}$ and hot methane continuum}}.
\bjtitle{\aap}
\bvolume{605},
\bfpage{95}
(\byear{2017})
\doiurl{10.1051/0004-6361/201731026}
{\href{https://arxiv.org/abs/1706.05724}{{arXiv:1706.05724}}}
{[astro-ph.EP]}
\end{barticle}
\endbibitem

%%% 101
\bibitem[\protect\citeauthoryear{{Chubb} et~al.}{2020}]{Chubb2020}
\begin{barticle}
\bauthor{\bsnm{{Chubb}}, \binits{K.L.}},
\bauthor{\bsnm{{Tennyson}}, \binits{J.}},
\bauthor{\bsnm{{Yurchenko}}, \binits{S.N.}}:
\batitle{{ExoMol molecular line lists - XXXVII. Spectra of acetylene}}.
\bjtitle{\mnras}
\bvolume{493}(\bissue{2}),
\bfpage{1531}--\blpage{1545}
(\byear{2020})
\doiurl{10.1093/mnras/staa229}
{\href{https://arxiv.org/abs/2001.04550}{{arXiv:2001.04550}}}
{[astro-ph.SR]}
\end{barticle}
\endbibitem

%%% 102
\bibitem[\protect\citeauthoryear{{Mant} et~al.}{2018}]{Mant2018}
\begin{barticle}
\bauthor{\bsnm{{Mant}}, \binits{B.P.}},
\bauthor{\bsnm{{Yachmenev}}, \binits{A.}},
\bauthor{\bsnm{{Tennyson}}, \binits{J.}},
\bauthor{\bsnm{al.}}:
\batitle{{ExoMol molecular line lists - XXVII. Spectra of C$_{2}$H$_{4}$}}.
\bjtitle{\mnras}
\bvolume{478}(\bissue{3}),
\bfpage{3220}--\blpage{3232}
(\byear{2018})
\doiurl{10.1093/mnras/sty1239}
{\href{https://arxiv.org/abs/1806.03469}{{arXiv:1806.03469}}}
{[astro-ph.EP]}
\end{barticle}
\endbibitem

%%% 103
\bibitem[\protect\citeauthoryear{{Hargreaves} et~al.}{2019}]{Hargreaves2019}
\begin{barticle}
\bauthor{\bsnm{{Hargreaves}}, \binits{R.J.}},
\bauthor{\bsnm{{Gordon}}, \binits{I.E.}},
\bauthor{\bsnm{{Rothman}}, \binits{L.S.}},
\bauthor{\bsnm{al.}}:
\batitle{{Spectroscopic line parameters of NO, NO$_{2}$, and N$_{2}$O for the HITEMP database}}.
\bjtitle{\jqsrt}
\bvolume{232},
\bfpage{35}--\blpage{53}
(\byear{2019})
\doiurl{10.1016/j.jqsrt.2019.04.040}
{\href{https://arxiv.org/abs/1904.02636}{{arXiv:1904.02636}}}
{[astro-ph.EP]}
\end{barticle}
\endbibitem

%%% 104
\bibitem[\protect\citeauthoryear{{Feng} et~al.}{2018}]{Feng2018}
\begin{barticle}
\bauthor{\bsnm{{Feng}}, \binits{Y.K.}},
\bauthor{\bsnm{{Robinson}}, \binits{T.D.}},
\bauthor{\bsnm{{Fortney}}, \binits{J.J.}},
\bauthor{\bsnm{al.}}:
\batitle{{Characterizing Earth Analogs in Reflected Light: Atmospheric Retrieval Studies for Future Space Telescopes}}.
\bjtitle{\aj}
\bvolume{155}(\bissue{5}),
\bfpage{200}
(\byear{2018})
\doiurl{10.3847/1538-3881/aab95c}
{\href{https://arxiv.org/abs/1803.06403}{{arXiv:1803.06403}}}
{[astro-ph.EP]}
\end{barticle}
\endbibitem

%%% 105
\bibitem[\protect\citeauthoryear{{Changeat} et~al.}{2019}]{Changeat2019}
\begin{barticle}
\bauthor{\bsnm{{Changeat}}, \binits{Q.}},
\bauthor{\bsnm{{Edwards}}, \binits{B.}},
\bauthor{\bsnm{{Waldmann}}, \binits{I.P.}},
\bauthor{\bsnm{{Tinetti}}, \binits{G.}}:
\batitle{{Toward a More Complex Description of Chemical Profiles in Exoplanet Retrievals: A Two-layer Parameterization}}.
\bjtitle{\apj}
\bvolume{886}(\bissue{1}),
\bfpage{39}
(\byear{2019})
\doiurl{10.3847/1538-4357/ab4a14}
{\href{https://arxiv.org/abs/1903.11180}{{arXiv:1903.11180}}}
{[astro-ph.EP]}
\end{barticle}
\endbibitem

%%% 106
\bibitem[\protect\citeauthoryear{{Feroz} et~al.}{2009}]{Feroz2009}
\begin{barticle}
\bauthor{\bsnm{{Feroz}}, \binits{F.}},
\bauthor{\bsnm{{Hobson}}, \binits{M.P.}},
\bauthor{\bsnm{{Bridges}}, \binits{M.}}:
\batitle{{MULTINEST: an efficient and robust Bayesian inference tool for cosmology and particle physics}}.
\bjtitle{\mnras}
\bvolume{398}(\bissue{4}),
\bfpage{1601}--\blpage{1614}
(\byear{2009})
\doiurl{10.1111/j.1365-2966.2009.14548.x}
{\href{https://arxiv.org/abs/0809.3437}{{arXiv:0809.3437}}}
{[astro-ph]}
\end{barticle}
\endbibitem

%%% 107
\bibitem[\protect\citeauthoryear{{Buchner}}{2023}]{Buchner2023}
\begin{barticle}
\bauthor{\bsnm{{Buchner}}, \binits{J.}}:
\batitle{{Nested Sampling Methods}}.
\bjtitle{Statistics Surveys}
\bvolume{17},
\bfpage{169}--\blpage{215}
(\byear{2023})
\doiurl{10.1214/23-SS144}
{\href{https://arxiv.org/abs/2101.09675}{{arXiv:2101.09675}}}
{[stat.CO]}
\end{barticle}
\endbibitem

%%% 108
\bibitem[\protect\citeauthoryear{{Bocchieri} et~al.}{2024}]{Bocchieri2024}
\begin{botherref}
\oauthor{\bsnm{{Bocchieri}}, \binits{A.}},
\oauthor{\bsnm{{Booth}}, \binits{L.}},
\oauthor{\bsnm{{Mugnai}}, \binits{L.V.}}:
{abocchieri/exploring-synergies-ariel-twinkle- plots: v1.0.0}.
Zenodo
(2024).
\doiurl{10.5281/zenodo.10563853} .
\url{https://doi.org/10.5281/zenodo.10563853}
\end{botherref}
\endbibitem

%%% 109
\bibitem[\protect\citeauthoryear{{Wakeford} et~al.}{2018}]{MZTrend_Wakeford2018}
\begin{barticle}
\bauthor{\bsnm{{Wakeford}}, \binits{H.R.}},
\bauthor{\bsnm{{Sing}}, \binits{D.K.}},
\bauthor{\bsnm{{Deming}}, \binits{D.}},
\bauthor{\bsnm{al.}}:
\batitle{{The Complete Transmission Spectrum of WASP-39b with a Precise Water Constraint}}.
\bjtitle{\aj}
\bvolume{155}(\bissue{1}),
\bfpage{29}
(\byear{2018})
\doiurl{10.3847/1538-3881/aa9e4e}
{\href{https://arxiv.org/abs/1711.10529}{{arXiv:1711.10529}}}
{[astro-ph.EP]}
\end{barticle}
\endbibitem

%%% 110
\bibitem[\protect\citeauthoryear{{Welbanks} et~al.}{2019}]{MZTrend_Welbanks2019}
\begin{barticle}
\bauthor{\bsnm{{Welbanks}}, \binits{L.}},
\bauthor{\bsnm{{Madhusudhan}}, \binits{N.}},
\bauthor{\bsnm{{Allard}}, \binits{N.F.}},
\bauthor{\bsnm{al.}}:
\batitle{{Mass-Metallicity Trends in Transiting Exoplanets from Atmospheric Abundances of H$_{2}$O, Na, and K}}.
\bjtitle{\apjl}
\bvolume{887}(\bissue{1}),
\bfpage{20}
(\byear{2019})
\doiurl{10.3847/2041-8213/ab5a89}
{\href{https://arxiv.org/abs/1912.04904}{{arXiv:1912.04904}}}
{[astro-ph.EP]}
\end{barticle}
\endbibitem

%%% 111
\bibitem[\protect\citeauthoryear{{Edwards} et~al.}{2023}]{NoMZTrend_Edwards2022_HST70}
\begin{barticle}
\bauthor{\bsnm{{Edwards}}, \binits{B.}},
\bauthor{\bsnm{{Changeat}}, \binits{Q.}},
\bauthor{\bsnm{{Tsiaras}}, \binits{A.}},
\bauthor{\bsnm{al.}}:
\batitle{{Exploring the Ability of Hubble Space Telescope WFC3 G141 to Uncover Trends in Populations of Exoplanet Atmospheres through a Homogeneous Transmission Survey of 70 Gaseous Planets}}.
\bjtitle{\apjs}
\bvolume{269}(\bissue{1}),
\bfpage{31}
(\byear{2023})
\doiurl{10.3847/1538-4365/ac9f1a}
{\href{https://arxiv.org/abs/2211.00649}{{arXiv:2211.00649}}}
{[astro-ph.EP]}
\end{barticle}
\endbibitem

%%% 112
\bibitem[\protect\citeauthoryear{{Edwards} and {Tinetti}}{2022}]{ArielMRS_Edwards2022}
\begin{barticle}
\bauthor{\bsnm{{Edwards}}, \binits{B.}},
\bauthor{\bsnm{{Tinetti}}, \binits{G.}}:
\batitle{{The Ariel Target List: The Impact of TESS and the Potential for Characterizing Multiple Planets within a System}}.
\bjtitle{\aj}
\bvolume{164}(\bissue{1}),
\bfpage{15}
(\byear{2022})
\doiurl{10.3847/1538-3881/ac6bf9}
{\href{https://arxiv.org/abs/2205.05073}{{arXiv:2205.05073}}}
{[astro-ph.EP]}
\end{barticle}
\endbibitem

%%% 113
\bibitem[\protect\citeauthoryear{{Yoshida} et~al.}{2023}]{TOI-1420b_TESS_LowDensity_2023}
\begin{barticle}
\bauthor{\bsnm{{Yoshida}}, \binits{S.}},
\bauthor{\bsnm{{Vissapragada}}, \binits{S.}},
\bauthor{\bsnm{{Latham}}, \binits{D.W.}},
\bauthor{\bsnm{al.}}:
\batitle{{TESS Spots a Super-puff: The Remarkably Low Density of TOI-1420b}}.
\bjtitle{\aj}
\bvolume{166}(\bissue{5}),
\bfpage{181}
(\byear{2023})
\doiurl{10.3847/1538-3881/acf858}
{\href{https://arxiv.org/abs/2309.09945}{{arXiv:2309.09945}}}
{[astro-ph.EP]}
\end{barticle}
\endbibitem

%%% 114
\bibitem[\protect\citeauthoryear{{Hobson} et~al.}{2023}]{TOI-199b_TESS_CharacterisedwarmSaturn_2023}
\begin{barticle}
\bauthor{\bsnm{{Hobson}}, \binits{M.J.}},
\bauthor{\bsnm{{Trifonov}}, \binits{T.}},
\bauthor{\bsnm{{Henning}}, \binits{T.}},
\bauthor{\bsnm{al.}}:
\batitle{{TOI-199 b: A Well-characterized 100 day Transiting Warm Giant Planet with TTVs Seen from Antarctica}}.
\bjtitle{\aj}
\bvolume{166}(\bissue{5}),
\bfpage{201}
(\byear{2023})
\doiurl{10.3847/1538-3881/acfc1d}
{\href{https://arxiv.org/abs/2309.14915}{{arXiv:2309.14915}}}
{[astro-ph.EP]}
\end{barticle}
\endbibitem

%%% 115
\bibitem[\protect\citeauthoryear{{Powers} et~al.}{2023}]{Mdwarf_TOI-3785b_Powers_2023}
\begin{barticle}
\bauthor{\bsnm{{Powers}}, \binits{L.C.}},
\bauthor{\bsnm{{Libby-Roberts}}, \binits{J.}},
\bauthor{\bsnm{{Lin}}, \binits{A.S.J.}},
\bauthor{\bsnm{al.}}:
\batitle{{TOI-3785 b: A Low-density Neptune Orbiting an M2-dwarf Star}}.
\bjtitle{\aj}
\bvolume{166}(\bissue{2}),
\bfpage{44}
(\byear{2023})
\doiurl{10.3847/1538-3881/acd8bf}
{\href{https://arxiv.org/abs/2304.04730}{{arXiv:2304.04730}}}
{[astro-ph.EP]}
\end{barticle}
\endbibitem

%%% 116
\bibitem[\protect\citeauthoryear{{Harris} et~al.}{2023}]{Mdwarf_TOI-904bc_Harris_2023}
\begin{barticle}
\bauthor{\bsnm{{Harris}}, \binits{M.}},
\bauthor{\bsnm{{Dragomir}}, \binits{D.}},
\bauthor{\bsnm{{Mireles}}, \binits{I.}},
\bauthor{\bsnm{al.}}:
\batitle{{Separated Twins or Just Siblings? A Multiplanet System around an M Dwarf Including a Cool Sub-Neptune}}.
\bjtitle{\apjl}
\bvolume{959}(\bissue{1}),
\bfpage{1}
(\byear{2023})
\doiurl{10.3847/2041-8213/ad037d}
{\href{https://arxiv.org/abs/2310.15118}{{arXiv:2310.15118}}}
{[astro-ph.EP]}
\end{barticle}
\endbibitem

%%% 117
\bibitem[\protect\citeauthoryear{{Han} et~al.}{2024}]{Mdwarf_TOI-5344b_Han_2023}
\begin{barticle}
\bauthor{\bsnm{{Han}}, \binits{T.}},
\bauthor{\bsnm{{Robertson}}, \binits{P.}},
\bauthor{\bsnm{{Kanodia}}, \binits{S.}},
\bauthor{\bsnm{al.}}:
\batitle{{TOI-5344 b: A Saturn-like Planet Orbiting a Super-solar Metallicity M0 Dwarf}}.
\bjtitle{\aj}
\bvolume{167}(\bissue{1}),
\bfpage{4}
(\byear{2024})
\doiurl{10.3847/1538-3881/ad09c2}
{\href{https://arxiv.org/abs/2310.20634}{{arXiv:2310.20634}}}
{[astro-ph.EP]}
\end{barticle}
\endbibitem

%%% 118
\bibitem[\protect\citeauthoryear{{Osborn} et~al.}{2023}]{HIP9618_TESS+CHEOPS_2023}
\begin{barticle}
\bauthor{\bsnm{{Osborn}}, \binits{H.P.}},
\bauthor{\bsnm{{Nowak}}, \binits{G.}},
\bauthor{\bsnm{{H{\'e}brard}}, \binits{G.}},
\bauthor{\bsnm{al.}}:
\batitle{{Two warm Neptunes transiting HIP 9618 revealed by TESS and Cheops}}.
\bjtitle{\mnras}
\bvolume{523}(\bissue{2}),
\bfpage{3069}--\blpage{3089}
(\byear{2023})
\doiurl{10.1093/mnras/stad1319}
{\href{https://arxiv.org/abs/2306.04450}{{arXiv:2306.04450}}}
{[astro-ph.EP]}
\end{barticle}
\endbibitem

%%% 119
\bibitem[\protect\citeauthoryear{{Ulmer-Moll} et~al.}{2023}]{TOI-5678b_TESS+CHEOPS+HARPS_2023}
\begin{barticle}
\bauthor{\bsnm{{Ulmer-Moll}}, \binits{S.}},
\bauthor{\bsnm{{Osborn}}, \binits{H.P.}},
\bauthor{\bsnm{{Tuson}}, \binits{A.}},
\bauthor{\bsnm{al.}}:
\batitle{{TOI-5678b: A 48-day transiting Neptune-mass planet characterized with CHEOPS and HARPS}}.
\bjtitle{\aap}
\bvolume{674},
\bfpage{43}
(\byear{2023})
\doiurl{10.1051/0004-6361/202245478}
{\href{https://arxiv.org/abs/2306.04295}{{arXiv:2306.04295}}}
{[astro-ph.EP]}
\end{barticle}
\endbibitem

%%% 120
\bibitem[\protect\citeauthoryear{{Mireles} et~al.}{2023}]{TOI-4600bc_TESS+ground-based_2023}
\begin{barticle}
\bauthor{\bsnm{{Mireles}}, \binits{I.}},
\bauthor{\bsnm{{Dragomir}}, \binits{D.}},
\bauthor{\bsnm{{Osborn}}, \binits{H.P.}},
\bauthor{\bsnm{al.}}:
\batitle{{TOI-4600 b and c: Two Long-period Giant Planets Orbiting an Early K Dwarf}}.
\bjtitle{\apjl}
\bvolume{954}(\bissue{1}),
\bfpage{15}
(\byear{2023})
\doiurl{10.3847/2041-8213/aceb69}
{\href{https://arxiv.org/abs/2308.15572}{{arXiv:2308.15572}}}
{[astro-ph.EP]}
\end{barticle}
\endbibitem

\end{thebibliography}

%%%%%%%%%%%%%%%%% APPENDICES %%%%%%%%%%%%%%%%%%%%%
% \appendix

\end{document}